# The Pierre Auger Observatory IV: Operation and Monitoring

THE PIERRE AUGER COLLABORATION

*Observatorio Pierre Auger, Av. San Martín Norte 304, 5613 Malargüe, Argentina*

# The Pierre Auger Collaboration

P. Abreu[74], M. Aglietta[57], E.J. Ahn[93], I.F.M. Albuquerque[19], D. Allard[33], I. Allekotte[1], J. Allen[96], P. Allison[98], J. Alvarez Castillo[67], J. Alvarez-Muñiz[84], M. Ambrosio[50], A. Aminaei[68], L. Anchordoqui[109], S. Andringa[74], T. Antičić[27], A. Anzalone[56], C. Aramo[50], E. Arganda[81], F. Arqueros[81], H. Asorey[1], P. Assis[74], J. Aublin[35], M. Ave[41], M. Avenier[36], G. Avila[12], T. Bäcker[45], M. Balzer[40], K.B. Barber[13], A.F. Barbosa[16], R. Bardenet[34], S.L.C. Barroso[22], B. Baughman[98], J. Bäuml[39], J.J. Beatty[98], B.R. Becker[106], K.H. Becker[38], A. Bellétoile[37], J.A. Bellido[13], S. BenZvi[108], C. Berat[36], X. Bertou[1], P.L. Biermann[42], P. Billoir[35], F. Blanco[81], M. Blanco[82], C. Bleve[38], H. Blümer[41, 39], M. Boháčová[29, 101], D. Boncioli[51], C. Bonifazi[25, 35], R. Bonino[57], N. Borodai[72], J. Brack[91], P. Brogueira[74], W.C. Brown[92], R. Bruijn[87], P. Buchholz[45], A. Bueno[83], R.E. Burton[89], K.S. Caballero-Mora[99], L. Caramete[42], R. Caruso[52], A. Castellina[57], O. Catalano[56], G. Cataldi[49], L. Cazon[74], R. Cester[53], J. Chauvin[36], S.H. Cheng[99], A. Chiavassa[57], J.A. Chinellato[20], A. Chou[93, 96], J. Chudoba[29], R.W. Clay[13], M.R. Coluccia[49], R. Conceição[74], F. Contreras[11], H. Cook[87], M.J. Cooper[13], J. Coppens[68, 70], A. Cordier[34], U. Cotti[66], S. Coutu[99], C.E. Covault[89], A. Creusot[33, 79], A. Criss[99], J. Cronin[101], A. Curutiu[42], S. Dagoret-Campagne[34], R. Dallier[37], S. Dasso[8, 4], K. Daumiller[39], B.R. Dawson[13], R.M. de Almeida[26], M. De Domenico[52], C. De Donato[67, 48], S.J. de Jong[68, 70], G. De La Vega[10], W.J.M. de Mello Junior[20], J.R.T. de Mello Neto[25], I. De Mitri[49], V. de Souza[18], K.D. de Vries[69], G. Decerprit[33], L. del Peral[82], O. Deligny[32], H. Dembinski[41], N. Dhital[95], C. Di Giulio[47, 51], J.C. Diaz[95], M.L. Díaz Castro[17], P.N. Diep[110], C. Dobrigkeit [20], W. Docters[69], J.C. D'Olivo[67], P.N. Dong[110, 32], A. Dorofeev[91], J.C. dos Anjos[16], M.T. Dova[7], D. D'Urso[50], I. Dutan[42], J. Ebr[29], R. Engel[39], M. Erdmann[43], C.O. Escobar[20], A. Etchegoyen[2], P. Facal San Luis[101], I. Fajardo Tapia[67], H. Falcke[68, 71], G. Farrar[96], A.C. Fauth[20], N. Fazzini[93], A.P. Ferguson[89], A. Ferrero[2], B. Fick[95], A. Filevich[2], A. Filipčič[78, 79], S. Fliescher[43], C.E. Fracchiolla[91], E.D. Fraenkel[69], U. Fröhlich[45], B. Fuchs[16], R. Gaior[35], R.F. Gamarra[2], S. Gambetta[46], B. García[10], D. García Gámez[83], D. Garcia-Pinto[81], A. Gascon[83], H. Gemmeke[40], K. Gesterling[106], P.L. Ghia[35, 57], U. Giaccari[49], M. Giller[73], H. Glass[93], M.S. Gold[106], G. Golup[1], F. Gomez Albarracin[7], M. Gómez Berisso[1], P. Gonçalves[74], D. Gonzalez[41], J.G. Gonzalez[41], B. Gookin[91], D. Góra[41, 72], A. Gorgi[57], P. Gouffon[19], S.R. Gozzini[87], E. Grashorn[98], S. Grebe[68, 70], N. Griffith[98], M. Grigat[43], A.F. Grillo[58], Y. Guardincerri[4], F. Guarino[50], G.P. Guedes[21], A. Guzman[67], J.D. Hague[106], P. Hansen[7], D. Harari[1], S. Harmsma[69, 70], J.L. Harton[91], A. Haungs[39], T. Hebbeker[43], D. Heck[39], A.E. Herve[13], C. Hojvat[93], N. Hollon[101], V.C. Holmes[13], P. Homola[72], J.R. Hörandel[68], A. Horneffer[68], M. Hrabovský[30, 29], T. Huege[39], A. Insolia[52], F. Ionita[101], A. Italiano[52], C. Jarne[7], S. Jiraskova[68], M. Josebachuili[2], K. Kadija[27], K.-H. Kampert[38], P. Karhan[28], P. Kasper[93], B. Kégl[34], B. Keilhauer[39], A. Keivani[94], J.L. Kelley[68], E. Kemp[20], R.M. Kieckhafer[95], H.O. Klages[39], M. Kleifges[40], J. Kleinfeller[39], J. Knapp[87], D.-H. Koang[36], K. Kotera[101], N. Krohm[38], O. Krömer[40], D. Kruppke-Hansen[38], F. Kuehn[93], D. Kuempel[38], J.K. Kulbartz[44], N. Kunka[40], G. La Rosa[56], C. Lachaud[33], P. Lautridou[37], M.S.A.B. Leão[24], D. Lebrun[36], P. Lebrun[93], M.A. Leigui de Oliveira[24], A. Lemiere[32], A. Letessier-Selvon[35], I. Lhenry-Yvon[32], K. Link[41], R. López[63], A. Lopez Agüera[84], K. Louedec[34], J. Lozano Bahilo[83], A. Lucero[2, 57], M. Ludwig[41], H. Lyberis[32], M.C. Maccarone[56], C. Macolino[35], S. Maldera[57], D. Mandat[29], P. Mantsch[93], A.G. Mariazzi[7], J. Marin[11, 57], V. Marin[37], I.C. Maris[35], H.R. Marquez Falcon[66], G. Marsella[54], D. Martello[49], L. Martin[37], H. Martinez[64], O. Martínez Bravo[63],

H.J. Mathes[39], J. Matthews[94,100], J.A.J. Matthews[106], G. Matthiae[51], D. Maurizio[53], P.O. Mazur[93], G. Medina-Tanco[67], M. Melissas[41], D. Melo[2,53], E. Menichetti[53], A. Menshikov[40], P. Mertsch[85], C. Meurer[43], S. Mićanović[27], M.I. Micheletti[9], W. Miller[106], L. Miramonti[48], S. Mollerach[1], M. Monasor[101], D. Monnier Ragaigne[34], F. Montanet[36], B. Morales[67], C. Morello[57], E. Moreno[63], J.C. Moreno[7], C. Morris[98], M. Mostafá[91], C.A. Moura[24,50], S. Mueller[39], M.A. Muller[20], G. Müller[43], M. Münchmeyer[35], R. Mussa[53], G. Navarra[57,†], J.L. Navarro[83], S. Navas[83], P. Necesal[29], L. Nellen[67], A. Nelles[68,70], J. Neuser[38], P.T. Nhung[110], L. Niemietz[38], N. Nierstenhoefer[38], D. Nitz[95], D. Nosek[28], L. Nožka[29], M. Nyklicek[29], J. Oehlschläger[39], A. Olinto[101], V.M. Olmos-Gilbaja[84], M. Ortiz[81], N. Pacheco[82], D. Pakk Selmi-Dei[20], M. Palatka[29], J. Pallotta[3], N. Palmieri[41], G. Parente[84], E. Parizot[33], A. Parra[84], R.D. Parsons[87], S. Pastor[80], T. Paul[97], M. Pech[29], J. Pękala[72], R. Pelayo[84], I.M. Pepe[23], L. Perrone[54], R. Pesce[46], E. Petermann[105], S. Petrera[47], P. Petrinca[51], A. Petrolini[46], Y. Petrov[91], J. Petrovic[70], C. Pfendner[108], N. Phan[106], R. Piegaia[4], T. Pierog[39], P. Pieroni[4], M. Pimenta[74], V. Pirronello[52], M. Platino[2], V.H. Ponce[1], M. Pontz[45], P. Privitera[101], M. Prouza[29], E.J. Quel[3], S. Querchfeld[38], J. Rautenberg[38], O. Ravel[37], D. Ravignani[2], B. Revenu[37], J. Ridky[29], S. Riggi[84,52], M. Risse[45], P. Ristori[3], H. Rivera[48], V. Rizi[47], J. Roberts[96], C. Robledo[63], W. Rodrigues de Carvalho[84,19], G. Rodriguez[84], J. Rodriguez Martino[11,52], J. Rodriguez Rojo[11], I. Rodriguez-Cabo[84], M.D. Rodríguez-Frías[82], G. Ros[82], J. Rosado[81], T. Rossler[30], M. Roth[39], B. Rouillé-d'Orfeuil[101], E. Roulet[1], A.C. Rovero[8], C. Rühle[40], F. Salamida[47,39], H. Salazar[63], G. Salina[51], F. Sánchez[2], M. Santander[11], C.E. Santo[74], E. Santos[74], E.M. Santos[25], F. Sarazin[90], B. Sarkar[38], S. Sarkar[85], R. Sato[11], N. Scharf[43], V. Scherini[48], H. Schieler[39], P. Schiffer[43], A. Schmidt[40], F. Schmidt[101], O. Scholten[69], H. Schoorlemmer[68,70], J. Schovancova[29], P. Schovánek[29], F. Schröder[39], S. Schulte[43], D. Schuster[90], S.J. Sciutto[7], M. Scuderi[52], A. Segreto[56], M. Settimo[45], A. Shadkam[94], R.C. Shellard[16,17], I. Sidelnik[2], G. Sigl[44], H.H. Silva Lopez[67], A. Śmiałkowski[73], R. Šmída[39,29], G.R. Snow[105], P. Sommers[99], J. Sorokin[13], H. Spinka[88,93], R. Squartini[11], S. Stanic[79], J. Stapleton[98], J. Stasielak[72], M. Stephan[43], E. Strazzeri[56], A. Stutz[36], F. Suarez[2], T. Suomijärvi[32], A.D. Supanitsky[8,67], T. Šuša[27], M.S. Sutherland[94,98], J. Swain[97], Z. Szadkowski[73], M. Szuba[39], A. Tamashiro[8], A. Tapia[2], M. Tartare[36], O. Taşcău[38], C.G. Tavera Ruiz[67], R. Tcaciuc[45], D. Tegolo[52,61], N.T. Thao[110], D. Thomas[91], J. Tiffenberg[4], C. Timmermans[70,68], D.K. Tiwari[66], W. Tkaczyk[73], C.J. Todero Peixoto[18,24], B. Tomé[74], A. Tonachini[53], P. Travnicek[29], D.B. Tridapalli[19], G. Tristram[33], E. Trovato[52], M. Tueros[84,4], R. Ulrich[99,39], M. Unger[39], M. Urban[34], J.F. Valdés Galicia[67], I. Valiño[84,39], L. Valore[50], A.M. van den Berg[69], E. Varela[63], B. Vargas Cárdenas[67], J.R. Vázquez[81], R.A. Vázquez[84], D. Veberič[79,78], V. Verzi[51], J. Vicha[29], M. Videla[10], L. Villaseñor[66], H. Wahlberg[7], P. Wahrlich[13], O. Wainberg[2], D. Walz[43], D. Warner[91], A.A. Watson[87], M. Weber[40], K. Weidenhaupt[43], A. Weindl[39], S. Westerhoff[108], B.J. Whelan[13], G. Wieczorek[73], L. Wiencke[90], B. Wilczyńska[72], H. Wilczyński[72], M. Will[39], C. Williams[101], T. Winchen[43], L. Winders[109], M.G. Winnick[13], M. Wommer[39], B. Wundheiler[2], T. Yamamoto[101,a], T. Yapici[95], P. Younk[45], G. Yuan[94], A. Yushkov[84,50], B. Zamorano[83], E. Zas[84], D. Zavrtanik[79,78], M. Zavrtanik[78,79], I. Zaw[96], A. Zepeda[64], M. Zimbres-Silva[20,38] M. Ziolkowski[45]

[1] *Centro Atómico Bariloche and Instituto Balseiro (CNEA- UNCuyo-CONICET), San Carlos de Bariloche, Argentina*
[2] *Centro Atómico Constituyentes (Comisión Nacional de Energía Atómica/CONICET/UTN-FRBA), Buenos Aires, Argentina*
[3] *Centro de Investigaciones en Láseres y Aplicaciones, CITEFA and CONICET, Argentina*
[4] *Departamento de Física, FCEyN, Universidad de Buenos Aires y CONICET, Argentina*
[7] *IFLP, Universidad Nacional de La Plata and CONICET, La Plata, Argentina*
[8] *Instituto de Astronomía y Física del Espacio (CONICET- UBA), Buenos Aires, Argentina*
[9] *Instituto de Física de Rosario (IFIR) - CONICET/U.N.R. and Facultad de Ciencias Bioquímicas y Farmacéuticas U.N.R., Rosario, Argentina*
[10] *National Technological University, Faculty Mendoza (CONICET/CNEA), Mendoza, Argentina*
[11] *Observatorio Pierre Auger, Malargüe, Argentina*
[12] *Observatorio Pierre Auger and Comisión Nacional de Energía Atómica, Malargüe, Argentina*
[13] *University of Adelaide, Adelaide, S.A., Australia*
[16] *Centro Brasileiro de Pesquisas Fisicas, Rio de Janeiro, RJ, Brazil*
[17] *Pontifícia Universidade Católica, Rio de Janeiro, RJ, Brazil*

[18] *Universidade de São Paulo, Instituto de Física, São Carlos, SP, Brazil*
[19] *Universidade de São Paulo, Instituto de Física, São Paulo, SP, Brazil*
[20] *Universidade Estadual de Campinas, IFGW, Campinas, SP, Brazil*
[21] *Universidade Estadual de Feira de Santana, Brazil*
[22] *Universidade Estadual do Sudoeste da Bahia, Vitoria da Conquista, BA, Brazil*
[23] *Universidade Federal da Bahia, Salvador, BA, Brazil*
[24] *Universidade Federal do ABC, Santo André, SP, Brazil*
[25] *Universidade Federal do Rio de Janeiro, Instituto de Física, Rio de Janeiro, RJ, Brazil*
[26] *Universidade Federal Fluminense, EEIMVR, Volta Redonda, RJ, Brazil*
[27] *Rudjer Bošković Institute, 10000 Zagreb, Croatia*
[28] *Charles University, Faculty of Mathematics and Physics, Institute of Particle and Nuclear Physics, Prague, Czech Republic*
[29] *Institute of Physics of the Academy of Sciences of the Czech Republic, Prague, Czech Republic*
[30] *Palacky University, RCATM, Olomouc, Czech Republic*
[32] *Institut de Physique Nucléaire d'Orsay (IPNO), Université Paris 11, CNRS-IN2P3, Orsay, France*
[33] *Laboratoire AstroParticule et Cosmologie (APC), Université Paris 7, CNRS-IN2P3, Paris, France*
[34] *Laboratoire de l'Accélérateur Linéaire (LAL), Université Paris 11, CNRS-IN2P3, Orsay, France*
[35] *Laboratoire de Physique Nucléaire et de Hautes Energies (LPNHE), Universités Paris 6 et Paris 7, CNRS-IN2P3, Paris, France*
[36] *Laboratoire de Physique Subatomique et de Cosmologie (LPSC), Université Joseph Fourier, INPG, CNRS-IN2P3, Grenoble, France*
[37] *SUBATECH, École des Mines de Nantes, CNRS-IN2P3, Université de Nantes, Nantes, France*
[38] *Bergische Universität Wuppertal, Wuppertal, Germany*
[39] *Karlsruhe Institute of Technology - Campus North - Institut für Kernphysik, Karlsruhe, Germany*
[40] *Karlsruhe Institute of Technology - Campus North - Institut für Prozessdatenverarbeitung und Elektronik, Karlsruhe, Germany*
[41] *Karlsruhe Institute of Technology - Campus South - Institut für Experimentelle Kernphysik (IEKP), Karlsruhe, Germany*
[42] *Max-Planck-Institut für Radioastronomie, Bonn, Germany*
[43] *RWTH Aachen University, III. Physikalisches Institut A, Aachen, Germany*
[44] *Universität Hamburg, Hamburg, Germany*
[45] *Universität Siegen, Siegen, Germany*
[46] *Dipartimento di Fisica dell'Università and INFN, Genova, Italy*
[47] *Università dell'Aquila and INFN, L'Aquila, Italy*
[48] *Università di Milano and Sezione INFN, Milan, Italy*
[49] *Dipartimento di Fisica dell'Università del Salento and Sezione INFN, Lecce, Italy*
[50] *Università di Napoli "Federico II" and Sezione INFN, Napoli, Italy*
[51] *Università di Roma II "Tor Vergata" and Sezione INFN, Roma, Italy*
[52] *Università di Catania and Sezione INFN, Catania, Italy*
[53] *Università di Torino and Sezione INFN, Torino, Italy*
[54] *Dipartimento di Ingegneria dell'Innovazione dell'Università del Salento and Sezione INFN, Lecce, Italy*
[56] *Istituto di Astrofisica Spaziale e Fisica Cosmica di Palermo (INAF), Palermo, Italy*
[57] *Istituto di Fisica dello Spazio Interplanetario (INAF), Università di Torino and Sezione INFN, Torino, Italy*
[58] *INFN, Laboratori Nazionali del Gran Sasso, Assergi (L'Aquila), Italy*
[61] *Università di Palermo and Sezione INFN, Catania, Italy*
[63] *Benemérita Universidad Autónoma de Puebla, Puebla, Mexico*
[64] *Centro de Investigación y de Estudios Avanzados del IPN (CINVESTAV), México, D.F., Mexico*
[66] *Universidad Michoacana de San Nicolas de Hidalgo, Morelia, Michoacan, Mexico*
[67] *Universidad Nacional Autonoma de Mexico, Mexico, D.F., Mexico*
[68] *IMAPP, Radboud University Nijmegen, Netherlands*
[69] *Kernfysisch Versneller Instituut, University of Groningen, Groningen, Netherlands*
[70] *Nikhef, Science Park, Amsterdam, Netherlands*
[71] *ASTRON, Dwingeloo, Netherlands*
[72] *Institute of Nuclear Physics PAN, Krakow, Poland*

[73] *University of Łódź, Łódź, Poland*
[74] *LIP and Instituto Superior Técnico, Lisboa, Portugal*
[78] *J. Stefan Institute, Ljubljana, Slovenia*
[79] *Laboratory for Astroparticle Physics, University of Nova Gorica, Slovenia*
[80] *Instituto de Física Corpuscular, CSIC-Universitat de València, Valencia, Spain*
[81] *Universidad Complutense de Madrid, Madrid, Spain*
[82] *Universidad de Alcalá, Alcalá de Henares (Madrid), Spain*
[83] *Universidad de Granada & C.A.F.P.E., Granada, Spain*
[84] *Universidad de Santiago de Compostela, Spain*
[85] *Rudolf Peierls Centre for Theoretical Physics, University of Oxford, Oxford, United Kingdom*
[87] *School of Physics and Astronomy, University of Leeds, United Kingdom*
[88] *Argonne National Laboratory, Argonne, IL, USA*
[89] *Case Western Reserve University, Cleveland, OH, USA*
[90] *Colorado School of Mines, Golden, CO, USA*
[91] *Colorado State University, Fort Collins, CO, USA*
[92] *Colorado State University, Pueblo, CO, USA*
[93] *Fermilab, Batavia, IL, USA*
[94] *Louisiana State University, Baton Rouge, LA, USA*
[95] *Michigan Technological University, Houghton, MI, USA*
[96] *New York University, New York, NY, USA*
[97] *Northeastern University, Boston, MA, USA*
[98] *Ohio State University, Columbus, OH, USA*
[99] *Pennsylvania State University, University Park, PA, USA*
[100] *Southern University, Baton Rouge, LA, USA*
[101] *University of Chicago, Enrico Fermi Institute, Chicago, IL, USA*
[105] *University of Nebraska, Lincoln, NE, USA*
[106] *University of New Mexico, Albuquerque, NM, USA*
[108] *University of Wisconsin, Madison, WI, USA*
[109] *University of Wisconsin, Milwaukee, WI, USA*
[110] *Institute for Nuclear Science and Technology (INST), Hanoi, Vietnam*
[†] *Deceased*
[a] *at Konan University, Kobe, Japan*

# Long Term Performance of the Surface Detectors of the Pierre Auger Observatory.

RICARDO SATO[1] FOR THE PIERRE AUGER COLLABORATION[1]
[1]*Observatorio Pierre Auger, Av. San Martin Norte 304, 5613 Malargüe, Argentina*
*(Full author list: http://www.auger.org/archive/authors_2011_05.html)*
*auger_spokespersons@fnal.gov*

**Abstract:** The Surface Array Detector of the Pierre Auger Observatory consists of about 1600 water Cherenkov detectors. The operation of each station is continuosly monitored with respect to its individual components like batteries and solar panels, aiming at the diagnosis and the anticipation of failures. In addition, the evolution with time of the response and of the trigger rate of each station is recorded. The behavior of the earliest deployed stations is used to predict the future performance of the full array.

**Keywords:** Long term, surface detector, Pierre Auger Observatory

## 1 Introduction

The Surface Detector (SD) of the Pierre Auger Observatory [1] consists of about 1600 stations based on cylindrical tanks of 1.2 m $\times$ 10 m$^2$ volume filled with ultra pure water of 8 to 10 MΩ-cm [1]. Each station is autonomous and uses two 12 V batteries and two solar panels.

Particles of extensive air showers generated by primary cosmic rays produce Cherenkov radiation in the tank water. This light is reflected by a material (Tyvek®[1]) which covers the inside of the water-containing liner and is observed by 3 photomultiplier tubes (PMT) of 9" diameter. The nominal operating gain of the PMTs is $2 \times 10^5$ and can be extended to $10^6$. Stations in the main array are distributed in a triangular grid of 1.5 km spacing, covering about 3000 km$^2$. This design has a full efficiency for primary cosmic rays with energies above about $3\times10^{18}$eV [2] and is intended to be operational for at least 20 years.

An important issue is the signal stability which is related to the PMT gain, water transparency and the reflection coefficient of the Tyvek®.

It is important for the station to be able to measure both the current I and the charge Q (time-integrated current) produced by the PMTs in response to an extensive air shower. The charge is used to determine the energy deposited in the tank by the shower, and the time distribution of the current is used to form the trigger in each station. The charge and maximum current due to a single vertical muon, $Q_{VEM}$ and $I_{VEM}$, respectively, referred to in this paper as Area or $A$ and Peak or $P$, respectively, are constantly monitored by the calibration and monitoring system and provide the basis for calibration of each station [2, 3].

These quantities together with others such as the baseline values and the dynode/anode ratio (the ratio of the output signal from the last PMT dynode to that of the anode) are available to evaluate the behavior of the stations. Although the calibration system provides continuously updated values of all these signals, it is important to model the underlying changes in detector performance in order to determine the long term effectiveness of the performance and calibration of the detectors. In this work we provide a method for the phenomenological understanding of the signal evolution allowing us to predict the long term performance of the detector. The model has been shown to be reliable previously [4, 5]. In this work we review the model presented earlier after more years of operational experience and apply it to the full SD array, which was completed in 2008. This allows us to predict the array lifetime.

In section 2 we will examine the power system of the stations as it is an important system for the stable operation of the stations. In section 3 we quantify and predict how much the signal properties will change in the next decade of operation, mainly through the Area over Peak ratio of the muon signals ($A/P$) as will be described. In section 4 we show the evolution of the the trigger rate of the array and of individual stations.

## 2 Power system

Each station has its power supply running autonomously with solar panels and batteries. Two important issues then are the battery lifetime and solar panel efficiency loss over time. The main power system design consists of two solar

1. Tyvek® is a registered trademark of DuPont corporation.

panels of 53 Wp each connected in series and two batteries[2] of 12 V and 100 Ah also connected in series. A station with fully charged batteries can operate 7-10 days without further charging during a cloudy period. During all the operation of the observatory there has not been any general loss of operation due to extended cloudiness.

The current provided by the solar panels is monitored constantly and the information obtained is useful to determine when solar panels need attention. Because of modulation of the solar panel current by the solar power regulator it is challenging to remotely measure performance of solar panels that are working properly. So far, we do not identify a significant solar panel efficiency loss, though we have found apparent cell damage to many of the solar panels due to some not-yet-understood manufacturing problem. We are currently studying these solar panels to estimate whether or not this will adversely affect long term performance.

As the daily discharge is quite small (about 10% of the rated capacity), we estimate the end of battery life in this work to be when the battery voltage drops below 11V if the drop is not generated by a very long cloudy period or an apparent problem related to other part of the system. Note that it is quite different from the definition normally used in the industry which considers the lifetime to have been reached when the battery can not accumulate more that 80% of its rated capacity.

Figure 1 is a histogram of the time interval between initial battery operation and the time at which the battery voltage goes below 11V. In total, 808 pairs of batteries have satisfied this criterion. Some failures are observed in operation before reaching the expected lifetime in one of the two batteries, mostly for newer ones, populating the lower values of the histogram. From our experience we find that the quality of the batteries have not been constant.

Many batteries in the array have operated for more than 3 years with no sign of failure. As a consequence, they are not included in the histogram of figure 1 and their inclusion would have raised the overall apparent lifetime.

In most cases, a station can still operate for more than 3 months without data acquisition interruption even though we have considered the battery dead in this way. Therefore, the battery lifetime might be considered to be a little higher than obtained here. The average lifetime is then between 4.5 and 6 years.

## 3 VEM Signal: Area over Peak

The output signal from the PMTs of a single vertical muon has a fast rise and decays exponentially with time. The fast rise is dominated by the Cherenkov radiation which is only reflected once at the Tyvek®, while the exponential decay is dominated by multiple reflections. As a consequence, the exponential decay has a strong dependence on the reflection coefficient of the tank wall and the transparency of the water.

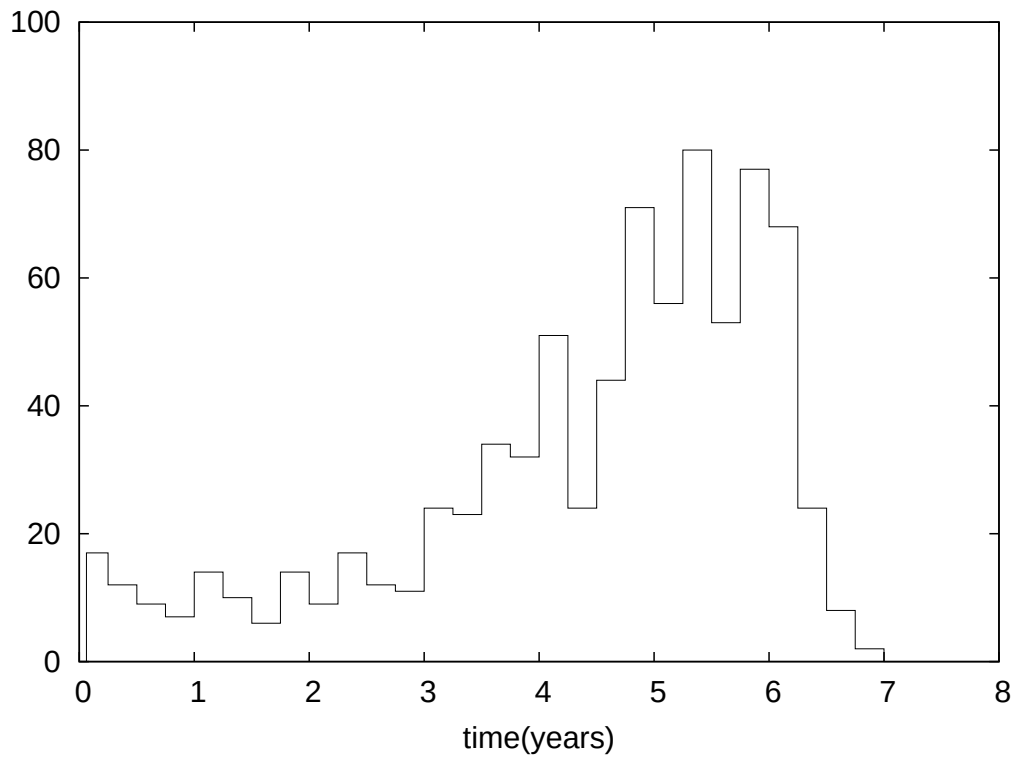

Figure 1: Histogram of the battery lifetime (see text).

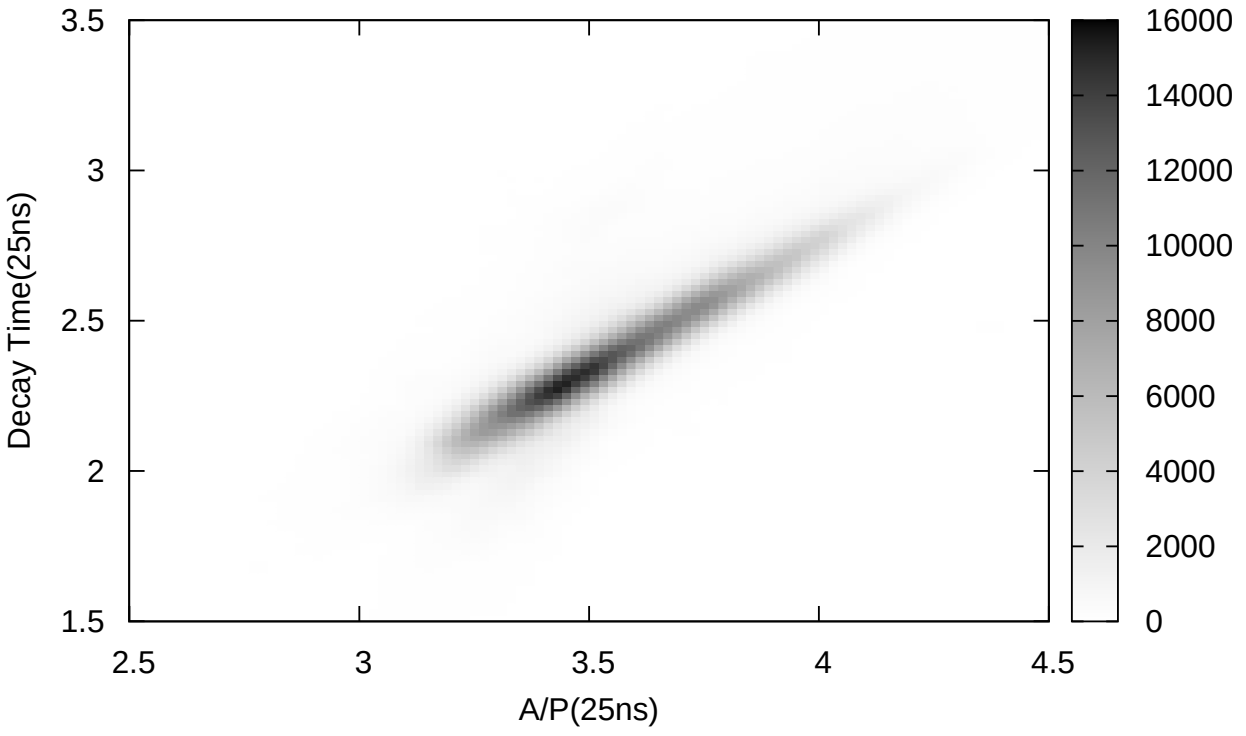

Figure 2: Histogram of correlation between the area to peak ratio ($A/P$) and signal decay constant for muon signals in SD array.

In figure 2 we can see a good correlation between the exponential decay constant and the parameter area/peak ($A/P$) ratio. As the $A$ and $P$ are directly available in the online monitoring of each station, we are going to look the $A/P$ ratio, instead of the exponential decay constant.

The proper description of the $A/P$ evolution with time might be very complicated, taking into account the daily and seasonal temperature variation, maintenance and hardware replacement of the stations for example. In this work we examine the main long term trend of the $A/P$. We consider that it might be described by an exponential behavior as:

$$\frac{A}{P} = s(t) \times \left[1 - p_1 \cdot (1 - e^{\frac{-t}{p_2}})\right] \qquad (1)$$

where $s(t)$ takes into account the seasonal variations and initial value, $p_1$ is the fractional loss and is a dimensionless quantity that varies between 0 and 1, and $p_2$ is the characteristic time in units of years. This decay assumes that the $A/P$ will stabilize at $1 - p_1$ combined with a seasonal variations.

We propose the $A/P$ seasonal variation as:

2. Moura Clean model 12MC105, a flooded lead acid battery with a selectively permeable membrane to reduce water loss. www.moura.com.br

$$s(t) = p_0 \times \Big[1 + p_3 \cdot \sin(2\pi(\frac{t}{T} - \phi))\Big] \qquad (2)$$

where $p_0$ is the overall normalization factor, $p_3$ quantifies the strength of the seasonal variation and is a dimensionless quantity that varies between 0 and 1. The $T$ will be considered to be 1 year and the $\phi$ is just a phase parameter to adjust the annual temperature variation.

As the analog signal from the PMT is digitized at a 40 MHz rate (one sample every 25 ns), which is fast enough to have a good idea of the muon signal shape, $A$ is basically calculated as the sum of digitized information around the region where the main signal appears and the $P$ is the maximum value of this signal. To simplify this analysis, the ratio $A/P$ as well as the parameter $p_0$ will be given in units of 25 ns. We calculate the mean and deviation of $A/P$ over 7 days. We use these values to find the parameters of equation 1 using a least square fit.

For long term operation station maintenance may be required which might involve PMTs or general electronics replacement. This might adjust the voltage of the PMTs [3] and generate a slight gain change and, consequently, the values of Peak and Area. However, it is expected that most of these changes generate an almost unchanged $A/P$ ratio. Some residual effects may remain and, in many of the cases, it is a little difficult to treat them properly.

The parameters which are expected to have big effects on $A/P$ are mostly the water transparency and the coefficient of reflection of the tank wall. In particular what we are most interested in is the variation of the $A/P$ with time and its correlation with possible degradation of the station. The analysis is thus rather complex and, to try to avoid bias, we considered only well operating stations that were installed before 2007 so as to have a long term operational period, and PMTs which also pass the following restriction: $1 \leq p_0 \leq 5.5$, in units of 25ns; $0 \leq p_2 \leq 500$ yr; $\chi^2/\nu \leq 2000$, where $\nu > 40$ is the number of degree of freedom.

The last constraint is much weaker than acceptable statistically. This is because there are many short term effects in the data which are not taken into account in a simple expression as considered in equation 1, although it describes quite well the general behavior, as shown in figure 3. In the local winter of 2007 we observed a deviation from the steady trend due to extreme low temperatures (below $-15^oC$). This weather generated a 10 cm thick ice layer in the stations, which produced an extra drop, at a level of 1-3%, in $A/P$. Reasons for that drop are being studied.

In total we found approximately 1500 PMTs which pass the above restrictions. We obtain the characteristic time around few years, an overall normalization $p_0 \approx 3.5 \times 25$ ns and less than 1% for the seasonal amplitude.

In figure 4 we show an example histogram for the parameters $p_1$ of equation 1. We can see that the fractional loss factor ($p_1$) is below 20% in general.

In figure 5 there is an estimation of the $A/P$ loss using equation 1 and the parameters predicted for the next 10 years. We can see that the final $A/P$ will be larger than 85% in most cases. There are a few cases for which this value is much smaller that may require some intervention in the near future. However, they are few and would not greatly affect the general operation of the surface detector array.

The $A/P$ would be affected by growth of microorganisms in the water which could produce some turbidity. Bacteriological testing of the water and the surface of the Tyvek® is carried out regularly in some stations, but until now there has been no identification of relevant microorganism growth.

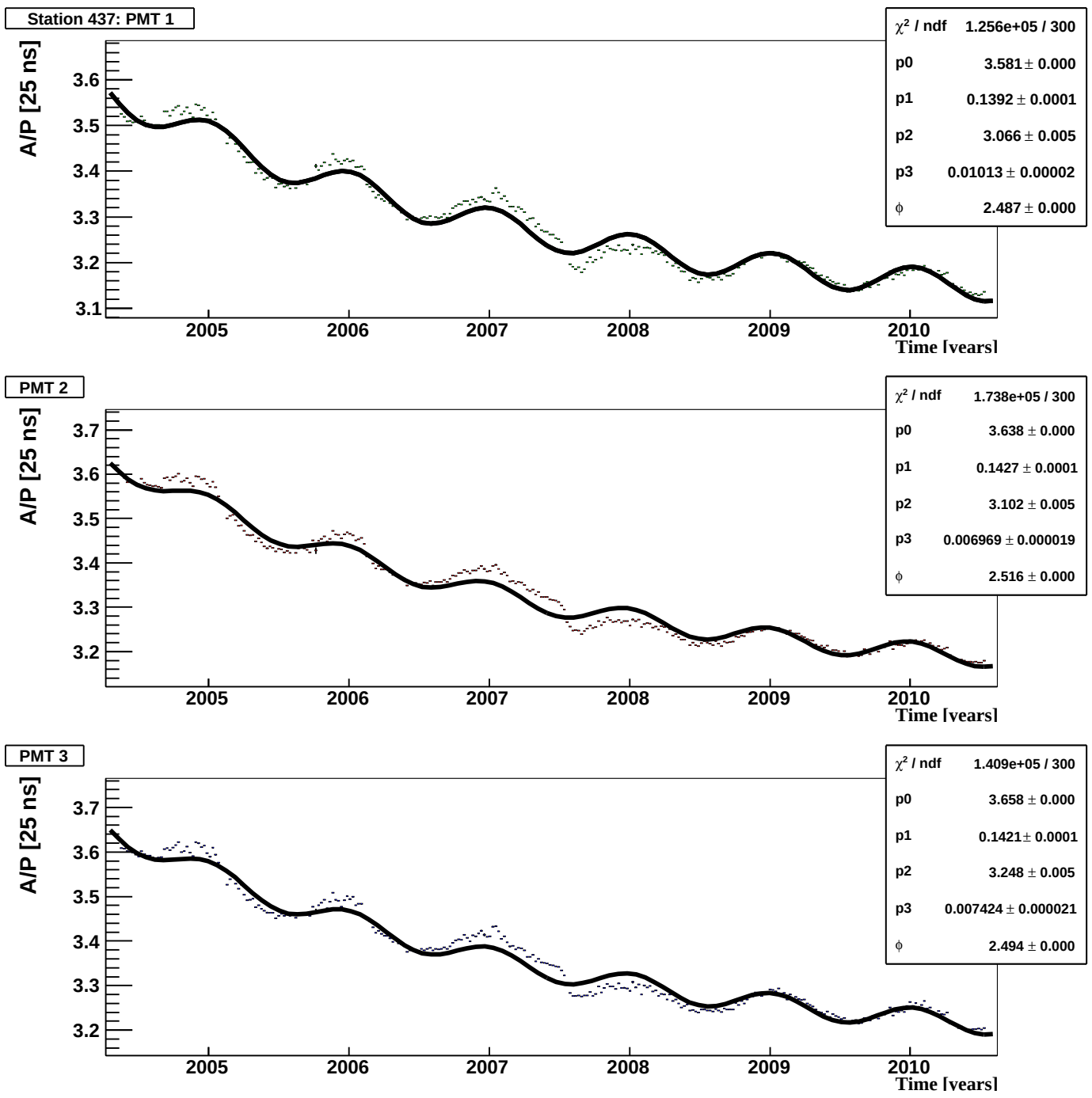

Figure 3: A/P as a function of time for station 437. The dots are the average of the A/P over 7 day and the continuous line is the fit of the equation 1.

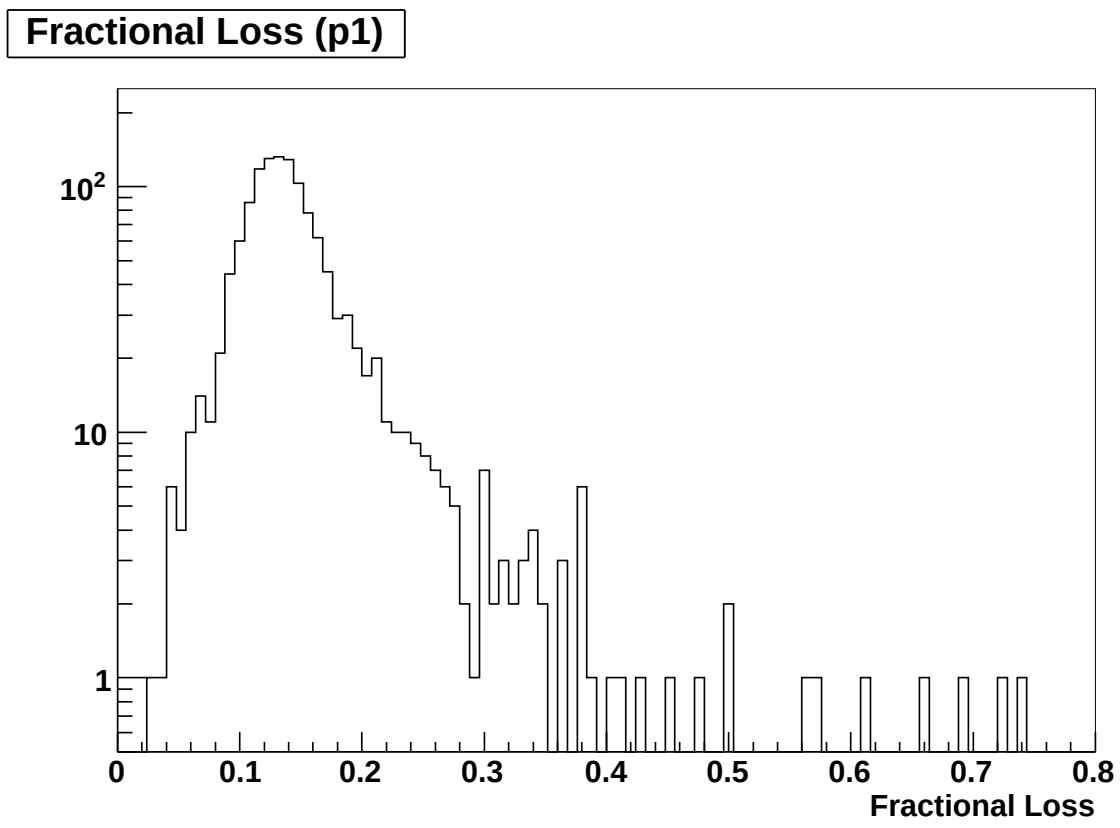

Figure 4: Values of the fractional loss $p_1$ 1.

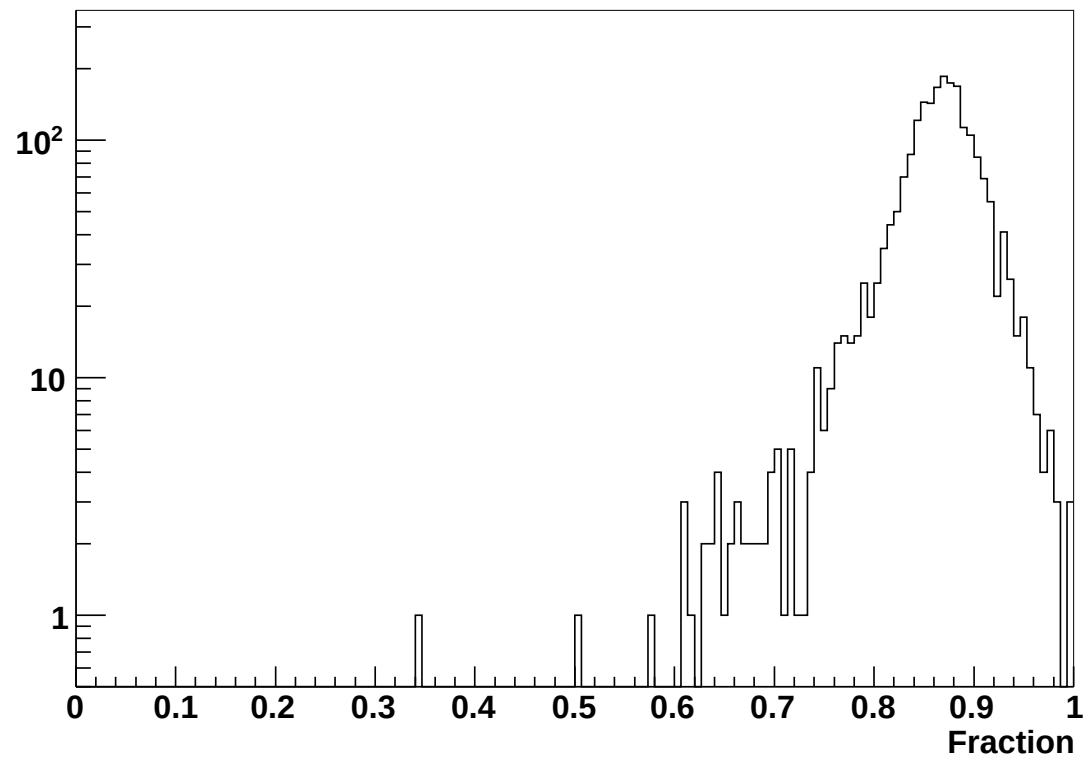

Figure 5: Estimated relative values (Fraction) of $A/P$ after 10 years of operation with respect to its initial value.

# 4 Trigger

It is also important to monitor the trigger rates of the array. As an example we show the trigger rate of one particular station (see figure 6). The T1 and T2 triggers [2], which are just simple threshold triggers, are quite stable with time. On the contrary, the ToT (Time over Threshold) rate [2] which follows the $A/P$ evolution, with an initial decay time followed by a stable operation in time. The ToT trigger requires thirteen 25 nsec FADC bins in a larger time window of 3 $\mu$s to be above a 0.2 VEM threshold, so this trigger is sensitive to a broad time distribution of low energy showers and sensitive to the individual pulse width.

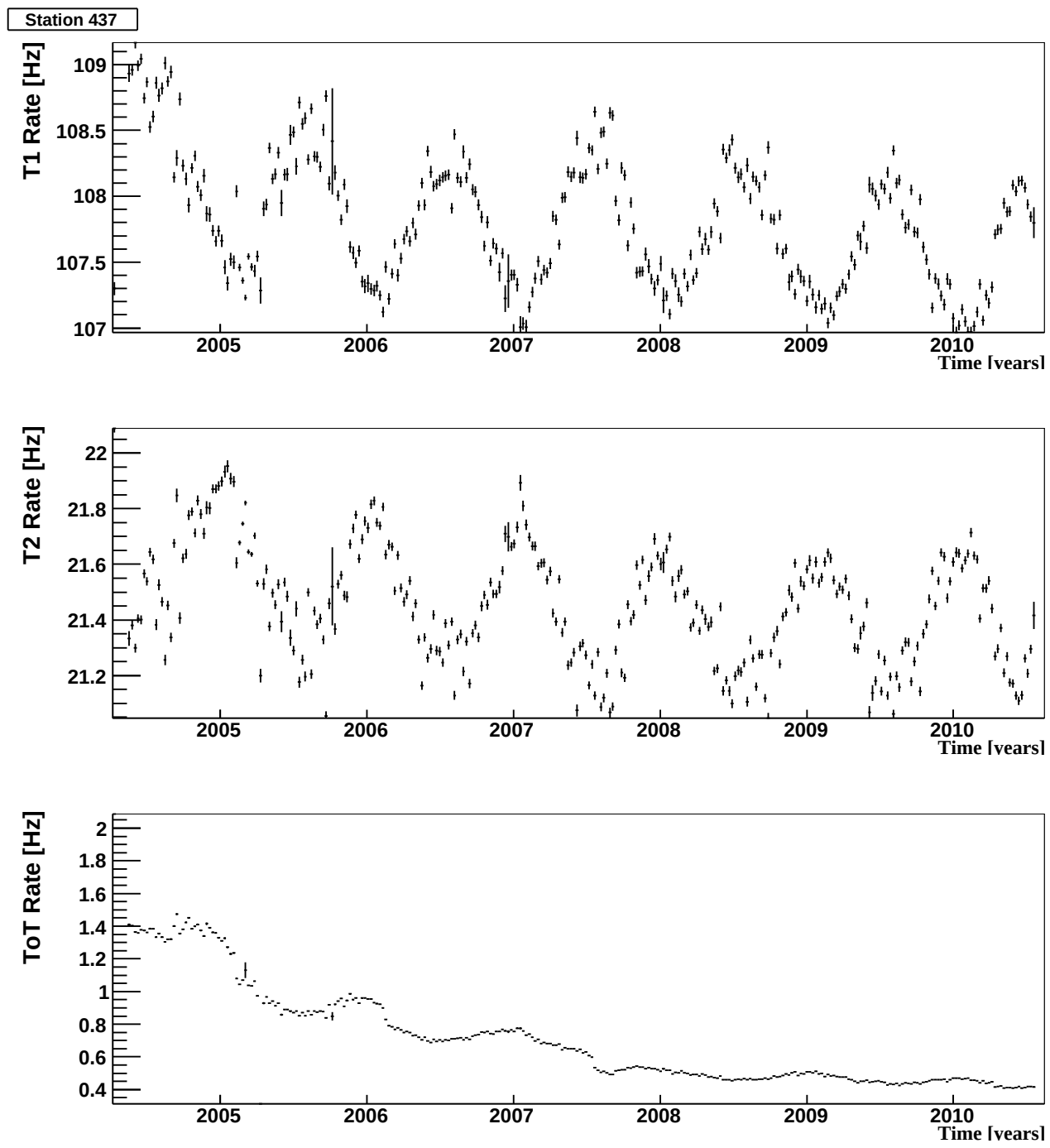

Figure 6: Trigger rate T1, T2 and ToT for the station 437 as function of time.

The figure 7 shows the highest SD level trigger (T5) event rate normalized by the number of active hexagons in the array. T5 which is sometimes also called as 6T5 request that the station with the largest signal is surrounded by 6 working stations at the time of shower impact and have already passed the previous trigger levels (T3 and T4) [2]. We can see that the physical event rate above the threshold for SD full efficiency, was unaffected by the decrease of ToT trigger rate of individual stations.

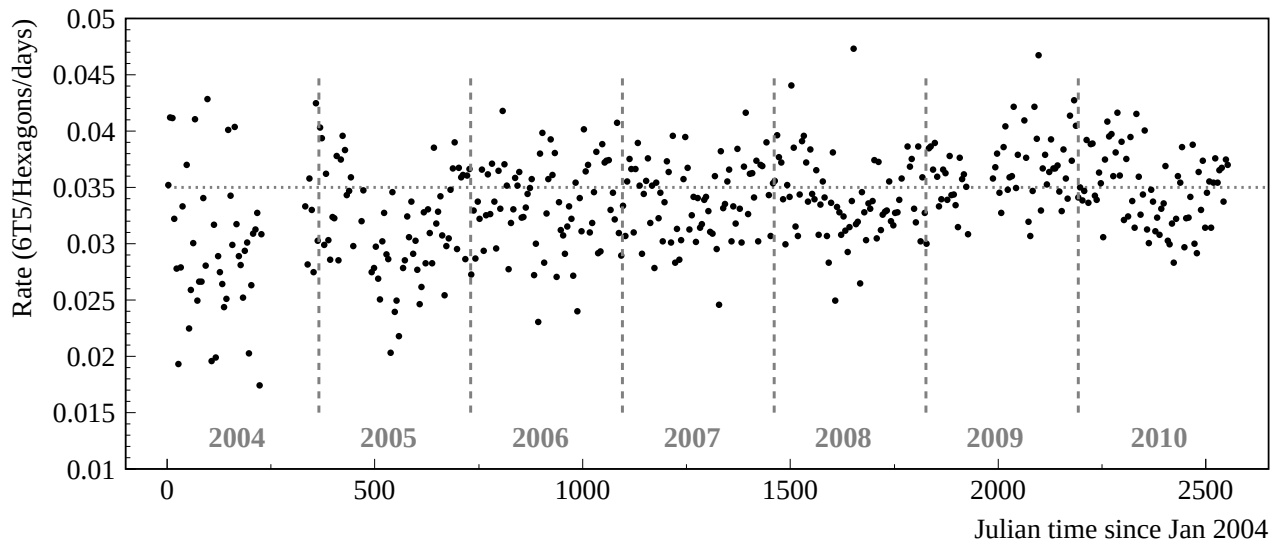

Figure 7: Event rate as function of time.

# 5 Conclusions

With the experience of more than 6 years of operation of the detector, the studies of power system and single muons signals has shown a perfectly normal behavior.

As the lifetime of the batteries obtained in the present analysis confirms initial expectations, the maintenance cost should be consistent with the programmed one.

The study carried out on single muons shows that the Area over Peak reduction will be less than 15% in the next decade. The reasons for the decay of A/P with time are a convolution of water transparency, Tyvek® reflection and electronic response of the detectors. The proportion of each of these three causes has not yet been determined. The overall event rate above the threshold for SD full efficiency have not been so far affected by the evolution of the signals described in this work.

# Remote operation of the Pierre Auger Observatory

JULIAN RAUTENBERG[1] FOR THE PIERRE AUGER COLLABORATION[2]
[1]*Bergische Universität Wuppertal, Gaußstr. 20, D-42119 Wuppertal, Germany*
[2]*Observatorio Pierre Auger, Av. San Martín Norte 304, 5613 Malargüe, Argentina*
*(Full author list: http://www.auger.org/archive/authors_2011_05.html)*
*auger_spokespersons@fnal.gov*

**Abstract:** The different components of the Pierre Auger Observatory, the surface detectors (SD) and the fluorescence telescopes (FD), are operated and maintained mainly by operators on site. In addition, the FD data-acquisition has to be supervised by a shift crew on site to guarantee a smooth operation. To provide access to the detector-systems for experts from remote sites not only increases the knowledge available for the maintenance, but opens the possibility to operate the detector from remote sites. Establishing remote shift operation has the benefit of saving substantial travelling time and cost, but also offers the possibility of remote support for shifters, increasing the quality of the data and the safety of the detector. The monitoring of the Pierre Auger Observatory has been designed with the server and replication scheme for remote availability. In addition, grid based technology has been used to implement the access to the control of the detector to make remote shift operation possible.

**Keywords:** Pierre Auger Observatory, UHECR, detector operation, monitoring, remote control, remote shift

## 1 Introduction

The Pierre Auger Observatory measures cosmic rays at the highest energies. The southern site in the province of Mendoza, Argentina, was completed during the year 2008. The instrument [1] was designed to measure extensive air showers with energies ranging from $1 - 100\,\mathrm{EeV}$ and beyond. It combines two complementary observational techniques, the detection of particles on the ground using an array of 1660 water Cherenkov detectors distributed on an area of 3000 $\mathrm{km}^2$ and the observation of fluorescence light generated in the atmosphere above the ground by a total of 27 wide-angle Schmidt telescopes positioned at four sites on the border around the ground array. Routine operation of the detectors has started in 2002.

## 2 Shift operation

The data-acquisition system of the surface detector array is not operated manually. It runs continuously without starting and stopping discrete runs. The duty cycle of the SD reaches almost 100%. The fluorescence telescopes operate on clear, moonless, nights and are sensitive to environmental factors such as rain, strong winds and lightning. Therefore, the telescopes have to be operated manually and the data-acquisition is organized in runs. The operation of the FD is controlled by a shift-crew from a control room within the Pierre Auger campus building. A total of 61 shifters per year are required to cover the shift operation of up to 13 hours per night in dark periods of up to 18 days per lunar cycle. These shifters have to travel long distances to be on site.

## 3 Monitoring

A monitoring system [2] has been developed to help the shifter judge the operation of the FD on the basis of the available information. The overview page for one FD-site is shown in fig.1. An alarm-system has been implemented to notify the shifter in case of occurrences that require immediate action. The monitoring system overviews the operation and maintenance of the SD. Daily checks on the monitoring data of the single surface detectors can identify the onset of failures. This starts a maintenance process which typically leads to an intervention of a crew visiting the surface detector in the field. The maintenance and intervention system realized within the monitoring system of the Pierre Auger Observatory covers the whole work-flow from the alarm being raised to the intervention in the field and finally resolving the alarm. It represents a tailored ticketing system which has been developed for the SD, but is extended to other components like the monitoring system itself.

Technically, the monitoring system is based on a set of databases that store all monitoring information available and a web-interface that is used to display the information using mainly PHP, JavaScript, JPGraph and gnuplot. In the case of the FD, the databases are partially filled locally at the FD sites. The mysql build-in mechanism of replication is used to transport the data to the central database server on the campus. Replication guarantees the completeness of the data in the case of lost connections between the campus and the FD-sites.

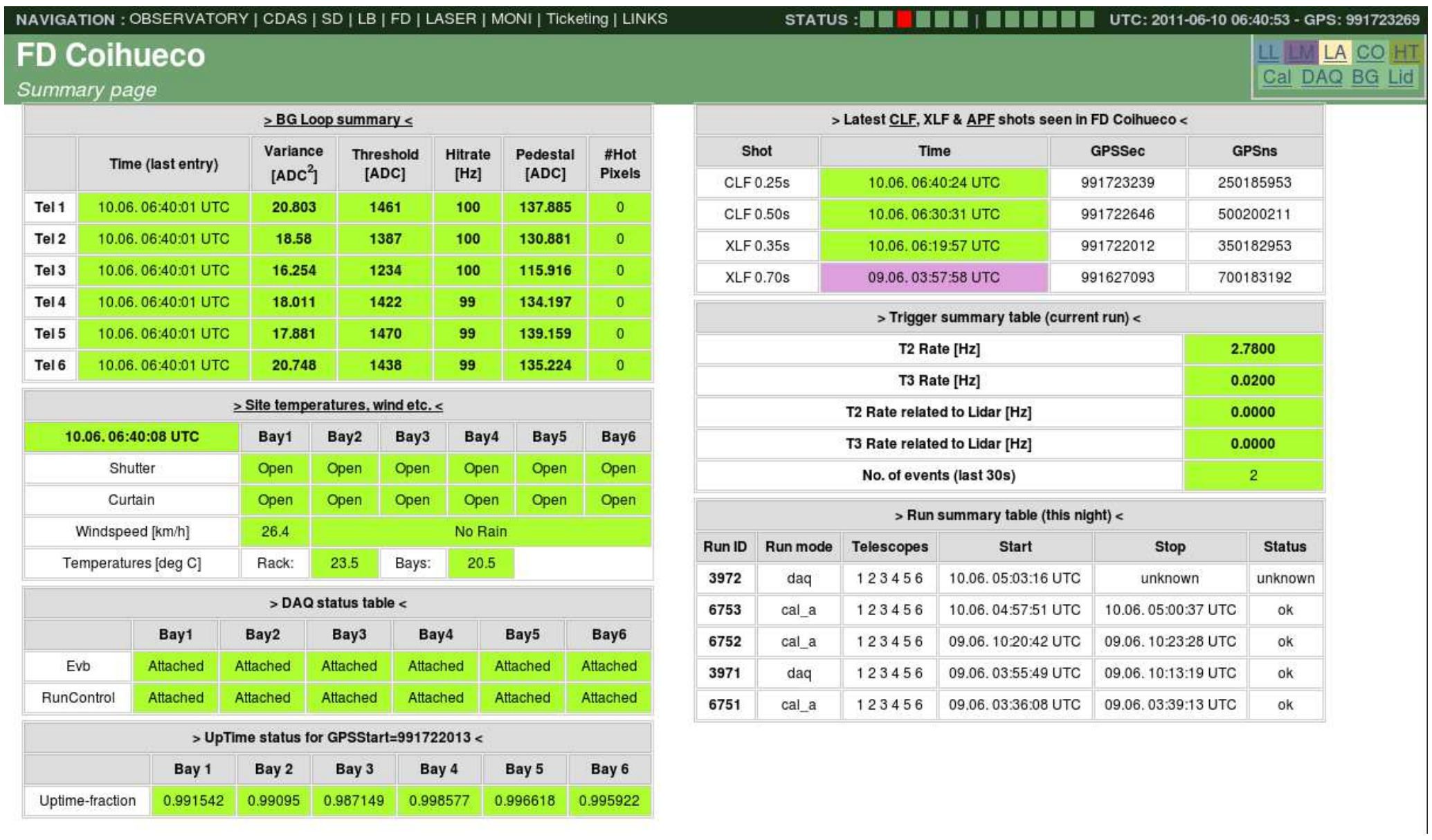

Figure 1: Overview page of the monitoring showing the status of one FD-site, Coihueco.

An authentication schema and a sophisticated role model allow the user to interact with the monitoring system according to their privileges. These interactions include not only the acknowledgement of an alarm, but also administrative tasks like the configuration of alarms or the assignment of roles to users. The maintenance and intervention system is highly interactive and thus relies on the proper assignment of roles to users.

In addition, replication is used to transport the information to a database on a server in Europe. This server contains the monitoring information in quasi real-time, as long as the internet-connection between the observatory and Europe is stable. The mirror site in Europe can provide the web-interface without additional traffic to the observatory. The problem of an unstable internet connection with limited bandwidth has been addressed by the AugerAccess project [3] that involved the installation of an optical fibre connecting the observatory with the internet backbone.

## 4 Remote control

The possibility of connecting via internet to the inner control systems of the detector allows the expert for a specific system to inspect it, in case of failure, from all over the world. This supports the local staff who are trained for the operation of the systems, but which cannot have all the knowledge of the experts that developed it. Previously, the low bandwidth of the internet connection to the observatory prevented the knowledge of experts being available on site, leading in the worst case to expensive and time consuming travel to the detector with consequential severe delay in the processing of problems. With AugerAccess the internet connection now provides the required reliability to connect remotely to the system for debugging purposes. Experts (e.g. from Europe) can inspect the system and share their knowledge in understanding the symptom of a problem and its possible cure.

The operation of the FD [4] is secured by a slow-control system. The slow-control system works autonomously and continuously monitors detector and weather conditions. Commands from operators are accepted only if they do not violate safety rules. Data-acquisition takes place within the run-control. These two systems, the slow-control and the run-control, are the main components of the operation of the FD.

The security of a connection to the sensitive inner system of the observatory is established by using grid technologies for the authentication and encrypted protocols. For the access an X.509 certificate obtaining by a national certificate authority is used. These certificates are valid for only one year. Both, a valid certificate and a password are required for authentication, and the user has to be registered on site at the observatory through authorization by an administrator. The remote client alleviate certificate handling includes a single-sign-on with the passphrase to be valid only 24 hours. The graphical user interface allows the renewal of the decryption. The decrypted certificate is checked on every operation, on the DAQ as well as the slow-control. The system handles the slow-control for operation of the FD system by connecting to the slow-control server on the campus via a Grid secured SSH connection and port forwarding. This server in turn is connected to the systems at each FD-site. In this way, the remote operator sees the same interface as the operator in the control room on the campus. An example is given in fig.2. The topology of the services and the connections is illustrated in fig.3.

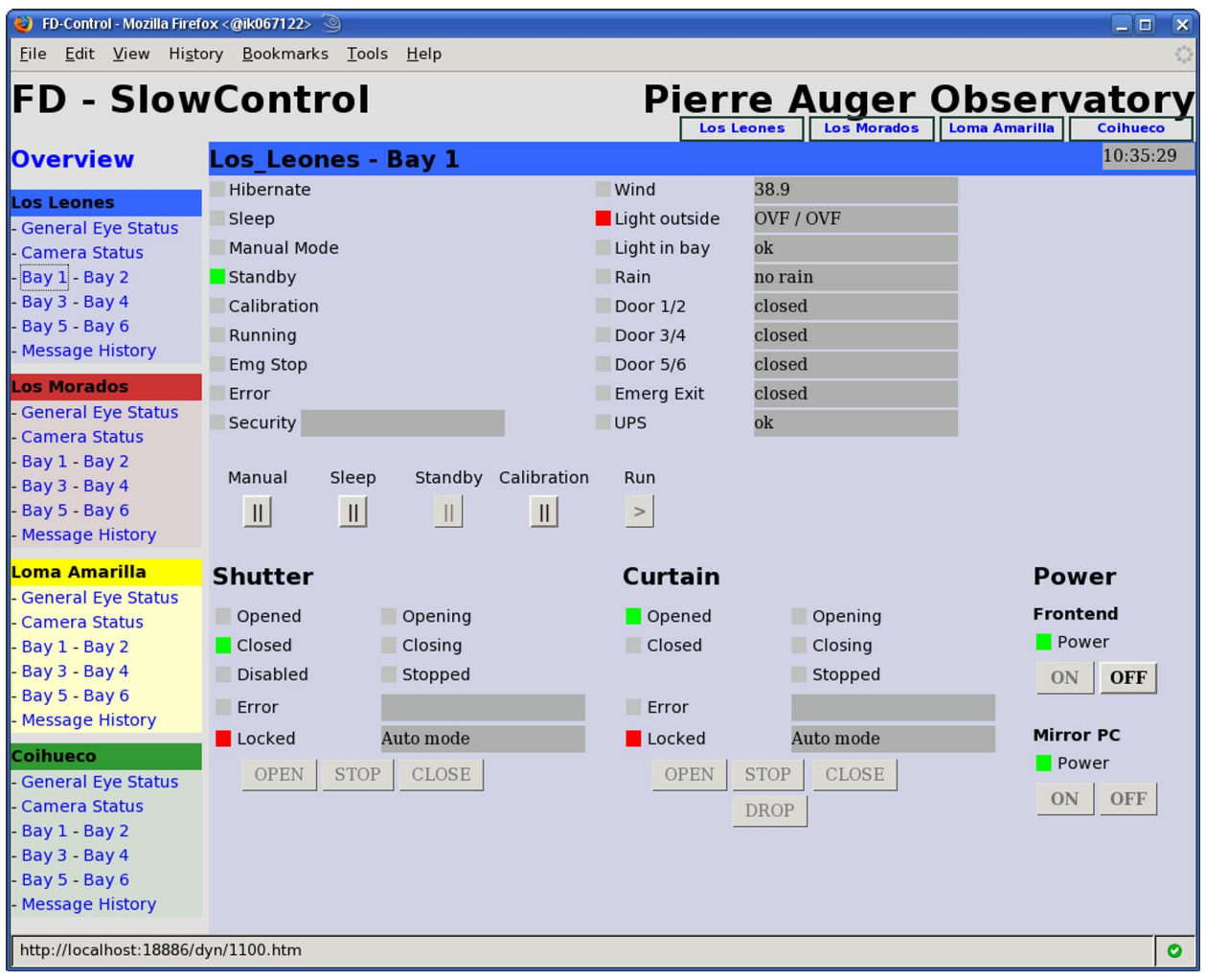

Figure 2: Example of the slow-control as it is displayed the same way on the campus as in the remote control room.

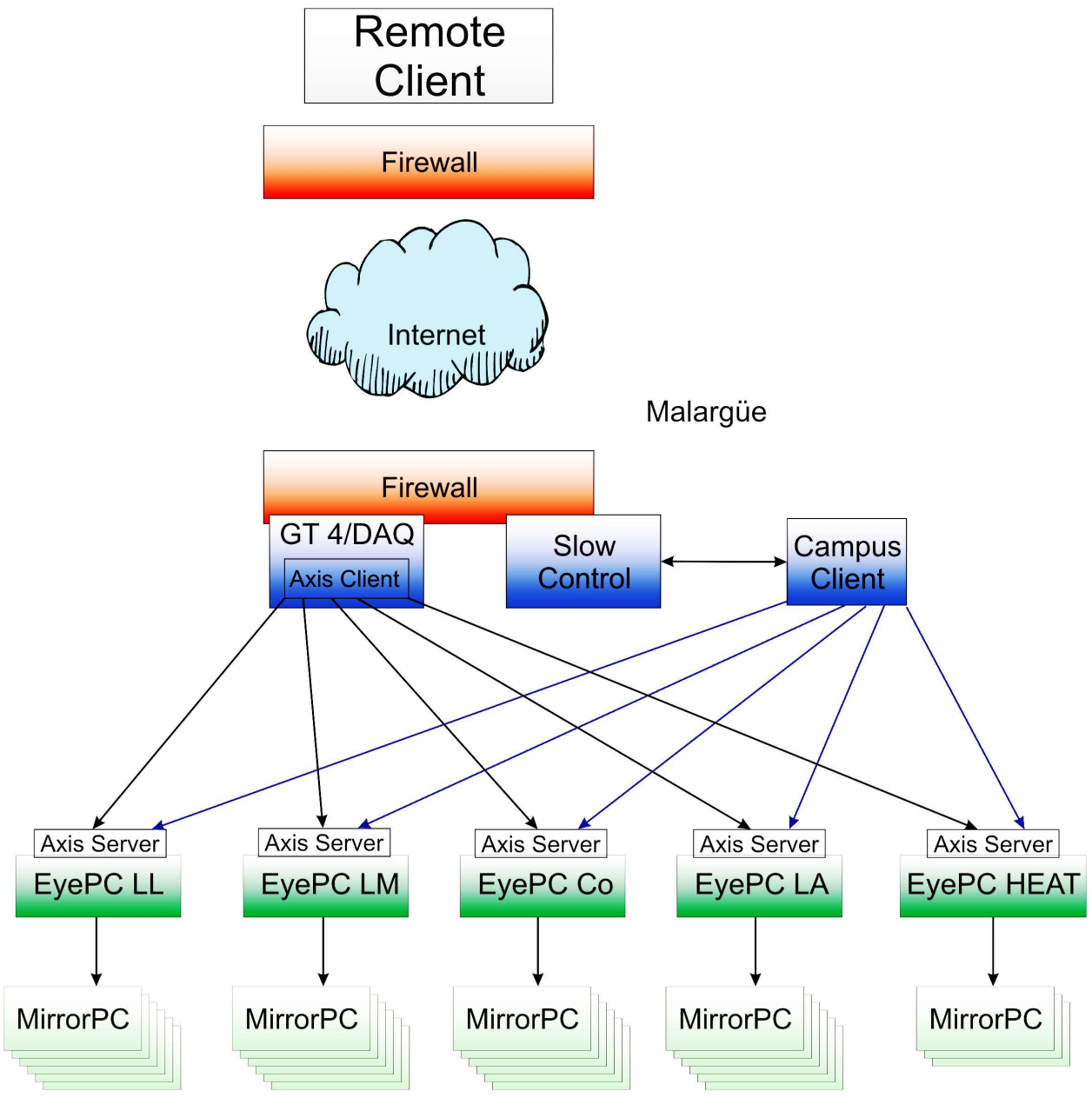

Figure 3: Topology of the services and the connections for the remote access of the DAQ and the slow-control.

## 5 Remote shift

The Pierre Auger Collaboration established a task force to study the feasibility of operating the observatory remotely. The task force was especially concerned with the operation of the FD shifts from a remote control room. This is not necessary for the SD, which operates continuously. In that case, off-site work uses monitoring information to detect malfunctioning detectors as part of the maintenance process. No special arrangements are then needed for off-site SD work, which is part of the monitoring program and can be performed from any site at any time. Since the Auger Collaboration has not had a program of regular on-site shifts for SD, off-site SD shifts do not reduce the workload but do improve maintenance efficiency and, thus, detector performance.

The remote FD-shift aims at reducing the load on the collaborators, since travelling to the site is time-consuming and expensive. Even with the more reliable internet connection via the optical fibre installed as the main part of AugerAccess the connection to the observatory from a remote site, even including other cities in Argentina, is not guaranteed. Therefore, even with shifters operating the observatory remotely, we need to have two shifters on site for safety reasons. Those shifters need not stay focused all the night and can rest while being on call in case of alarms. This way, the number of shifters needed for operation might be reduced in the end by 60%. This is not only a relief for the collaborators, but could also open the possibility of running shifts on nights with even smaller fractions of observing time, thus increasing the scientific output of the observatory.

Shift operation is a good experience, especially for new or young collaborators to get familiar with the detector. Running the shifts remotely prevents the collaborators from getting on-site experience of shifts. On the other hand, it opens the possibility of “dropping in” for just some hours or nights, if no travelling is needed. In addition, the shift might even be partially in normal working time instead of night time making it more attractive to follow the operation. Therefore, remote control rooms open the prospect of getting more people in close contact with the operation of the observatory.

For the operation of the FD, i. e. the data-acquisition, the run-control of the FD has been extended to a client server configuration where the communication is done through grid-authenticated connections using SOAP, a platform and language independent specification that allows one to couple the existing DAQ software [5] with the new remote software components via a message based communication. The client has been developed to cover the same functionality as the previous stand alone run-control running on a central server on the campus. For the development and evaluation phase, before the internet connection of AugerAccess is available, a virtual testbed has been set up at the Karlsruhe Institute of Technology in Germany. This testbed simulates the real systems including firewalls. Only the long connection to Malargüe and its reliability cannot be simulated realistically.

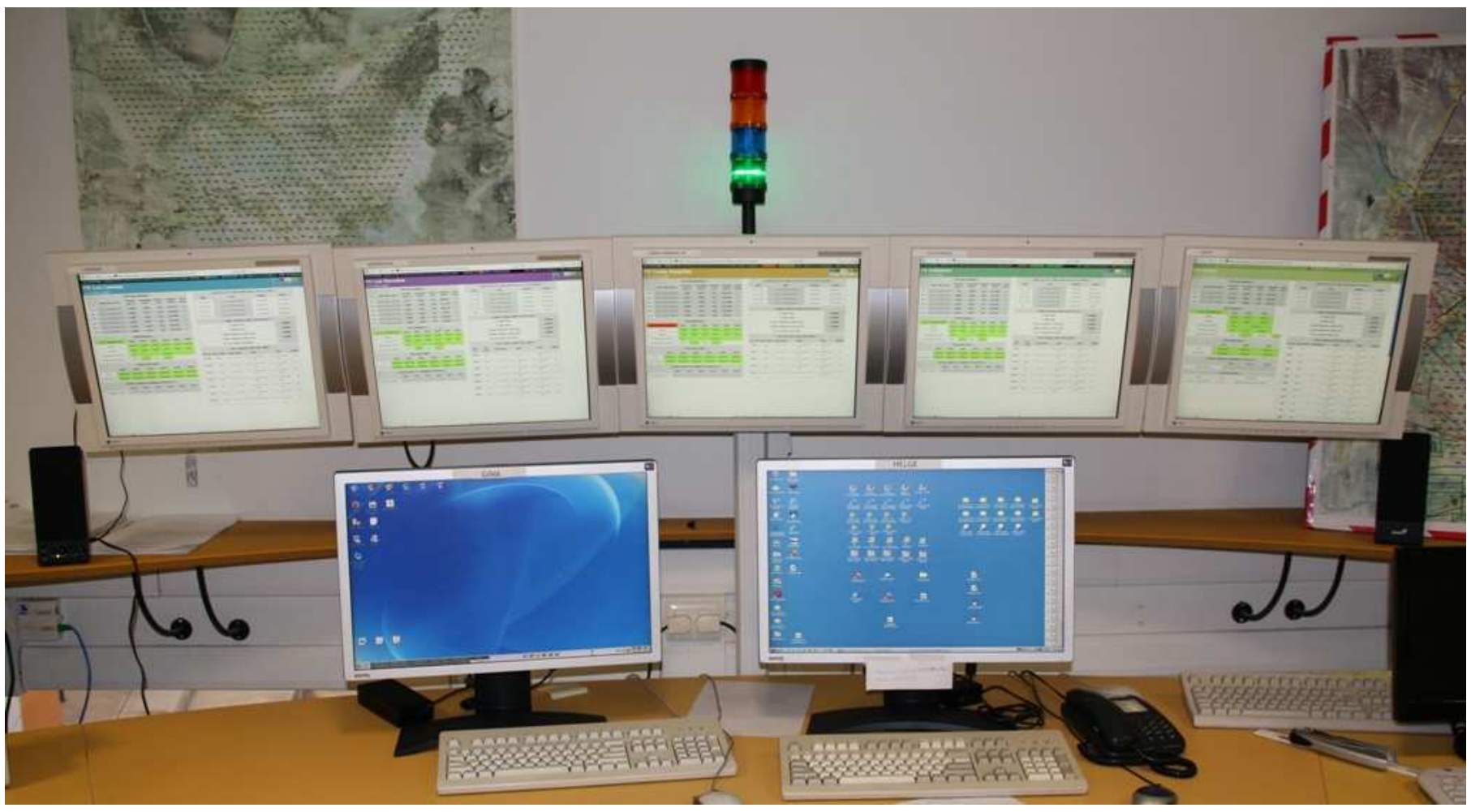

Figure 4: The control room at the observatory.

## 6 Remote control rooms

Ideally a remote control room offers the same functionality as the control room at the observatory, shown in fig.4. But, with the introduction of remote operation, additional communication measures have to be taken in the control room as well. The task force established requirements for making the remote control rooms functional. As at the observatory, two desktop systems, together with one spare system, have to be available for operation of the FD and the Lidar system. In addition, five screens on the wall show the status of each FD site with its telescopes. With one additional screen summarizing the status of the SD, we require at least six screens on the wall to present an overview of the detector systems. The remote control room has to have a prioritized internet connection. This can also be used for video-conferencing with the observatory control room. EVO [6] has been established to be used for video-conferencing within the Auger Collaboration, and we follow its recommendations for the necessary room microphone and video camera. If a network connection is lost, a regular phone with the ability to call international to Argentina has to be available in the remote control room. An alarm will be raised at the observatory if the connection to the remote control room is broken, thus notifying the on-site shifter on call to take over responsibility for the operation. As a first test, a remote control room has been installed at the Bergische Universität Wuppertal, Germany. From here, the first tests of passive shifts, i.e. initially without intervention from the remote control room, are being made. Once the technique is established, it is foreseen to install several remote control rooms distributed all over the world at the major collaborator sites.

## 7 Summary

The Pierre Auger Observatory and especially the FD data-acquisition is operated on site to guarantee a smooth operation. Access to the detector-systems for experts at remote sites increases the knowledge available for the observatory maintenance. Further, establishing remote shift operation can save substantial travelling time and cost by reducing the number of shifters by up to 60% and offers the possibility of supervising shifters by persons off site, increasing the quality of the data and the safety of the detector. The monitoring program of the Pierre Auger Observatory has been designed for a remote availability. Grid based technology has been used to implement the access to the run-control. Requirements for remote shift rooms have been established and the first passive tests are being performed.

# Atmospheric Monitoring at the Pierre Auger Observatory – Status and Update

KARIM LOUEDEC[1] FOR THE PIERRE AUGER COLLABORATION[2]
[1]*Laboratoire de l'Accélérateur Linéaire, Univ Paris Sud, CNRS/IN2P3, Orsay, France*
[2]*Observatorio Pierre Auger, Av. San Martín Norte 304, 5613 Malargüe, Argentina*
*(Full author list: http://www.auger.org/archive/authors_2011_05.html)*
*auger_spokespersons@fnal.gov*

**Abstract:** Calorimetric measurements of extensive air showers are performed with the fluorescence detector of the Pierre Auger Observatory. To correct these measurements for the effects introduced by atmospheric fluctuations, the Observatory operates several instruments to record atmospheric conditions across and above the detector site. New developments have been made in the study of the aerosol optical depth, the aerosol phase function and cloud identification. Also, for cosmic ray events meeting certain criteria, a rapid monitoring program has been developed to improve the accuracy of the reconstruction. We present an updated overview of performed measurements and their application to air shower reconstruction.

**Keywords:** Pierre Auger Observatory, ultra-high energy cosmic rays, air fluorescence technique, atmospheric monitoring, aerosols, clouds

## 1 Introduction

The Pierre Auger Observatory detects the highest energy cosmic rays with over 1600 water-Cherenkov detectors. It is surrounded by the fluorescence detector (FD) which consists of 27 telescopes grouped at four locations. The telescopes measure UV light emitted by atmospheric nitrogen molecules after having been excited by electrons produced in the extensive air showers. Since the fluorescence light is proportional to the energy deposited by the shower, the primary cosmic ray energy can be estimated if the fluorescence yield is known [1]. The FD telescopes are also used to reconstruct the slant depth of shower maximum ($X_{\max}$) which is sensitive to the mass composition of cosmic rays.

The Auger Observatory uses the atmosphere as a giant calorimeter. Light is produced and transmitted to the FD detector through an atmosphere with properties which change through the day. Thus, it is necessary to develop a sophisticated atmospheric monitoring program [2]. The production of fluorescence and Cherenkov photons in a shower depends on the atmospheric state variables such as temperature, pressure and humidity. When a photon travels from the shower to the observing telescopes, it can be scattered from its original path by molecules (*Rayleigh scattering*) and/or aerosols (*Mie scattering*).

In Fig. 1, the different experimental setups installed at Malargüe to monitor the atmosphere are listed. The state variables of the atmosphere are recorded at ground level using five weather stations. Above the Pierre Auger Observatory, the height-dependent profiles have been measured using meteorological radio-sondes launched from a helium balloon station. The balloon flight program ended in December 2010 after having been operated 331 times. The most recent monthly models of atmospheric state variables derived from these flights were developed from data between August 2002 and December 2008. Additionaly, a meteorological model has been implemented by the Auger Collaboration for air shower reconstruction [3] based on the Global Data Assimilation System (GDAS) developed by the National Oceanic and Atmospheric Administration (NOAA) which combines observations with results from a numerical weather prediction model.

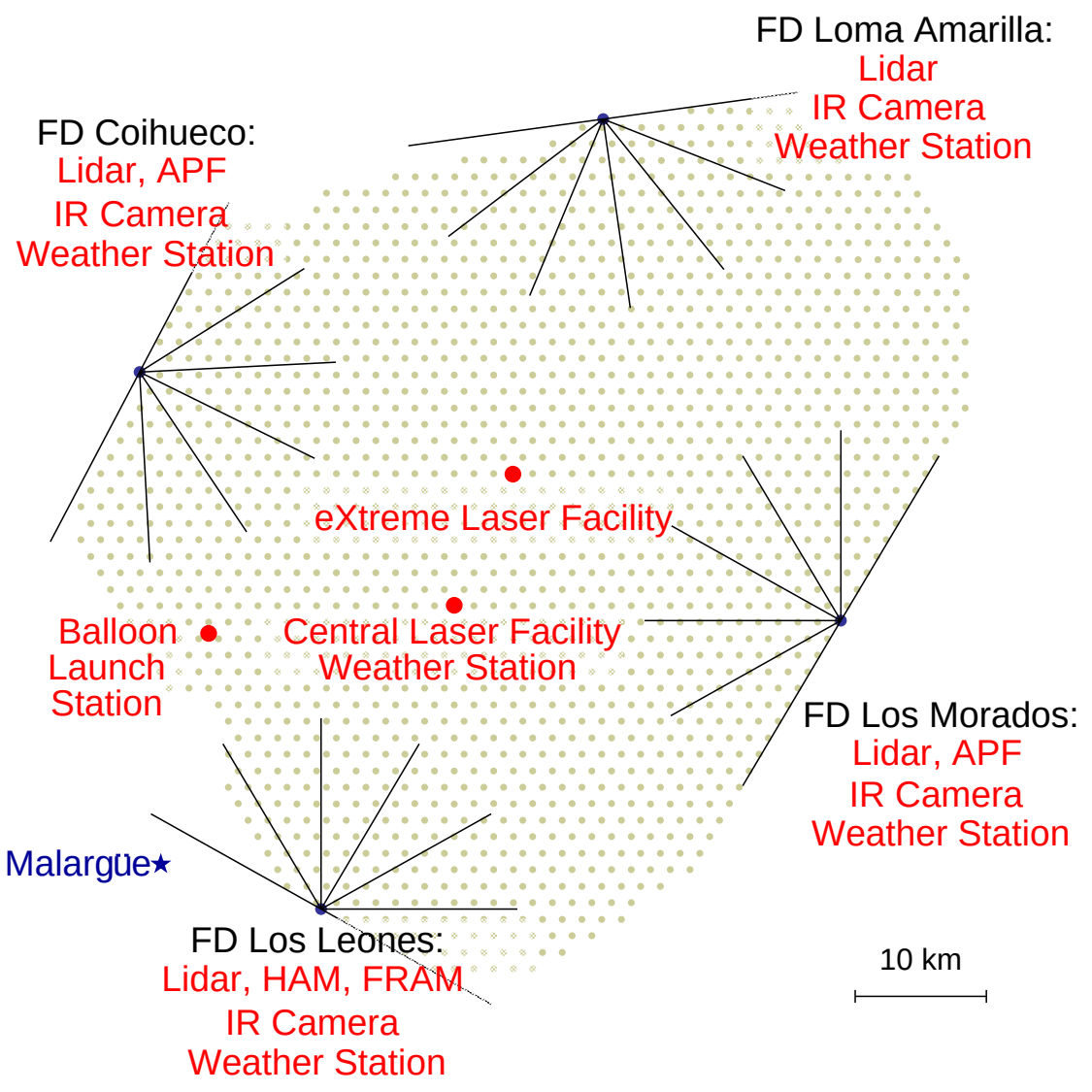

Figure 1: **Map of the Pierre Auger Observatory located close to Malargüe, in Argentina.** Each FD site hosts several atmospheric monitoring facilities.

Aerosol monitoring is performed using two central lasers (CLF / XLF), four elastic scattering lidar stations, two aerosol phase function monitors (APF) and two optical telescopes (HAM / FRAM). Also, a Raman lidar currently tested in Colorado (USA) is scheduled to be moved to the Auger Observatory for the Super-Test-Beam project [4]. For cloud detection, a Raytheon 2000B infrared cloud camera (IRCC) is installed on the roof of each FD building.

## 2 Extracting the Aerosol properties

Most of the aerosols are present only in the first few kilometers above the ground level. The aerosol component is highly variable in time and location. Two main physical quantities have to be estimated to correct the effect of the aerosols on the number of photons detected by the telescopes. These are the aerosol attenuation length, linked to the aerosol optical depth, and the aerosol scattering phase function.

### 2.1 Aerosol attenuation

Unlike molecular scattering, aerosol attenuation does not have an analytical solution. Aerosol optical depths are measured in the field at a fixed wavelength $\lambda_0$, chosen more or less in the centre of the nitrogen fluorescence spectrum. To evaluate the aerosol extinction at another incident wavelength, we use the power law

$$\tau_a(h,\lambda) = \tau_a(h,\lambda_0) \times (\lambda_0/\lambda)^{\gamma}\,, \qquad (1)$$

parameterized empirically, where $\tau_a(h,\lambda)$ is the vertical aerosol optical depth between the ground level and an altitude $h$, and $\gamma$ is known as the $\mathring{A}$ngström coefficient. Its value was estimated at the Auger Observatory by two facilities. The Horizontal Attenuation Monitor (HAM) provides a wavelength dependence with $\gamma = 0.7 \pm 0.5$ [5]. The small value of the exponent suggests a large component of large aerosols, i.e. aerosols larger than about 1 $\mu$m at least. This result is confirmed by the FRAM, the (F/Ph)otometric Robotic Atmospheric Monitor, a robotic optical telescope located about 30 m from the FD building at Los Leones [6]. In addition, an aerosol sampling program at ground level is being developed to study chemical composition and size distribution [7]. When enough statistics are accumulated, crosschecks between optical and direct measurements will be possible.

During FD operating shifts, vertical aerosol optical depth profiles are measured hourly by two lasers, the CLF and the XLF, located at sites towards the centre of the Auger array (see Fig. 2(a)). The incident wavelength is fixed at $\lambda_0 = 355$ nm and the mean energy per pulse is around 7 mJ, more or less the amount of fluorescence light produced by an air shower with an energy of $10^{20}$ eV. Only during FD data taking, more than six years of hourly data accumulated with the CLF is currently used to correct events for aerosol attenuation. The four lidars can also be used to estimate the optical depth and the horizontal attenuation for the four FD sites.

### 2.2 Angular dependence of aerosol scattering

The FD reconstruction of the cosmic ray energy must account not only for light attenuation between the shower and the telescopes, but also for direct and indirect Cherenkov light contributing to the recorded signal. Therefore, the scattering properties of the atmosphere need to be well estimated. The angular dependence of scattering is described by a phase function $P(\theta)$, defined as the probability of scattering per unit solid angle out of the beam path through an angle $\theta$. Whereas the molecular component is described analytically by the Rayleigh scattering theory, the Mie scattering cannot be described by a basic equation for the aerosol component. At the Auger Collaboration, the aerosol phase function (APF) is usually parameterized by the Henyey-Greenstein function

$$P_a(\theta|g) = \frac{1-g^2}{4\pi}\frac{1}{(1+g^2-2\,g\,\cos\theta)^{3/2}}, \qquad (2)$$

where $g = \langle\cos\theta\rangle$ is the asymmetry parameter. It quantifies the scattered light in the forward direction: a larger $g$ value corresponds to a stronger forward-scattered light.

At the Auger Observatory, the goal is to monitor the APF by estimating the $g$ parameter. Up to now, the phase function was measured by the APF monitors located at Coihueco and Los Morados [8]. Recently, a new method based on very inclined shots fired by the CLF was developed (laser shots with zenith angles higher than $86^{\rm o}$). Following the same idea as before, knowing the geometry of the laser shot and the signal recorded by the pixels, it is possible to extract the $g$ parameter. The advantage of this technique is that a $g$ parameter can be estimated for each FD site, and it can cover lower scattering angles (the angular range where larger aerosols could be detected). The two techniques give, on average, a similar value for the $g$ parameter, around $0.55$ (see Fig. 2(b)).

## 3 Cloud Detection

Cloud coverage has an influence on the FD measurements: it biases the estimation of the $X_{\rm max}$ by producing bumps or dips in the longitudinal profiles and it decreases the real flux of cosmic ray events. Thus, an event is reconstructed only if the cloud fraction is lower than $25\%$. Around $30\%$ of the events are rejected due to cloudy conditions. During the recent years, the Auger Collaboration has developed several methods to monitor the clouds all through the night.

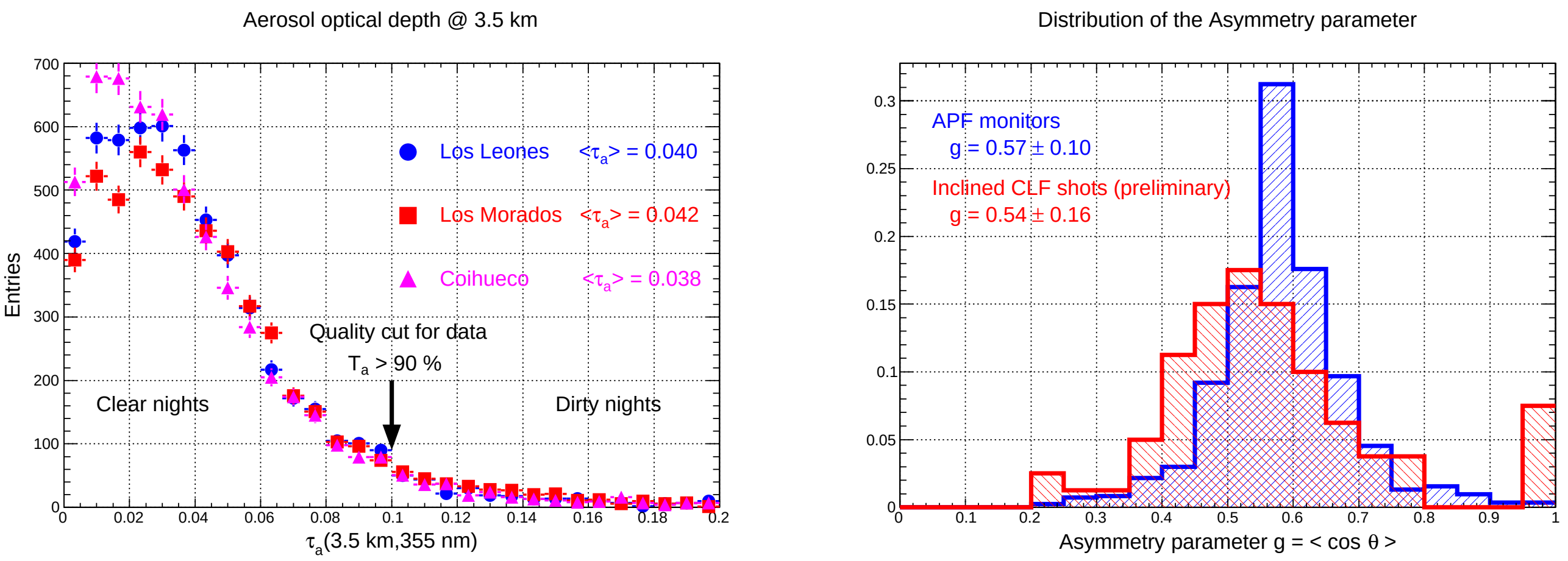

Figure 2: **Aerosol measurements.** *(a)* Vertical aerosol optical depth at 3.5 km above the fluorescence telescopes measured between January 2004 and December 2010. The transmission coefficient is defined as $T_a = \exp\,(-\tau_a)$. *(b)* Asymmetry parameter distribution measured by the APF monitor between June 2006 and June 2008, and the inclined CLF shots during 2008.

During FD data acquisition, each IRCC records 5 pictures of the FD field-of-view every 5 minutes: the raw image is converted into a binary image (white: cloudy / black: clear sky), then the fraction of cloud coverage for each pixel of the FD cameras is calculated producing the so-called FD pixels coverage mask. Different filters are applied in succession to remove camera artifacts and to get the clear sky background as uniform as possible. The cloud information for each pixel is updated every 5-15 minutes. These cloud masks are stored in a database and are now used as quality cuts after the air shower reconstruction. Fig. 3(a) gives an example of cloud mask ontop of on a telescope camera scheme, with the corresponding longitudinal profile showing dips and bumps typical for cloudy conditions.

A new method of identifying clouds over the Auger Observatory using infrared data from the imager instrument on the GOES-12/13 geostationary satellite is also used [9]. It obtains images using four infrared bands every 30 minutes. A brightness temperature $T_i$ is assigned to the $i$-th band. The whole array is described by 360 pixels: the infrared pixels projected on the ground have a spatial resolution of $\sim$ 2.4 km horizontally and $\sim$ 5.5 km vertically. The cloud identification algorithm uses the combination of $T_2 - T_4$ and $T_3$ to produce cloud probability maps (see Fig. 3(b)). Data from the satellite indicate clear conditions (cloud probability lower than 20%) during $\sim$ 50% of FD data acquisition and cloudy (cloud probability higher than 80%) during $\sim$ 20%.

Thanks to the IRCC and the data satellite, cloud coverage can be followed through the night. However, they cannot determine the cloud heights. At the Auger Observatory, this information is provided by the CLF / XLF and the lidars. The maximum height of clouds detected by these two techniques is between 12 km and 14 km, depending on the FD site. A cloud positioned along the vertical laser track scatters a higher amount of light, producing a peak in the recorded light profile. On the other hand, a cloud located between the laser and the FD site produces a local decrease in the laser light profile. Finally, the lidar telescopes sweep the sky during a 10-min scan every hour. Clouds are detected as strong light scatter regions in the backscattered light profiles recorded by the mirrors. The height of a cloud is deduced from the arrival time of the detected photons. These measurements have identified two cloud populations located at about 2.5 km and 8.0 km above sea level.

## 4 Rapid Atmospheric Monitoring

During FD data acquisition, showers meeting certain criteria are used to trigger dedicated measurements by the weather balloon, lidar and FRAM to get a detailed description of the atmosphere, partly in the vicinity of the shower track. The rapid monitoring system occurs as follows: a hybrid reconstruction using all the detectors and calibration data available is performed on shower data measured at most 10 min after their detection. Only events passing customized quality cuts activate subsystems of the rapid monitoring procedure.

The *Balloon-the-Shower* (BtS) program was dedicated to perform an atmospheric sounding within about three hours after the detection of a high-energy event. The measurements obtained by launching weather balloons provide altitude profiles of the air temperature, pressure and humidity up to about 23 km above sea level. Such a delay is expected to be compatible with the temporal variation of these atmospheric state variables. Between March 2009 and December 2010, 53 launches were performed covering 63 selected events. Using monthly models instead of BtS profiles introduces an uncertainty on the energy $\Delta E/E = (0.43 \pm 2.38)\%$ and on the position of the shower maximum $X_{\max} = (0.60 \pm 5.93)\,\mathrm{g\,cm^{-2}}$, for showers with energies between $10^{19.3}$ eV and $10^{19.9}$ eV [10]. The balloon program, including BtS, was terminated in December

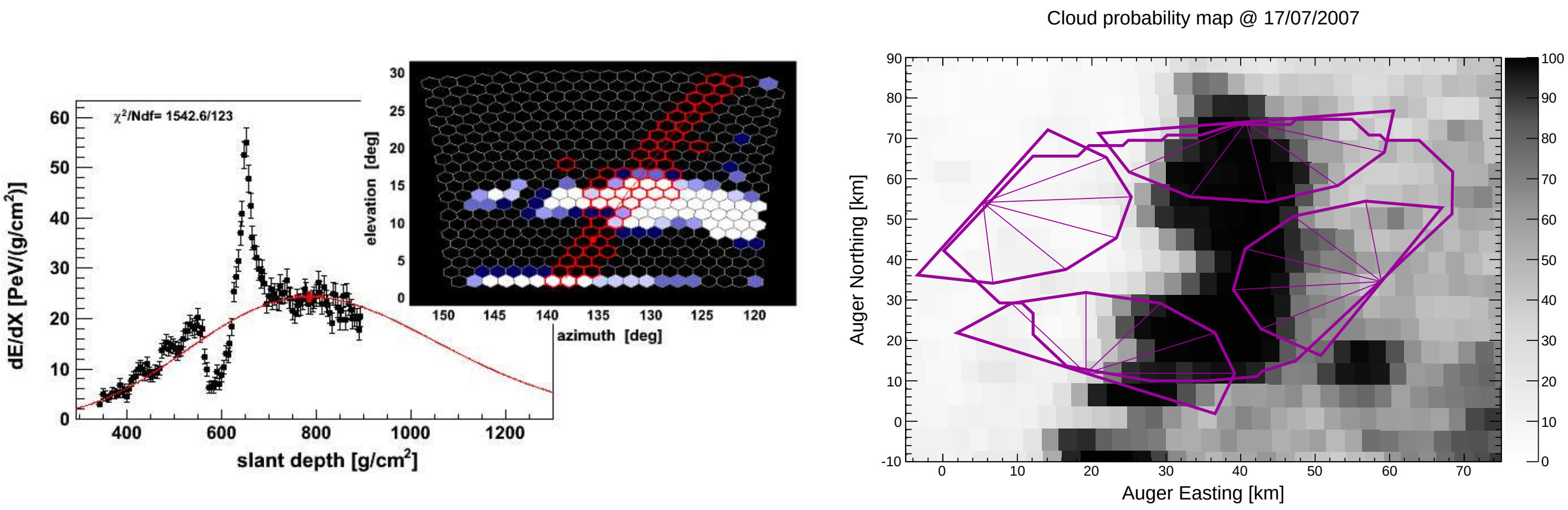

Figure 3: **Cloud coverage.** *(a)* Display of an event recorded by a FD camera, with index of cloud coverage for each pixel (lighter pixels mean higher cloud coverage). Pixels with no cloud are in black. The associated longitudinal profile is also shown. *(b)* Cloud probability map for 17/07/2007 at 01:09:24 UT. Pixels and their cloud probability are colored in accordance with the scale to the right of the map.

2010 and is now replaced by numerical meteorological profiles [3].

The motivation of the *Shoot-the-Shower* (StS) program is to identify non-uniformities – especially clouds or aerosol layers – that affect light transmission between the shower and detector. The StS sequence, or lidar scan, goes from the ground to the top of the FD field of view, all along the shower track. Each shooting direction is separated from the previous one by $1.5^{\circ}$. Between January 2009 and July 2010, $70$ hybrid events passed the online quality cuts and triggered a StS scan. $9$ out of $70$ StS were aborted because of various hardware issues, reducing the sample to $61$ events. The StS scans were analyzed and clouds were detected in $20$ of the $61$ events.

FRAM can be programmed to scan the shower path, recording images with a wide-field CCD camera mounted on the telescope. For each event passing the different cuts and being close to Los Leones, a sequence of $10$ to $20$ CCD images is produced. The CCD images can be analyzed automatically, and an atmospheric attenuation is obtained for each image. This goal is achieved using the photometric observations of selected standard, i.e. non-variable, stars. From January 2010 to July 2010, 173 successful observations were done. These observations permitted detection of the presence of clouds or aerosol layers and images corresponding to an attenuation coefficient higher than expected for a clear sky.

## 5 Conclusion & Future Plans

Thanks to a collection of atmospheric monitoring data, the Auger Collaboration has accumulated a large database of atmospheric measurements. This effort significantly reduced the systematic uncertainties in the air shower reconstruction. The rapid monitoring, focused on the highest energy events, also reduced uncertainties due to atmospheric effects. The program can be easily extended to incorporate new instruments as the Raman lidar, expected to be installed close to the CLF in 2011 for the Super-Test-Beam project. Also, a design study for new elastic scattering lidars has been undertaken. The goals are a more compact lidar with better mechanical stability and weatherproofing.

Recently, a public conference took place at Cambridge, UK, where interdisciplinary science at the Pierre Auger Observatory (IS@AO) was discussed [11]. During this meeting, scientists from a variety of disciplines talked about the potential of the Observatory site and to exchanged ideas exploiting it further. Among them, we can cite the possible connection between clouds, thunderstorms and cosmic rays, a larger aerosol sampling program and the detection of atmospheric gravity waves.

# Implementation of meteorological model data in the air shower reconstruction of the Pierre Auger Observatory

MARTIN WILL[1] FOR THE PIERRE AUGER COLLABORATION[2]
[1]*Karlsruher Institut für Technologie, Institut für Kernphysik, Karlsruhe, Germany*
[2]*Observatorio Pierre Auger, Av. San Martín Norte 304, 5613 Malargüe, Argentina*
*(Full author list: http://www.auger.org/archive/authors_2011_05.html)*
*auger_spokespersons@fnal.gov*

**Abstract:** The Global Data Assimilation System (GDAS) provides altitude-dependent profiles of the main state variables of the atmosphere. The original data and their application to the air shower reconstruction of the Pierre Auger Observatory are described. By comparisons with radiosonde and weather station measurements obtained on-site at the observatory and averaged monthly mean profiles, the informative value of the data is shown.

**Keywords:** cosmic rays, extensive air showers, atmospheric monitoring, atmospheric models

## 1 Introduction

The Pierre Auger Observatory [1, 2] is located near Malargüe in the province Mendoza, Argentina. Extensive air showers are measured using a hybrid detector, consisting of a Surface Detector (SD) array and five Fluorescence Detector (FD) buildings. For the reconstruction of air showers, the atmospheric conditions at the site have to be known quite well. This is particularly true for reconstructions based on data obtained by the FD [3]. Weather conditions near the ground and height-dependent profiles of temperature, pressure and humidity are relevant.

Atmospheric conditions over the observatory are measured by intermittent meteorological radio soundings. Ground-based weather stations measure surface data continuously. The profiles from the ascents of weather balloons were averaged to obtain local models, called (new) Malargüe Monthly Models (nMMM) [3]. However, performing radio soundings imposes a large burden on the collaboration.

Here, we investigate the possibility of using data from the Global Data Assimilation System (GDAS), a global atmospheric model, for the site of the Auger Observatory [4]. The data are publicly available free of charge via READY (Real-time Environmental Applications and Display sYstem). Each data set contains all the main state variables as a function of altitude.

## 2 Global Data Assimilation System

In the field of Numerical Weather Prediction, data assimilation is the process by which the development of a model incorporates the real behavior of the atmosphere as found in meteorological observations [5]. The atmospheric models describe the atmospheric state at a given time and position. Three steps are needed to perform a full data assimilation:

1. Collect data from meteorological measuring instruments placed all over the world.

2. Forecast the atmospheric state from the current state using numerical weather prediction.

3. Use data assimilation to adjust the model output to the measured atmospheric state, resulting in a 3-dimensional image of the atmosphere.

At a given time $t_0$, the observations provide the value of a state variable. A model forecast for this variable from a previous iteration exists for the same time. The data assimilation step combines observation and forecast. This analysis is the initial point for the weather prediction model to create the forecast for a later time $t_1$.

The Global Data Assimilation System [6] is an atmospheric model developed at NOAA's[1] National Centers for Environmental Prediction (NCEP). The numerical weather prediction model used is the Global Forecast System (GFS).

Data are available for every three hours at 23 constant pressure levels – from 1000 hPa ($\approx$ sea level) to 20 hPa ($\approx$ 26 km) – on a global 1°-spaced latitude-longitude grid (180° by 360°). Each data set is complemented by data for the surface level. The data are made available online [6].

1. National Oceanic and Atmospheric Administration

For the site of the observatory, applicable GDAS data are available starting June 2005. Because of the lateral homogeneity of the atmospheric variables across the Auger array [3], only one location is needed to describe the atmospheric conditions. The grid point at 35° S and 69° W was chosen, at the north-eastern edge of the SD array.

Our database used for air shower analyses describing the main state variables of the atmosphere contains values for temperature, pressure, relative humidity, air density, and atmospheric depth at several altitudes. The first three quantities are directly available in the GDAS data. Air density and atmospheric depth can be calculated. The surface data contain height and pressure at the ground, as well as relative humidity and temperature 2 m above the ground.

## 3 GDAS vs. Local Measurements

To validate the quality of GDAS data and to verify their applicability to air shower reconstructions for the Auger Observatory, we compare the GDAS data with local measurements – atmospheric soundings with weather balloons and ground-based weather stations. The nMMM are also shown in some comparisons as a reference since they were the standard profiles used in reconstructions until recently.

### 3.1 GDAS vs. Soundings with Weather Balloons

Local radio soundings are performed above the array of the Auger Observatory since 2002, but not on a regular basis. To provide a set of atmospheric data for every measured event, data were averaged to form monthly mean profiles.

The nMMM have been compiled using data until the end of 2008. The uncertainties for each variable are given by the standard error of the differences within each month together with the absolute uncertainties of the sensors measuring the corresponding quantity. For atmospheric depth profiles, a piecewise fitting procedure is performed to ensure a reliable application of these parameterizations to air shower simulation programs. An additional uncertainty is included which covers the quality of the fitting procedure.

Comparing the monthly models with ascent data until the end of 2008 shows, by construction, only small deviations [3]. In the comparison displayed in Fig. 1, only radiosonde data from 2009 and 2010 are used to illustrate the strength of the GDAS model data – the data set of local soundings is independent of the nMMM. The error bars denote the RMS of the differences at each height. These uncertainties are larger for the nMMM than for GDAS data, the latter describe the conditions of the years 2009 and 2010 better. In contrast, the GDAS data represent the local conditions much better and the intrinsic uncertainty is consistently small. For earlier years, the GDAS data fit the measured data equally well or better than the nMMM which were developed using the data from these years.

The GDAS data fit the radiosonde data in the upper part of the atmosphere, especially in the field of view of the fluorescence detectors. Possible inconsistencies between local measurements and GDAS data close to the surface are investigated using weather station data.

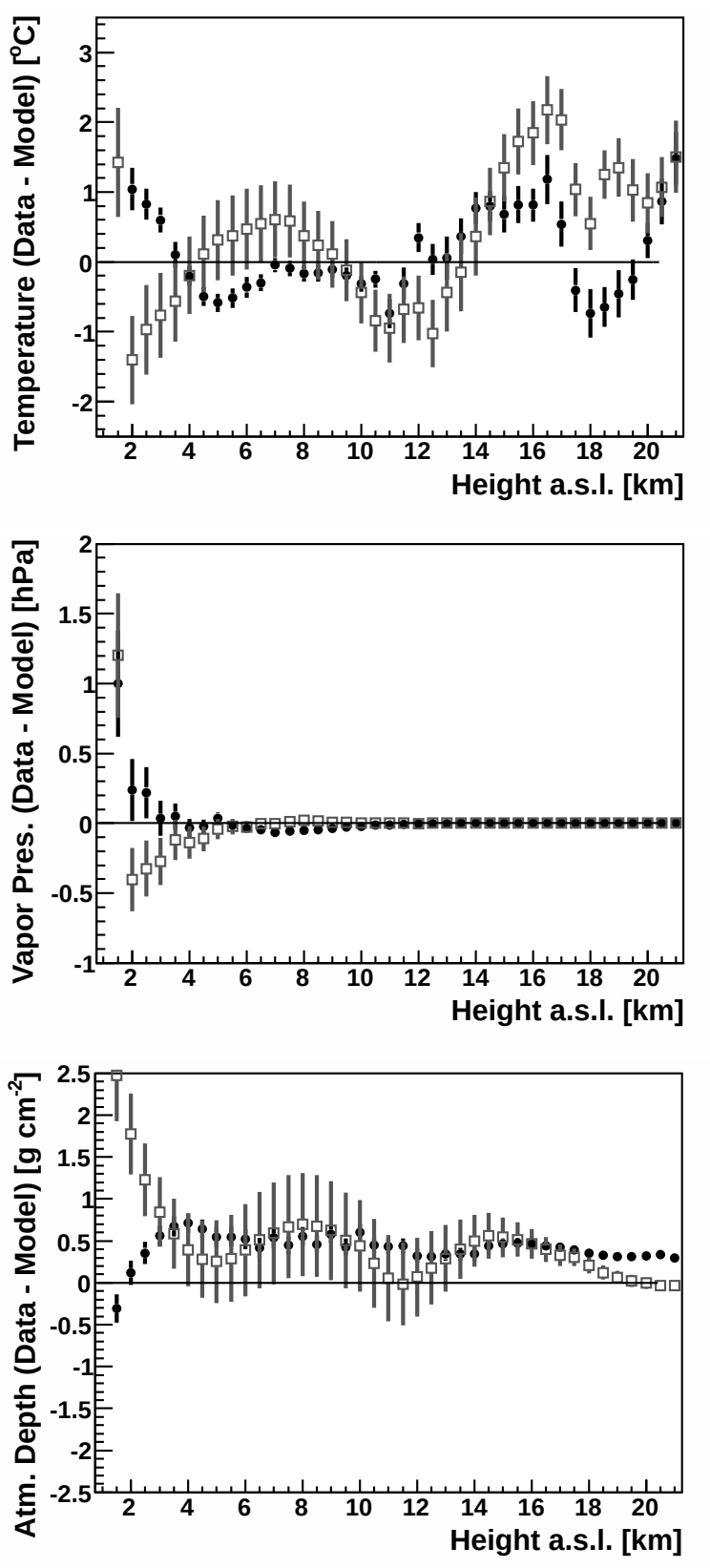

Figure 1: Difference between measured individual radiosonde data and the corresponding GDAS data (black dots) and nMMM (gray squares) versus height for all ascents performed in 2009 and 2010.

### 3.2 GDAS vs. Ground Weather Stations

Five ground weather stations continuously monitor atmospheric values, at about 2 to 4 m above surface level. Four are located at the FD stations, one was set up near the center of the array at the Central Laser Facility (CLF). To make sure that the GDAS data describe the conditions at the ground reasonably well, the values provided by the GDAS data set are compared to all available weather station data. The GDAS data are interpolated at the height of the station.

In Fig. 2, the differences between measured weather station data and GDAS data are shown for the stations close to the CLF and the FD station Loma Amarilla (LA). All data measured in 2009 were used. Temperature, pressure (not shown), and vapor pressure are in similar agreement as GDAS data with local sounding data close to ground. The mean difference in temperature for the CLF station is 1.3 K and $-0.3$ K for the LA station. For vapor pressure, the means are $-0.2$ hPa (CLF) and $-0.7$ hPa (LA). The differences between the GDAS and the weather station data are of the same order as the difference in data of two different stations [4]. The GDAS data fit the measured data at the observatory very well and are a suitable replacement for the nMMM and subsequently for radiosonde ascents.

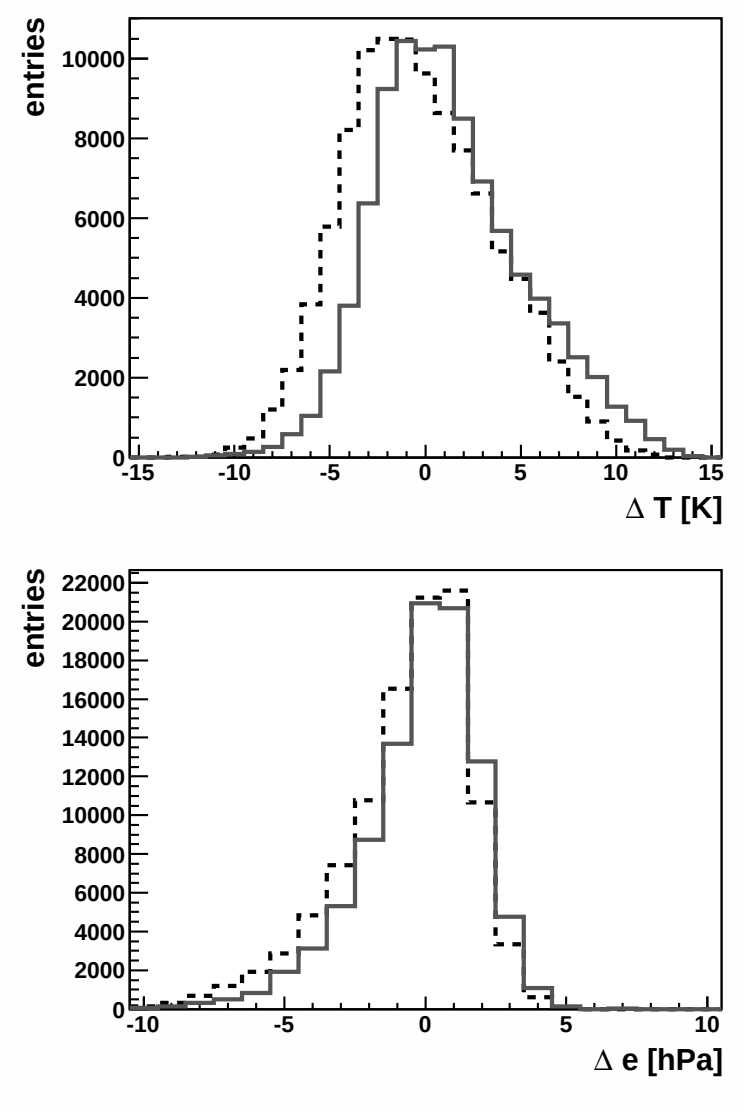

Figure 2: Difference between data measured at weather stations and from GDAS, all data of 2009 are used. The difference ('GDAS' minus 'weather station') in temperature and water vapor pressure is shown for the weather station at the CLF (dashed line) and the station at Loma Amarilla (solid line).

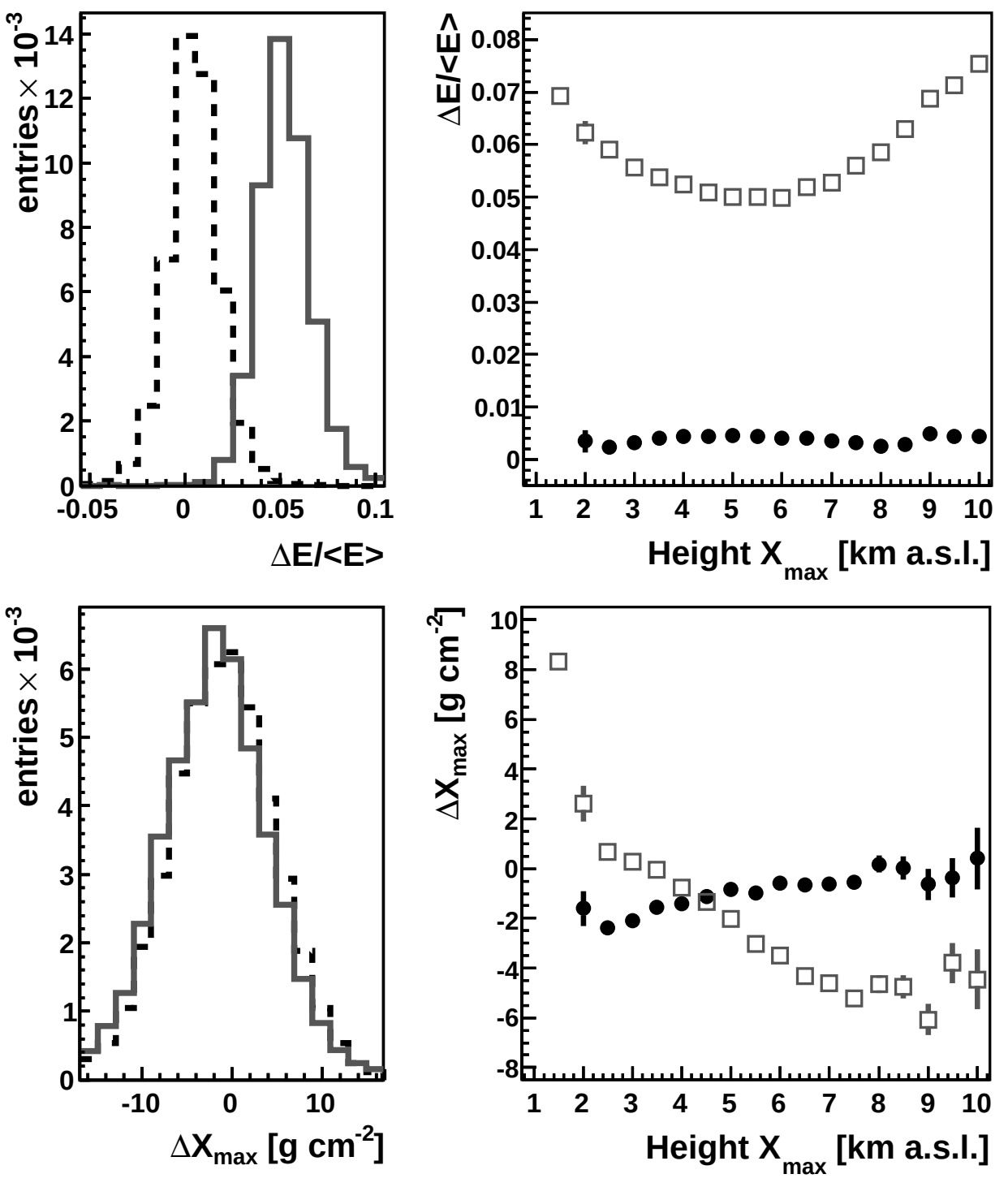

Figure 3: Difference of reconstructed $E$ (top) and $X_{\text{max}}$ (bottom), plotted versus geometrical height of $X_{\text{max}}$ in the right panels. Dashed black line or black dots for $\mathsf{FY}_{\text{mod}}^{\text{GDAS}}$ minus $\mathsf{FY}_{\text{mod}}$, and solid red line or open red squares for $\mathsf{FY}_{\text{mod}}^{\text{GDAS}}$ minus $\mathsf{FY}$.

## 4 Air Shower Reconstruction

To study the effects caused by using the GDAS data in the air shower reconstruction, all air shower data between June 1, 2005 and the end of 2010 were used. The change of the atmosphere's description will mainly affect the reconstruction of the fluorescence data. Varying atmospheric conditions alter the fluorescence light production and transmission [3]. The fluorescence model we use determines the fluorescence light as a function of atmospheric conditions [7], parameterized using results from the AIRFLY fluorescence experiment [8, 9].

### 4.1 Data Reconstruction

The following analysis is based on three sets of reconstructions. The first set, FY, is the reconstruction applying an atmosphere-dependent fluorescence yield calculation without temperature-dependent collisional cross sections and humidity quenching [10]. The nMMM are used in the calculations. For the second set, $\mathsf{FY}_{\text{mod}}$, all atmospheric effects in the fluorescence calculation are taken into account. Again, the nMMM are used. For the third set, $\mathsf{FY}_{\text{mod}}^{\text{GDAS}}$, the nMMM are exchanged with the new GDAS data in combination with the modified fluorescence calculation. Comparing the reconstruction sets with each other, the variation of the reconstructed primary energy $E$ and the position of shower maximum $X_{\text{max}}$ can be determined, see Fig. 3.

Using GDAS data in the reconstruction instead of nMMM affects $E$ only slightly. The mean of the difference $\mathsf{FY}_{\text{mod}}^{\text{GDAS}}$ minus $\mathsf{FY}_{\text{mod}}$ is 0.4% with an RMS of 1.4%. For the reconstructed $X_{\text{max}}$, only a small shift of $-1.1$ g cm$^{-2}$ is found with an RMS of 6.0 g cm$^{-2}$. Comparing the full atmosphere-dependent reconstruction $\mathsf{FY}_{\text{mod}}^{\text{GDAS}}$ with FY, a clear shift in $E$ can be seen: an increase in $E$ by 5.2% (RMS 1.5%) and a decrease of $X_{\text{max}}$ by $-1.9$ g cm$^{-2}$ (RMS 6.3 g cm$^{-2}$). These modified fluorescence settings are now used in the Auger reconstruction, in conjunction with other improvements to the procedure, see [11].

The difference in reconstructed $E$ vs. mean $E$ reveals a negligible effect for small energies, increasing slightly towards higher energies [4]. For $X_{\text{max}}$ differences, the dependence on mean $E$ is of minor importance. The description of atmospheric conditions close to ground is very difficult in monthly mean profiles since the fluctuations in temperature and humidity are larger below 4 km than in the upper layers of the atmosphere. Consequently, a more precise description of actual atmospheric conditions with GDAS than with nMMM will alter the reconstruction for those air showers which penetrate deeply into the atmosphere. The full atmosphere-dependent fluorescence calculation alters the light yield for conditions with very low temperatures, corresponding to higher altitudes. Showers reaching their maximum in the altitude range between 3 and 7 km show a difference in $E$ around 5%, see Fig. 3, upper right. However, showers with very shallow or very deep $X_{\text{max}}$ are reconstructed with a 7–8% higher energy than using the atmosphere-independent fluorescence calculation. The $X_{\rm max}$ sensitivity to the different parameterizations of the atmosphere and fluorescence yield (Fig. 3, lower right) is consistent to what has been reported in [12].

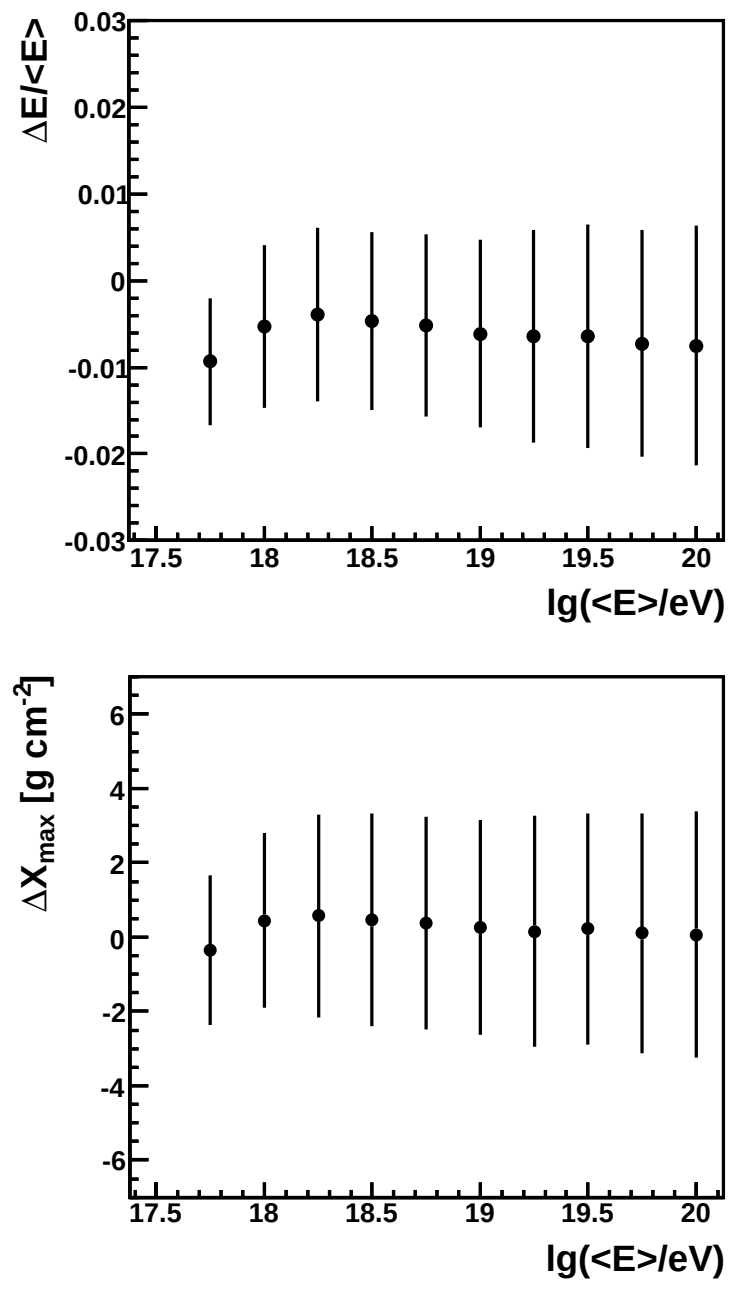

Figure 4: Energy difference (top) and $X_{\rm max}$ difference (bottom) vs. reconstructed FD energy for simulated showers. Error bars denote the true RMS spread.

### 4.2 Impact on Reconstruction Uncertainties

To study the effect that the GDAS data have on the uncertainties of air shower reconstructions, air showers induced by protons and iron nuclei are simulated with energies between $10^{17.5}$ eV and $10^{20}$ eV. The fluorescence light is generated using temperature-dependent cross sections and water vapor quenching. The times of the simulated events correspond to 109 radio soundings between August 2002 and December 2008 so that realistic atmospheric profiles can be used in the simulation. All launches were performed at night during cloud-free conditions. After the atmospheric transmission, the detector optics and electronics are simulated. The resulting data are reconstructed using the radiosonde data, as well as the GDAS data.

Some basic quality cuts are applied to the simulated showers. The same study has been performed to determine the uncertainties of the nMMM [13]. The systematic error due to different atmospheres was found to be less than 1% in $E$ and less than 2 g cm$^{-2}$ in $X_{\rm max}$. Between $10^{17.5}$ eV and $10^{20}$ eV, energy-dependent reconstruction uncertainties of $\pm$1% and $\pm$5 g cm$^{-2}$ for low energies and up to $\pm$2% and $\pm$7 g cm$^{-2}$ for high energies were found.

In Fig. 4, the influence on the reconstruction due to GDAS data is shown. A deviation from zero indicates a systematic error, the error bars denote the true RMS spread of all simulated events and are a measure of the reconstruction uncertainty due to this atmospheric parameterization. The systematic shifts in $E$ are of the same order, below 1%, and the shifts in $X_{\rm max}$ are much smaller, less than 0.5 g cm$^{-2}$, than for nMMM. The RMS spread is considerably smaller, $\pm$0.9% and $\pm$2.0 g cm$^{-2}$ for low energies, $\pm$1.3% and $\pm$3.5 g cm$^{-2}$ for high energies. The $E$ uncertainty at low energies is comparable to that introduced by the nMMM. At high energies, the uncertainty is almost half. For $X_{\rm max}$, the uncertainties at all energies is halved.

This study of the reconstruction uncertainties using different atmospheric parameterizations further demonstrates the advantages of GDAS data over the nMMM.

## 5 Conclusion

The comparison of GDAS data for the site of the Auger Observatory in Argentina with local atmospheric measurements validated the adequate accuracy of GDAS data with respect to spatial and temporal resolution. An air shower reconstruction analysis confirmed the applicability of GDAS for Auger reconstructions and simulations, giving improved accuracy when incorporating GDAS data instead of nMMM. Also, the value of using an atmosphere-dependent fluorescence description has been demonstrated.

# Night Sky Background measurements by the Pierre Auger Fluorescence Detectors and comparison with simultaneous data from the UVscope instrument

ALBERTO SEGRETO[1] FOR THE PIERRE AUGER COLLABORATION[2]
[1]*Ist. Astrofisica Spaziale e Fisica Cosmica, IASF-Pa/INAF, Via La Malfa 153, 90146, Palermo, Italy*
[2]*Observatorio Pierre Auger, Av. San Martín Norte 304, 5613 Malargüe, Argentina*
*(Full author list: http://www.auger.org/archive/authors_2011_05.html)*
*auger_spokespersons@fnal.gov*

**Abstract:** Due to the AC coupling of PMTs in the Pierre Auger Fluorescence Detectors, the slowly varying current induced by the Night Sky Background cannot be directly measured. Estimate of the Night Sky Background is however indirectly obtained by the statistical analysis of the current fluctuations, whose variance is recorded every 30 s for each pixel. We present the procedure used to convert the raw background data of the Fluorescence Detector to an absolute calibrated Night Sky Background flux and compare the results with data taken simultaneously with the UVscope instrument (a single photon counting UV detector) placed on the roof of one of the Fluorescence Detector buildings. We also show how the measurements of Night Sky Background flux can effectively be used as a general non-invasive tool to verify the end-to-end calibration of a large aperture telescope with multi-pixel cameras.

**Keywords:** Pierre Auger Observatory,fluorescence detector, calibration, night sky background

## 1 Introduction

The Pierre Auger Observatory includes 27 Fluorescence Detector (FD) telescopes in 4 sites [1]. In brief, a single FD telescope is composed by an aperture system with a UV filter, a spherical mirror (13 $m^2$) and a camera formed by an array of 440 hexagonal photomultipliers (PMTs), arranged in a matrix of $20 \times 22$ pixels in the focal surface. Each pixel has a hexagonal field of view of 1.5 °.

The signals from the PMTs are amplified, filtered and continuously digitized by 10 MHz (20 MHz for the latest telescopes named HEAT) 12 bit Flash Analog to Digital Converters (FADCs). Signals from the PMTs are AC coupled to the analog electronics, so the slowly varying anode current proportional to the Night Sky Background (NSB) light cannot be directly observed in the FADC traces.

However, due to the random nature of the process involved, there is a direct relation between the average anode current and its statistical fluctuations that, being fast varying signals, are not entirely blocked by the AC coupling. For this reason, the variances of the FADC traces are continuously computed and recorded for each pixel, in background files.

In this paper we report on the procedure to convert the recorded FD variance signal to an absolute calibrated photon flux and compare the results with measurements performed with an independent detector, named UVscope, observing, in Single Photon Counting mode, the same region of the sky. We also show how the NSB flux measurements can effectively be used as a general non-invasive tool to verify the end-to-end calibration of large aperture telescopes with multi-pixel cameras.

## 2 The FADC Variance

The statistical analysis of the FADC traces has been implemented in the FPGA logic on the FD first level trigger boards. The FADC variance and pedestal for each pixel are usually recorded with a sampling time of 30 s. Indicating with $\sigma^2_{FADC}$ the variance of the FADC trace in ADC-counts$^2$, and with $I_{FADC}$ the mean anode current in ADC-counts, the direct proportionality existing between these two variables can be expressed by the constant $K_V$, given by [2] :

$$K_V \doteq \frac{I_{FADC}}{\sigma^2_{FADC}} = \frac{10}{2 \cdot G \cdot (1+\nu_G) \cdot F} \quad (1)$$

where G is the PMT gain (FADC-counts/photoelectron), $\nu_G$ is the gain variance factor of the PMT, and F is the noise equivalent bandwidth (MHz) from the complete analog signal chain. Of course the electronic noise gives a small additional contribution to the measured FADC variance. For sake of simplicity, we will omit this term in the formula.

In the following sections we explain the procedure used to convert the recorded FADC variance values to an absolute calibrated photon flux.

## 3 The FD absolute calibration costants

The FD absolute end-to-end calibrations [3] provide, for each camera pixel, the scaling factors to convert any pulsed signal observed in the FADC traces to the corresponding photon flux at the telescope aperture. This calibration is performed periodically by placing in front of the FD telescope diaphragm an extended (2.5 m diameter) uniform light source that, for its shape, is called "Drum".

The "Drum calibration constants", $K_D$, are provided in units of photons/FADC-counts, being conventionally defined so that, for a monochromatic source at the reference wavelength $\lambda_D$ = 375 nm, they enable one to convert the FADC signal to the number of photons arriving at the FD diaphragm in the FADC integration time. The flux in photons/m$^2$/sr/s of a pulsed monochromatic source at wavelength $\lambda_D$, fully illuminating the pixel, is then obtained by:

$$\Phi_{\lambda_D} = \frac{K_D}{\Delta T \cdot A \cdot \Omega} \cdot I_{FADC} \quad (2)$$

where $I_{FADC}$ is the pulse amplitude (in FADC-counts), and the parameters have the following fixed numerical values:

- $\Delta T$ = 100 ns (FADC Integration Time, 50 ns for the HEAT telescope)
- $A = 3.8\,m^2$ (Area of FD telescope diaphragm)
- $\Omega = 5.94 \times 10^{-4}$ sr (Pixel Solid Angle Acceptance)

For a non-monochromatic source with spectral distribution $\Phi_\lambda$, the number of equivalent photons at $\lambda_D$ must be obtained by:

$$\Phi_{\lambda_D} = \int_0^\infty \Phi_\lambda \cdot F_\lambda \cdot d\lambda \quad (3)$$

where $F_\lambda$ is the pixel spectral response normalized to the reference wavelength $\lambda_D$ ($F_{\lambda_D} = 1$).

Of course, Eq. 2 can be applied only to fast ($\mu$s time scale) pulsed signals observed in the FADC traces, while the slowly varying component of the illuminating photon flux can only be derived from the variance of the FADC trace as explained in the following section.

## 4 Calibration of the FADC variance

Combining Eq. 1, 2, and 3, the end-to-end conversion formula to obtain the average flux (in photons/m$^2$/sr/ns/nm) of a slow-varying, non-monochromatic source fully illuminating the pixel, is expressed by:

$$< \Phi_\lambda > \doteq \frac{\int_0^\infty \Phi_\lambda \cdot F_\lambda \, d\lambda}{\int_0^\infty F_\lambda \, d\lambda} = \frac{K_D \cdot K_V}{\Delta T \cdot A \cdot \Omega \cdot \int_0^\infty F_\lambda \, d\lambda} \cdot \sigma^2_{FADC} \quad (4)$$

This formula then allows one to convert the variance signal recorded in the FD background files (after subtraction of the electronic noise) to the absolute calibrated value of the diffuse component of the NSB flux.

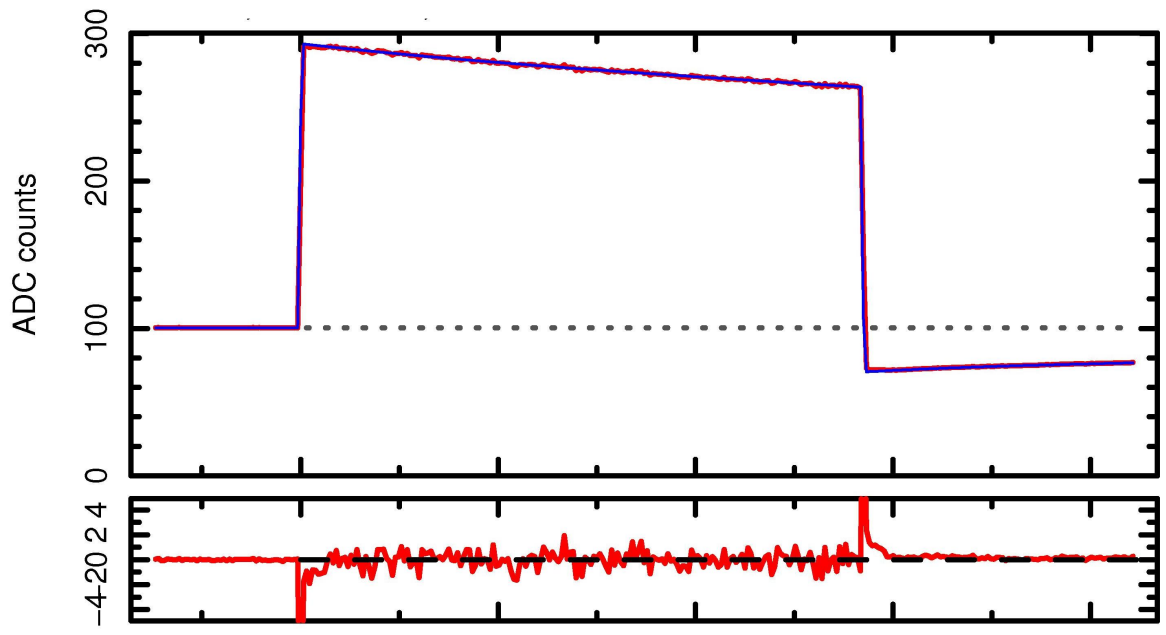

Figure 1: Upper panel: FADC average trace of 400 Cal. A pulses; Bottom panel: residuals after fitting an analytical model to the pulse. The delayed response of afterpulses (not included in the model) and the increase of fluctuations associated to the illuminating photons are evident.

In addition to the uncertainty associated with the absolute calibration constant $K_D$, the accuracy of the flux measurements obtained from this equation is also affected by pixel-to-pixel variations in the spectral response $F_\lambda$ and by the uncertainty in the variance scaling factor $K_V$.

Multi-wavelength calibration [4] has shown that pixel-to-pixel dispersion of the spectral response is of the order of few percent so, as a first approximation, we use the average spectral response of the FD pixels (reported in [4]), to obtain a common scaling factor:

$$\int_0^\infty F_\lambda \, d\lambda = 73.5\,\text{nm} \quad (5)$$

The uncertainty in the constant $K_V$, is affected (Eq. 1) by the cumulative indetermination in all the three physical parameters G, $\nu_G$ and F. Since the values of these parameters may change significatively from pixel to pixel and no accurate measurements are available for every pixel, the overall indeterminacy in $K_V$, (computed, e.g., by adopting the typical mean values) would be unacceptably high.

Rather than deriving $K_V$ from the pixel physical parameters, it is possible to directly obtain its value, with a much better statistical accuracy, from analysis of the relative calibration runs "Cal. A" as explained in the following section.

## 5 Cal. A calibrations

To follow the short term behavior of the FD photomultipliers, three relative calibrations, Cal. A, B, C are performed each night of data taking. In the Cal. A, a sequence of square-type light pulses (57 $\mu$s) are produced at a rate of 1/3 Hz with a very stable bright LED source (at 470 nm) and transmitted with light guides to a Teflon diffuser located in the center of the mirror, thus illuminating directly the camera photomultipliers [5].

In Fig. 1 we show an example of FADC trace obtained as an average of 400 Cal. A LED pulses. The analysis of the Cal. A FADC traces enables one to evaluate both the pulse amplitude (corrected for the AC coupling by fitting an exponential model to the pulse shape) and the associated variance increment (from statistical analysis of the fit residual). Their ratio, after subtraction of the contribution from the electronic noise, gives a direct measurement of the variance calibration factor $K_V$ of the pixel, without requiring any knowledge of the physical parameters on which it depends. The statistical uncertainty in the $K_V$ measurement is dominated by the statistical uncertainty in the computation of FADC trace variance associated with the light pulse. The relative standard deviation of $K_V$ can then be expressed by:

$$\frac{\Delta K_V}{K_V} \approx \frac{\Delta\sigma^2_{FADC}}{\sigma^2_{FADC}} = \sqrt{\frac{2}{N_p \cdot N_t}} \qquad (6)$$

where $N_p$ is the number of LED pulses and $N_t$ is the number of FADC time bins in the light pulse (with the exclusion of a few time bins close to the pulse start and stop times). For a typical Cal. A acquisition with $N_p$= 50 pulses and $N_t$= 560 time bins the statistical uncertainty on $K_V$ is 0.85%, which then enables one to obtain a good calibration of the pixel-to-pixel behavior.

To show that the method proposed for the FADC variance calibration effectively compensates for the pixel to pixel variability, we compare in Fig. 2 the NSB lightcurves as seen by a set of adjacent FD pixels in the same camera row (observing then the sky at a similar elevation angle). In the upper panel, we plot the raw variance values (after subtraction only of the electronic noise). The huge dispersion between lightcurves emphasizes the necessity to introduce different variance calibration factors for each pixel. In the lower panel, the lightcurves of the same set of pixels, converted to an absolute photon flux by means of Eq. 4, overlap as expected. In fact the diffuse component of the NSB does not usually show appreciable variation as a function of the azimuth.

As a final verification of the absolute value of the NSB photon flux, we performed a set of measurements with an independent photon detector, named UVscope, described in the following section.

## 6 The UVscope instrument

The UVscope is a portable photon detector designed to measure the NSB light in the UltraViolet wavelength range [6]. The UVscope photon detection unit is based on a multi-anode (8 × 8) photomultiplier (Hamamatsu, series R7600-03-M64), which is coupled to a 64-channel Front-End Electronic unit (developed at IASF Palermo) working in Single Photoelectron Counting (SPC) mode.

The UVscope Field of View is regulated by a cylindrical collimator with a square entrance pupil on top. The main advantage of using an imaging system based on a collimator (without e.g. a focusing lens), is that the instrumental angular response can be obtained by simple geometrical optics. Considering that the collimator geometry can be measured with high precision, the absolute calibration of the detector depends essentially on the calibration of its multi-anode PMT which is performed, as function of wavelength, in a dedicated laboratory, with a precision better than 10%.

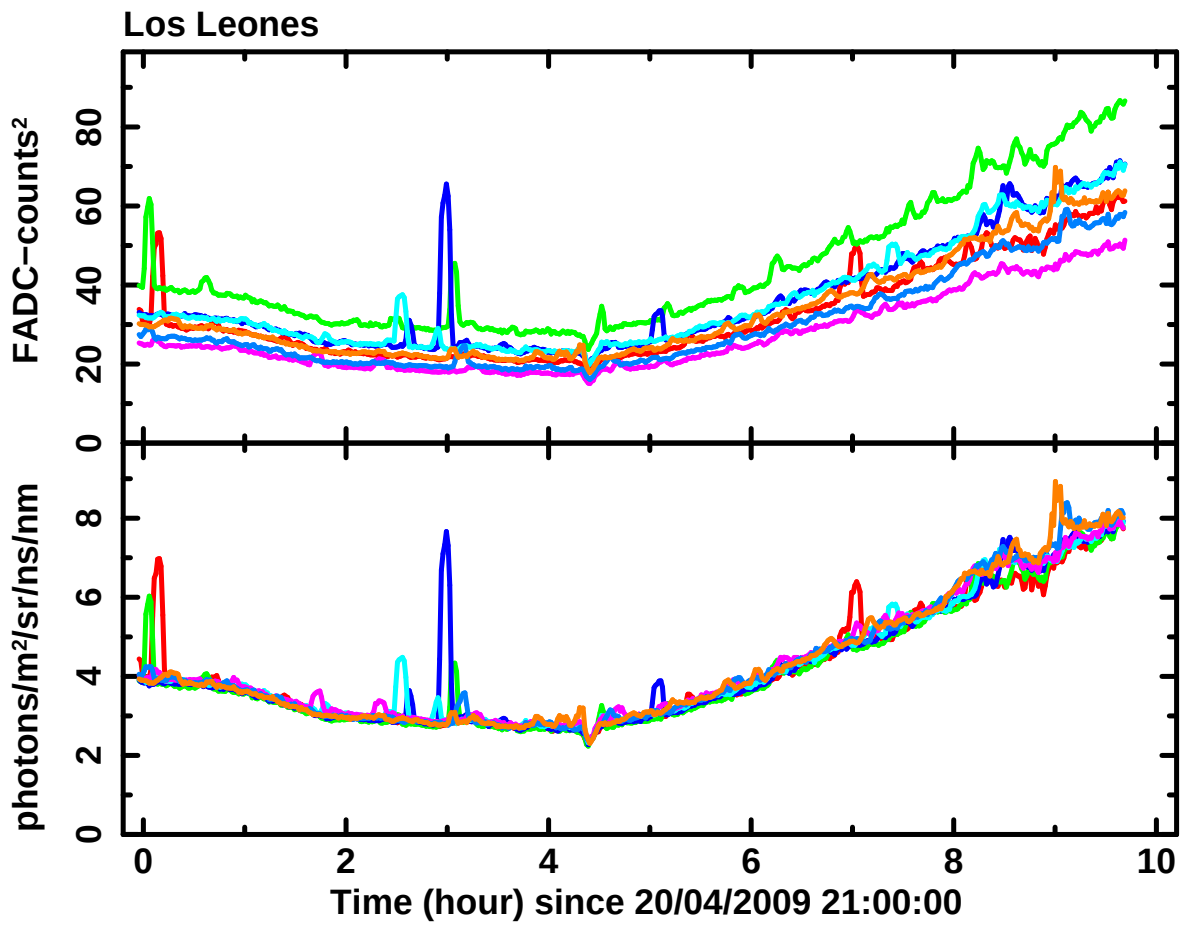

Figure 2: NSB light curves as seen by a set of seven FD pixels belonging to the same camera row (at an elevation of ≈ 25°). The upper panel shows the raw FADC variance values (electronic noise subtracted) while the lower panel shows the same light curves converted to photon flux physical units. The overlapping of the lightcurves demonstrates the good correction of the pixel-to-pixel variation.

In the following section we compare the NSB measurements performed with the UVscope at the Pierre Auger Observatory site, with the correspondent flux values obtained from the analysis of the FD variance background data.

## 7 NSB Comparison: UVscope vs. FD

To observe the same NSB flux as seen by the FD pixels, the UVscope instrument has been placed on top of the FD detector building at Los Leones, and the same filter (M-UG6) as the one used for FDs has been mounted. Moreover the UVscope collimator length (157.85 mm) and pupil size ($4.2 \times 4.2$ mm$^2$) have been designed to obtain a single pixel FoV of $1.52^\circ \times 1.52^\circ$, approximately matching the FoV of an FD pixel.

In Fig. 3 we show the orientation of the UVscope FoV with respect to the FoV of FD Bay 3 in Los Leones. Also shown is the path on the sky of the bright star Arcturus, that has been used to verify the UVscope pointing accuracy. By comparing the absolute NSB flux measured by UVscope with the one obtained from FD pixels looking in the same sky direction we have found a quite good agreement, with the UVscope flux slightly higher ($< 8\%$) than FD; this difference is consistent with the uncertainty in the absolute calibration of the two instruments.

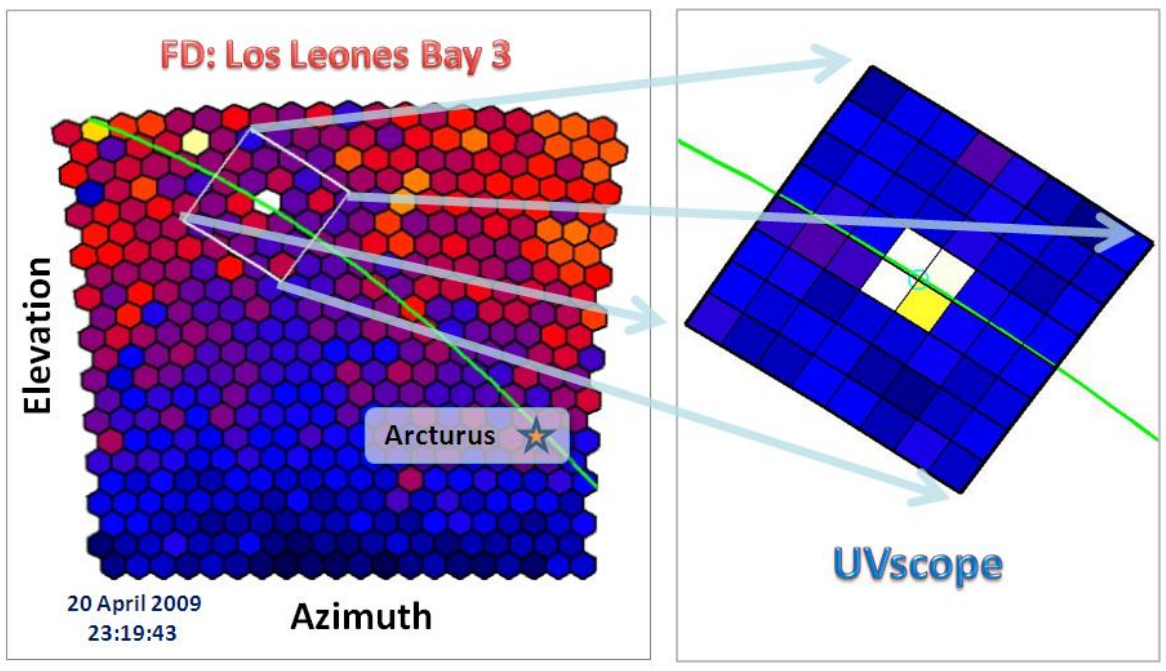

Figure 3: Left panel: orientation of the UVscope FoV with respect to the FD telescope in Los Leones Bay 3. Right panel: the UVscope Field of View with the bright star Arcturus in the center.

In Fig. 4 we compare (by using a scaling factor for the relative calibration) the NSB lightcurves as observed by a single UVscope pixel and the correspondent FD pixel. In the upper panel we have not applied any correction for the nightly evolution of the FD PMT gain, so a relative drift between the two lightcurves is expected. In fact, comparison between the Cal. A runs performed at the beginning and at the end of the night, always shows the presence of systematic nightly evolution (few percent) in the PMT gains.

In Fig. 4, lower panel, we have corrected the FD lightcurve by assuming a linear drift of the PMT gain between the values measured from Cal. A runs at the beginning and at the end of the night. In this case the match between the two NSB measurements is so good that the two lightcurves can hardly been distinguished.

Note that in the comparison of NSB lightcurves, the emission associated to point sources (stars and planets) are not expected to match exactly as the scaling factors used are valid only for the diffuse emission.

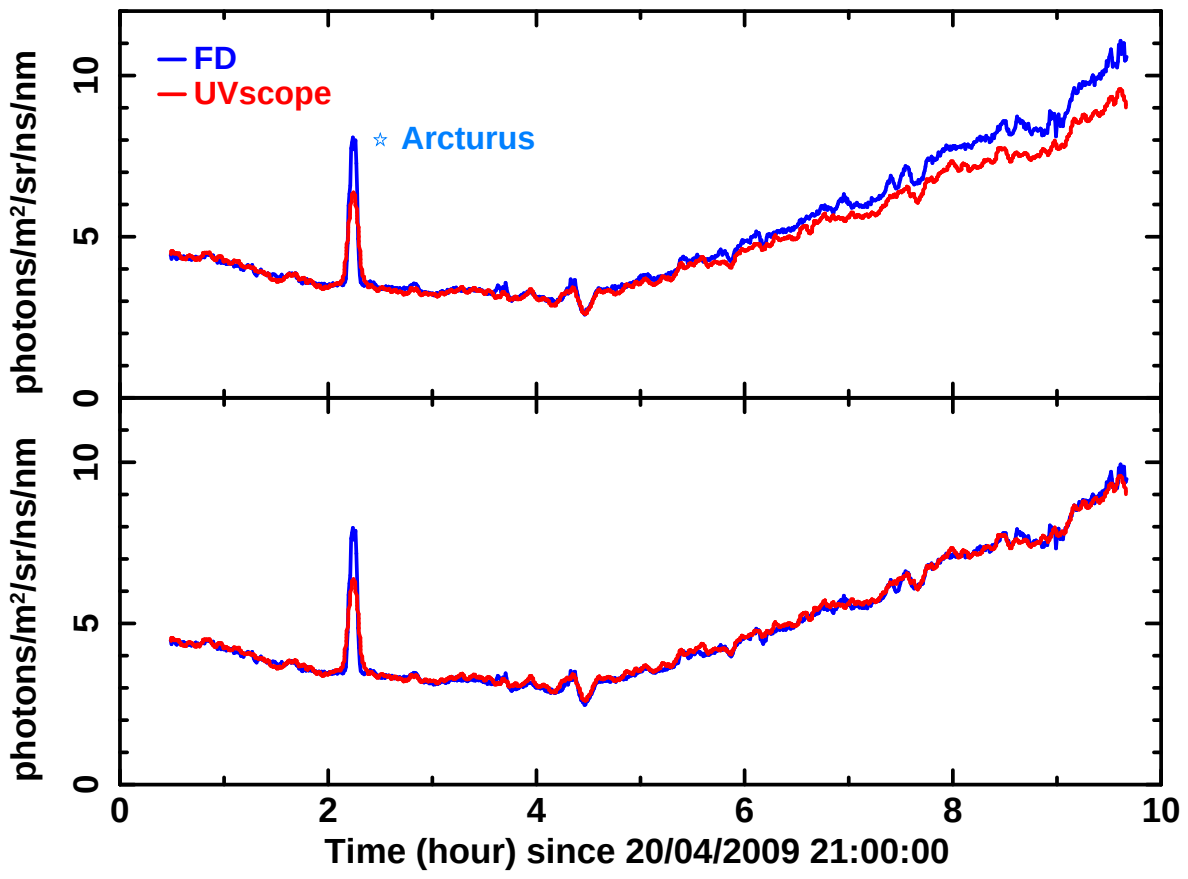

Figure 4: NSB lightcurves by single UVscope and FD pixels looking in the same sky direction. Upper panel: lightcurve obtained without correcting for the FD PMT gain drift. Lower panel: The FD PMT gain drift has been corrected assuming a linear evolution between the gain values measured from the Cal. A at start and stop times.

The example shows how independent measurements of NSB flux can be effectively used as a general non-invasive tool to verify the end-to-end calibration of telescopes with large fields of view and multi-pixel cameras without interfering at all with the normal operation of the telescopes. This gives the possibility of monitoring the gain of individual camera pixels in the real working condition (the background flux can induce a shift of the PMT gain with respect to dark condition). Moreover the method can be used to easily verify the flat-fielding of the telescope response by comparing all the telescope pixels, in turn, to the same reference detector.

## 8 Conclusions

We have shown how to convert the variance of the FADC traces, provided in the FD background files, to an absolute calibrated measurement of the NSB flux and that the resulting lightcurves are in excellent agreement with direct measurements obtained with an independent photon counting detector.

This opens up the possibility to use the huge FD Auger database for a scientific study of the NSB evolution as a function of time (short and long term) on an extremely large field of view. Moreover the spatial dependence (in particular with elevation) of the NSB flux can be correlated with atmospheric parameters provided by the several atmospheric monitoring devices installed at the Pierre Auger Observatory site.

We have also shown as the UVscope instrument can be used in support of the absolute end-to-end (camera + optics + electronics) calibration of telescopes with a camera composed by a mosaic of numerous independent photon detectors. This kind of application would be particularly useful when, due to the large telescope aperture, the use of a uniform extended illumination source, like the Drum, may not be feasible. This could make it an attractive option for future projects such as CTA [7] with a large number of telescopes distributed over a large area.

# Observation of Elves with the Fluorescence Detectors of the Pierre Auger Observatory

AURELIO S. TONACHINI[1] FOR THE PIERRE AUGER COLLABORATION[2]
[1] *Università degli Studi di Torino and INFN Sezione di Torino, Via Pietro Giuria 1, 10125 Torino, Italy*
[2] *Observatorio Pierre Auger, Av. San Martín Norte 304, 5613 Malargüe, Argentina*
*(Full author list: http://www.auger.org/archive/authors_2011_05.html)*
*auger_spokespersons@fnal.gov*

**Abstract:** We report the observation of elves using the Fluorescence Detectors of the Pierre Auger Observatory in Malargüe, Argentina. Elves are transient luminous phenomena originating in the D layer of the ionosphere, high above thunderstorm clouds, at an altitude of approximately 90 km. With a time resolution of 100 ns and a space resolution of about 1 degree, the Fluorescence Detectors can provide an accurate 3D measurement of elves for thunderstorms which are below the horizon. Prospects for the implementation of a dedicated trigger to improve detection efficiency and plans to perform multi-wavelength studies on these rare atmospheric phenomena will be given.

**Keywords:** elves, lightnings, ionosphere, thunderstorms, fluorescence detectors, Pierre Auger Observatory

## 1 Introduction

There is an electrodynamic coupling between electromagnetic fields produced by lightning discharges and the lower ionosphere. This coupling gives rise to distinct sets of observed phenomena, including various transient luminous events (TLEs) such as the so-called "Sprites" and "Elves". Sprites are luminous discharges located at altitudes between 40 and 90 km. They are due to the heating of ambient electrons, and last a few to tens of milliseconds. This characteristic makes them easily detectable with high speed cameras. Elves are optical flashes produced by heating, ionisation, and subsequent optical emissions due to intense electromagnetic pulses (EMPs) radiated by both positive and negative lightning discharges. Elves are confined to 80-95 km altitudes, and extend laterally up to 600 km [1]. Their duration, much shorter ($< 1$ ms) than that of sprites, made them somewhat harder to study. The first clear observation of elves was made using a high speed photometer pointed at altitudes above those of sprites [2]. More sophisticated instruments, such as "Fly's Eye" [3] and PIPER [4], consisting of linear arrays of horizontal and vertical photometers with a time resolution of $\sim 40\,\mu$s, have been used in the last decade to study the rapid lateral expansion of these high altitude optical emissions, and to test the excitation mechanism. Data from space on elves were acquired by the ISUAL/Formosat-2 mission, from 2004 to 2007 [5]. These data allowed one to conclude that elves develop on oceans or coastal regions ten times more frequently than on land. The satellite data were acquired with six PMTs and two 16-channel multi-anode PMTs, with time resolutions of 100 and 50 $\mu$s, respectively.

Further advancements in the understanding of these phenomena may be achieved using the fluorescence detector (FD) of the Pierre Auger Observatory [6]. The FD comprises four observation sites located atop small hills at the boundaries of the Auger surface array. Each FD building contains six independent telescopes, each with a field of view (FOV) of $30^\circ \times 30^\circ$ in azimuth and elevation. The combination of the FOV of the six telescopes covers $180^\circ$ in azimuth. Incoming light enters through a UV-transmitting filter window, and is focused by a mirror onto a camera, which is formed by $22 \times 20$ hexagonal photomultiplier tubes (PMTs). The wavelength of detected light ranges from 300 to 420 nm. Light pulses in each photomultiplier are digitized every 100 ns. The PMT processed data are passed through a flexible multi-stage trigger system, which is implemented in firmware and software. The resulting data are stored in $100\,\mu$s-long traces.

The FD geometry and time resolution are ideal for studying fast developing TLEs. However, the trigger chain contains a dedicated selection algorithm for rejecting lightning, which makes the FD a rather inefficient elve detector. Nevertheless, a few events which accidentally passed the rejection have been detected while searching for non-conventional cosmic ray shower events.

## 2 Observations

The first event was noticed serendipitously during an FD data taking shift. This unusual event presented a well defined space-time structure: a luminous ring starting from a cluster of pixels, and expanding in all directions.

A search for events with a similar space-time evolution in the data collected by the Pierre Auger Observatory since 2004 has identified two more events. These events are listed in the Table 1. The presence of dust and poor local weather conditions, recorded by local atmospheric monitoring devices [7], complicate the reconstruction of the first and the last event, but do not prevent one from recognizing the same overall features of the phenomenon. Most of the details which are given in this paper refer to the analysis of the second event. In Fig. 1, the photon time distributions for the three events are shown. Events which do not pass the whole trigger selection, but trigger a minimum of adjacent PMTs (second level trigger, or T2) leave some basic information in a log file, such as the GPS time and the number of PMTs hit. From these logs it is found that all the selected events last much longer than 70 $\mu$s, and are actually detected in adjacent FD bays, or even in other eyes, as summarized in Table 2. The number of buffered pages shows, in units of 0.1 ms, the time duration of the detected event.

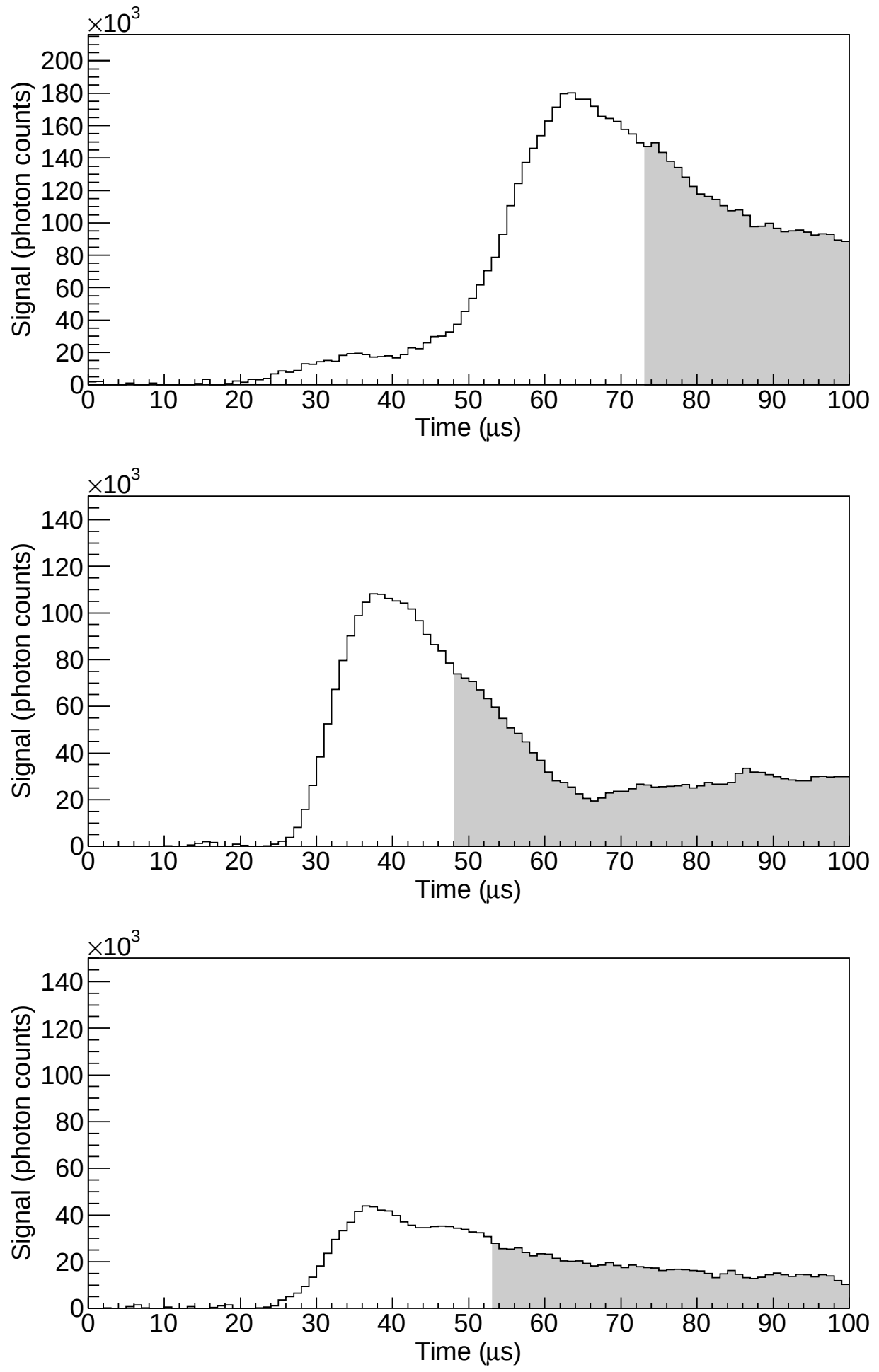

Figure 1: Photon counts at 370 nm obtained from the sum of all photomultiplier ADC traces of the three events in Table 1. Since these events are not totally contained inside the camera FOV, the gray areas denote a region with possible signal losses.

| Site-Bay | GPS time | GMT time |
|---|---|---|
| $LM-6$ | 800414142 | 18 May 2005 01:15:29 |
| $CO-3$ | 860806213 | 17 April 2007 00:49:59 |
| $LL-1$ | 861081389 | 20 April 2007 05:16:15 |

Table 1: Three elve candidates seen by Auger fluorescence detectors and their arrival times.

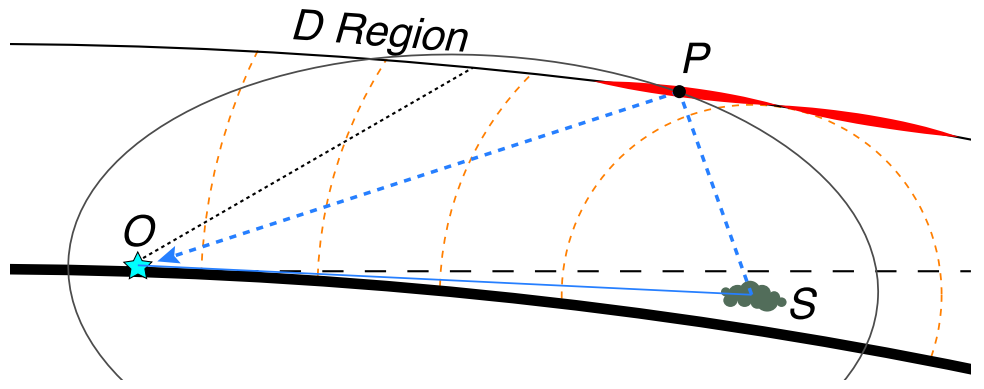

Figure 2: Schematic view of an EMP generated by a thunderstorm in $S$, which interacts with the D region of the ionosphere. The light emitted by the ionosphere (in red) is detected by the fluorescence detector at $O$. The observed signal time $t$ is the combination of the time needed by the pulse to move from $S$ to the interaction point $P$ and the time needed by the emitted light to travel from $P$ to $O$.

| Site-Bay | N buffered pages | Time delay ($\mu$ s) |
|---|---|---|
| LM-6 | 7 | 0 |
| LM-5 | 7 | 38 |
| CO-3 | 9 | 0 |
| CO-2 | 7 | 9.5 |
| LL-1 | 4 | 56.3 |
| LL-1 | 2 | 0 |
| LL-2 | 3 | 34.1 |
| CO-3 | 5 | 59.1 |

Table 2: FD telescope, number of T2 pages, and time delay for the three elve events.

## 3 Front propagation reconstruction

If the events observed are elves, the signals recorded correspond to the optical emission of the D region of the ionosphere, as a consequence of its interaction with a lightning-launched electromagnetic pulse (EMP).

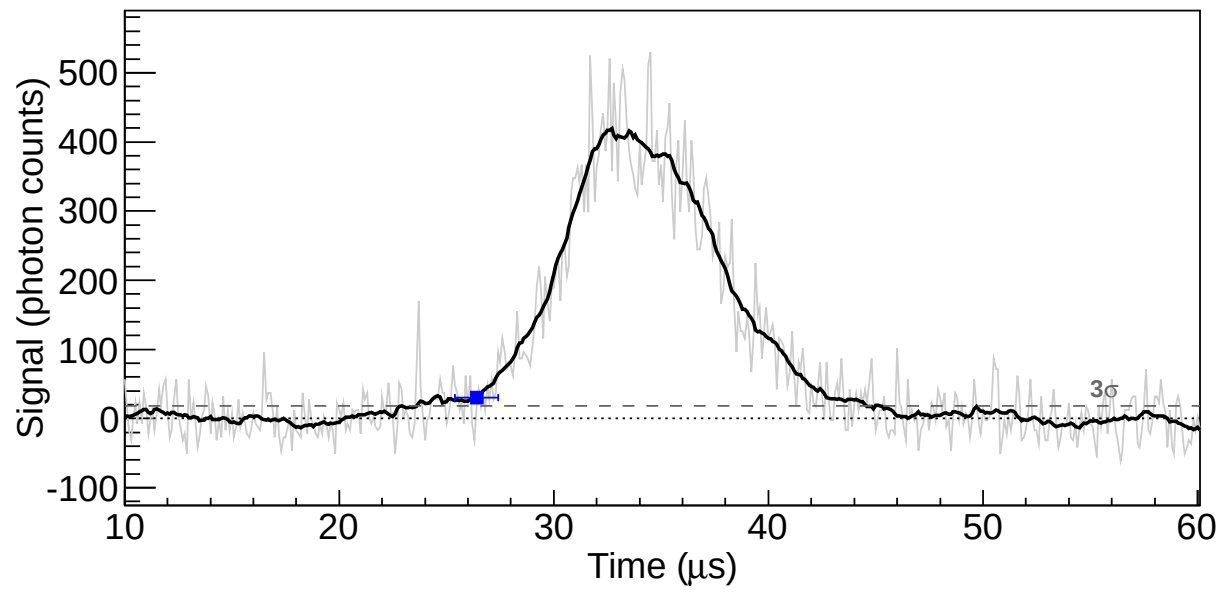

Figure 3: A signal measured with a single photomultiplier (gray graph). To reduce noise fluctuations, a moving average is performed on the original trace (black thick graph). The start point (blue square point) is defined from this graph when the signal is $5\sigma$ above the baseline (dotted line).

### 3.1 Geometrical model

The EMP source is confined inside the troposphere, while the optical emission takes place at 80-95 km altitudes. The observed light develops over times comparable with the time needed to travel from the source $S$ to a point in the D region, and from there to the observer $O$ at the speed of light (see Fig. 2). In fact, the light detected at time $t$ may come from any of the points belonging to the intersection of the D region with an ellipsoid whose foci are $O$ and $S$.

The first light arrives at a time $t_0$ defined by the ellipsoid tangent to the D region. The tangent point $P$ is found from observations, and puts constraints on the location of the source $S$. Indeed, it can be demonstrated geometrically that the line tangent at $P$ to an ellipse with foci $O$ and $S$ forms equal angles with the lines $OP$ and $PS$. Thus, once defined $P$, the locus of the foci $S_i$ is a line.

At a time $t_i > t_0$ the intersection of the ellipsoid with the D region corresponds to a closed curve: this is actually observed by the fluorescence detectors. The lateral expansion of this curve is expected to be symmetric, while the front moving towards the FD is expected to move faster than that moving in the opposite direction.

### 3.2 Signal treatment

The pixels considered in each event are the ones which have an FD first level trigger trace. Each trace is formed by 1000 time bins of 100 ns each. Signal bounds are searched in each trace by maximizing the signal to noise ratio. This allows one to roughly estimate the pulse start and stop times. Afterwards the signal is smoothed by applying a $2.1\ \mu$s running average in order to decrease short time signal fluctuations. The pulse start position is then moved back until the signal is less than $5\sigma$ above the noise. The error associated with this point is determined by searching the time where the signal is less than $3\sigma$, and then taking the time difference with respect to the start point (see Fig. 3).

The pulse start times measured by each photomultiplier are plotted in Fig. 4 as a function of PMT pointing directions. This development has been compared with the geometrical model discussed before. Once the pixel which recorded the minimum pulse start time was found, the direction of the start point $P$ was varied by $\pm 2°$ in both elevation and azimuth angles. For each $P$ the altitude of the D region and the direction of the EMP source with respect to the FD have been variated within 80 to 100 km and $+5°$ and $-5°$ respectively, in order to find the parameters which better fit the elve development.

## 4 Results

The best fit to the pulse start times of the event studied is obtained for a source elevation angle of $-1.15°$ and a D layer altitude of 92 km a.s.l.. The direction (elevation, azimuth) of the first light is $(14.6°, -52.1°)$. The source linear distance from the fluorescence detector of the Auger Observatory is about 580 km. A comparison of the times expected from a theoretical model with these parameters and the real data is shown in Fig. 5. The time residuals are plotted in Fig. 6.

The location of the event is strengthened by the presence of a large cloud perturbation seen by GOES geostationary satellites in the same region [8]. Moreover, a coincidence with a strong lightning pulse detected by the World Wide Lightning Location Network (WWLLN) [9] has been found.

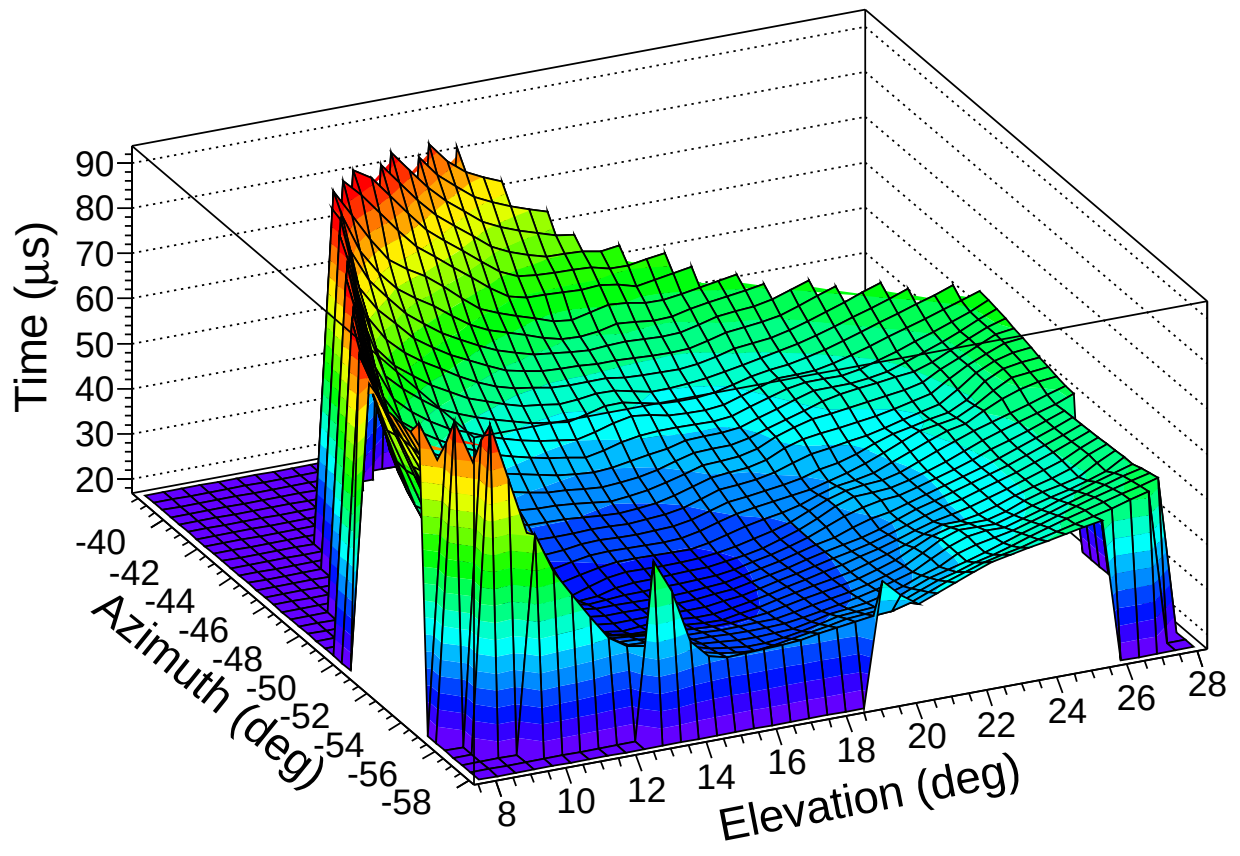

Figure 4: Interpolated tridimensional curve representing the time of arrival of photons at the FD diaphragm as a function of elevation and azimuth angle. Pulse start times belong to the event detected at GPS 860806213. This event triggered 143 pixels.

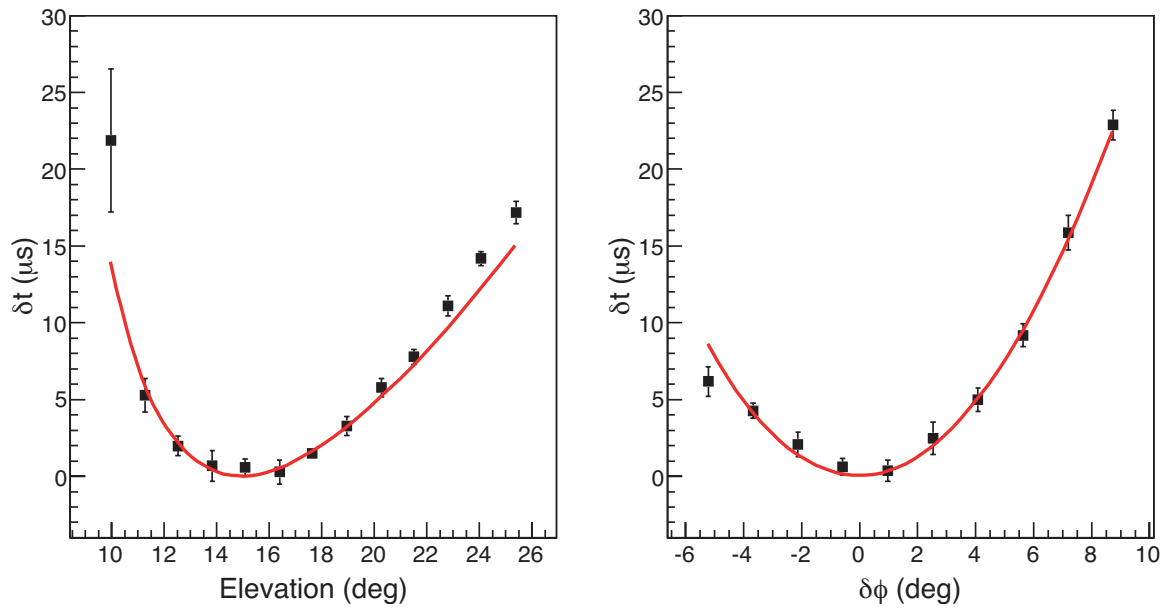

Figure 5: Best fit (red curve) compared to real data (black squares) for the column and the row of pixels passing through the centre of the event. $\delta\phi$ is the azimuth direction of the pixel with respect to the centre.

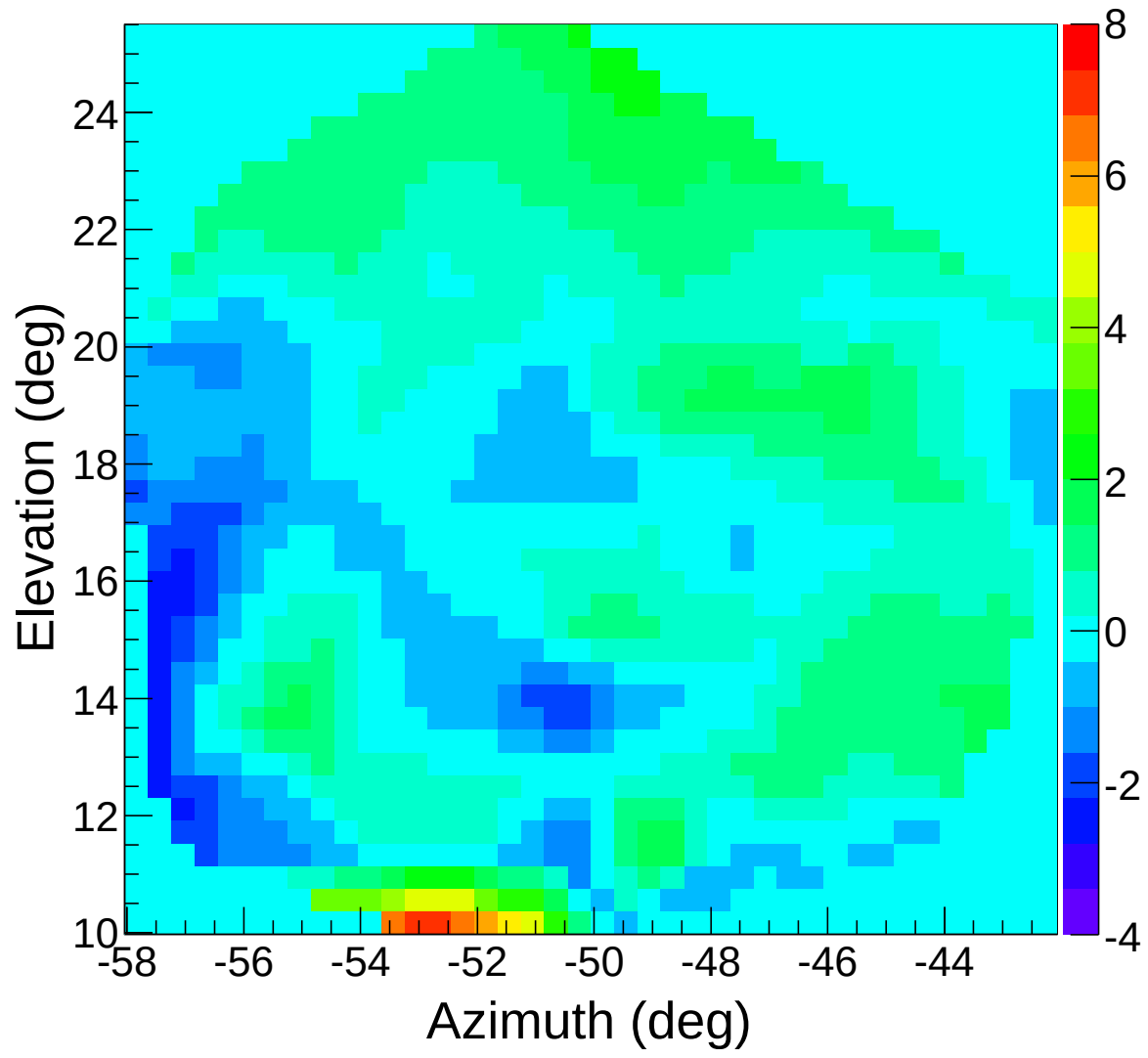

Figure 6: Difference between measured pulse start times and simulated ones as a function of the pixel pointing directions ($\mu$s). Differences are confined within $2\,\mu$s, with the exception of one pixel which recorded a trace delayed by $7\,\mu$s.

## 5 Final remarks

It has been shown that the fluorescence detector of the Pierre Auger Observatory may represent an interesting opportunity to study the elve evolution with an unprecedented time resolution. However, in order to transform the FD in an efficient elve detector it is necessary to design a dedicated software trigger. This would allow one not only to increase the FD efficiency, but to record subsequent signal traces up to the expected length of these optical flashes ($\sim$ 1 ms).

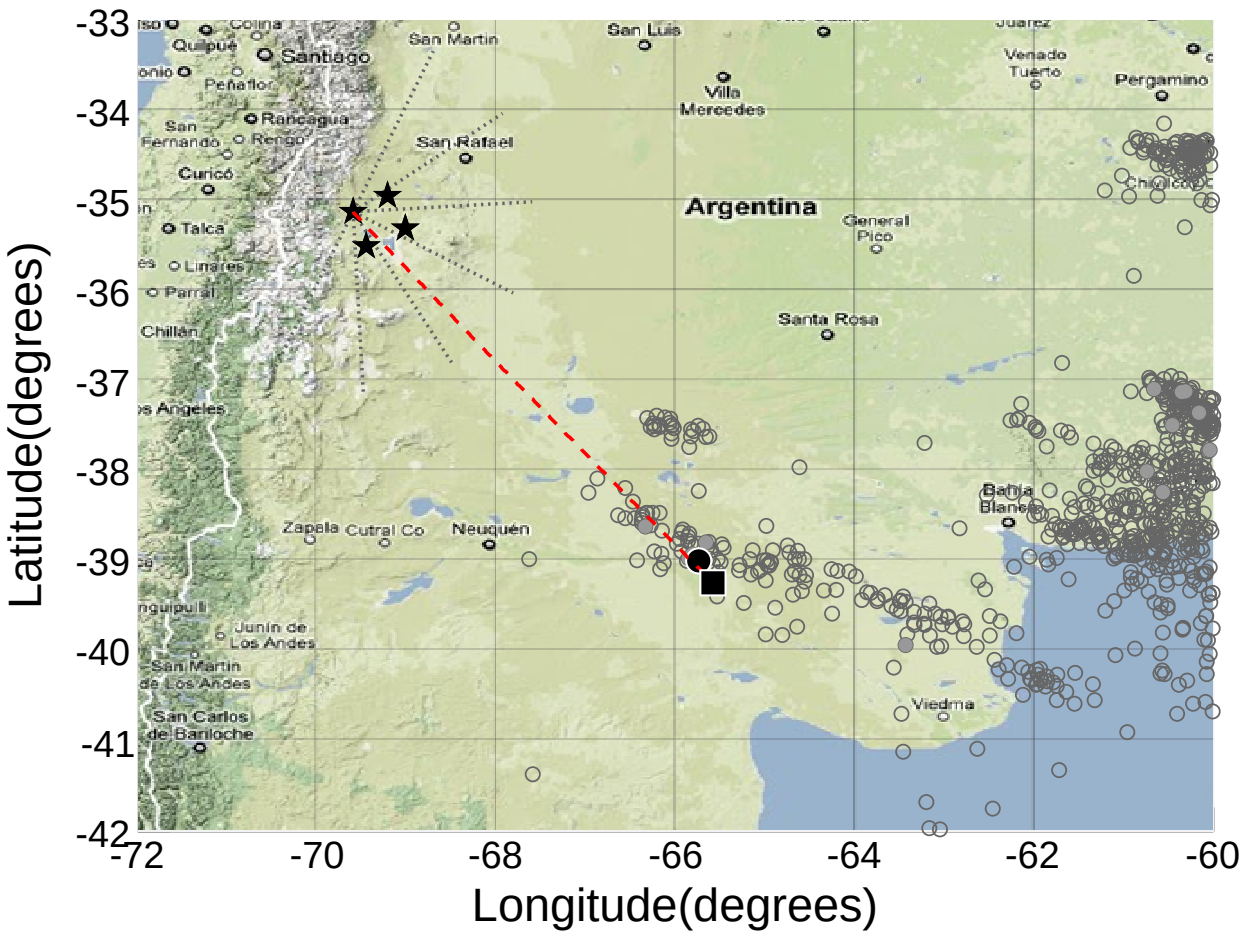

Figure 7: The position of the reconstructed elve (black square) is compared to that of lightning occurring at the same GPS second (black circle). Lightning recorded by WWLLN stations during the same day are plotted as gray circles. Lightning close in time (within 5 s) is represented by filled circles. Black stars mark the locations of Auger fluorescence detectors. Dotted lines define the six bays of the FD at Coihueco (CO).

## 6 Acknowledgements

We would like to thank Prof. R. Holzworth for having made available the list of lightnings recorded by WWLLN, and the GOES organization for having made public GOES satellite data.

# Atmospheric ”Super Test Beam” for the Pierre Auger Observatory

L. WIENCKE[1], FOR THE PIERRE AUGER COLLABORATION[2] AND A. BOTTS[1], C. ALLAN[1], M. CALHOUN[1], B. CARANDE[1], M. COCO[1,5], J. CLAUS[1], L. EMMERT[1], S. ESQUIBEL[1], L. HAMILTON[1], T.J. HEID[1], F. HONECKER[1,3], M. IARLORI[4], S. MORGAN[1], S. ROBINSON[1], D. STARBUCK[1], J. SHERMAN[1], M. WAKIN[5], O. WOLF[1]

[1] *Colorado School of Mines, Department of Physics, 1523 Illinois St., Golden CO, USA*
[2] *Observatorio Pierre Auger, Av. San Martin Norte 304, 5613 Malargüe, Argentina*
[3] *Karlsruhe Institute of Technology (KIT), 76131 Karlsruhe, Germany*
[4] *CETEMPS and INFN, Dep. di Fisica Università Degli Studi dell'Aquila, via Vetoio, I-67010, L'Aquila, Italy*
[5] *Colorado School of Mines, Division of Engineering, 1500 Illinois St., Golden CO, USA*
*(Full author list: http://www.auger.org/archive/authors_2011_05.html)*
*auger_spokespersons@fnal.gov*

**Abstract:** We present results from 200 hours of operation of an atmospheric super test beam system developed for the Pierre Auger Observatory. The approximate optical equivalence is that of a 100 EeV air shower. This new system combines a Raman backscatter LIDAR receiver with a calibrated pulsed UV laser system to generate a test beam in which the number of photons in the beam can be determined at ground level and as a function of height in the atmosphere where high energy air showers develop. The data have been recorded simultaneously by the Raman receiver and by a single mirror optical cosmic ray detector that tested the new system by measuring the side-scattered laser light across a horizontal distance of 39 km. The new test beam instrument will be moved from the R&D location in southeast Colorado to the Pierre Auger Observatory location in Argentina to effect a major upgrade of the central laser facility.

**Keywords:** Atmosphere, LIDAR, Pierre Auger Observatory, Calibration, Raman Scattering

## 1 Introduction

The Pierre Auger Observatory uses the atmosphere as a giant calorimeter to measure properties of the highest energy particles known to exist. Test beams of particles with these extreme energies (1-100 EeV) do not exist. However light scattered out of UV laser beams directed into the atmosphere from the Central and eXtreme Laser Facilities (CLF [1] & XLF) generate tracks that are recorded by the Auger Observatory fluorescence detector (FD) [2] telescopes that also record tracks from extensive air showers. There is an approximate effective optical equivalence between a 5 mJ UV laser track and that of a 100 EeV air shower.

Atmospheric clarity, specifically the aerosol optical depth profile, $\tau(z,t)$, is the largest and most variable calibration term, especially for the highest energy air showers. The method to obtain $\tau(z,t)$ that was pioneered by HiRes [3] and extended to the Auger Observatory uses FD measurements of side-scattered light from UV laser pulses [4] [5]. The relatively large light collecting power of the telescopes means that relatively few laser pulses are required. These pulses also provide a means to monitor detector calibration, performance, and aperture [6].

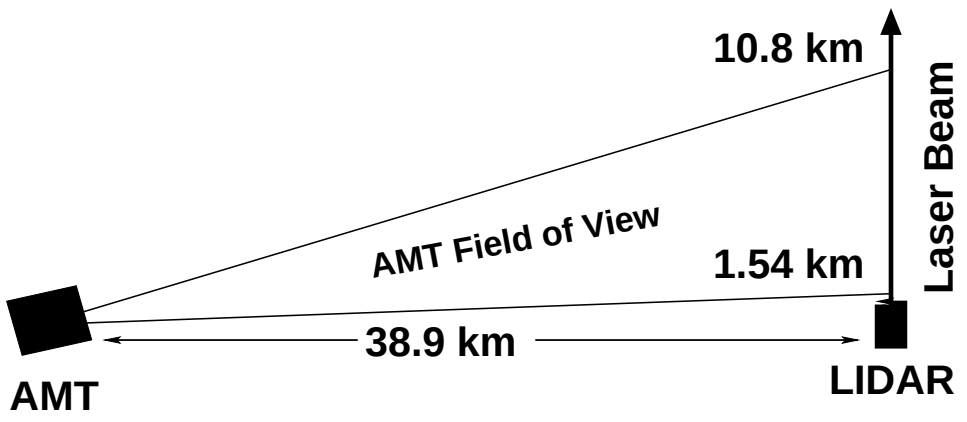

Figure 1: Geometrical arrangement, viewed from the side, of the laser and the two independent optical detectors.

To improve detector monitoring and $\tau(z,t)$ measurements, an upgrade is planned for the CLF. This will add a Raman LIDAR receiver, replace the flash lamp laser with a solid state laser, add an automated beam calibration system [8] as used at the XLF, and improve critical infrastructure.

Key components for the upgrade have been tested at the Pierre Auger North R&D site [9] in Southeast Colorado. Data collected have been used to measure $\tau(z,t)$ by two independent methods: elastic side scattering and inelastic (Raman) backscattering from $N_2$ molecules. The arrangement of instruments (Fig. 1) includes the solid state laser that generates a vertical pulsed beam (355 nm 7 ns pulse width), the collocated LIDAR receiver and a simplified FD telescope 38.9 km distant. Dubbed the atmospheric monitoring telescope (AMT), this instrument records side scattered light from the laser in the same way that the Auger Observatory FD telescopes record light from the CLF and the XLF in Argentina. Data collected also include temperature, pressure and humidity profiles recorded by 27 radiosonde weather balloons launched from the LIDAR site during 2009 to 2011.

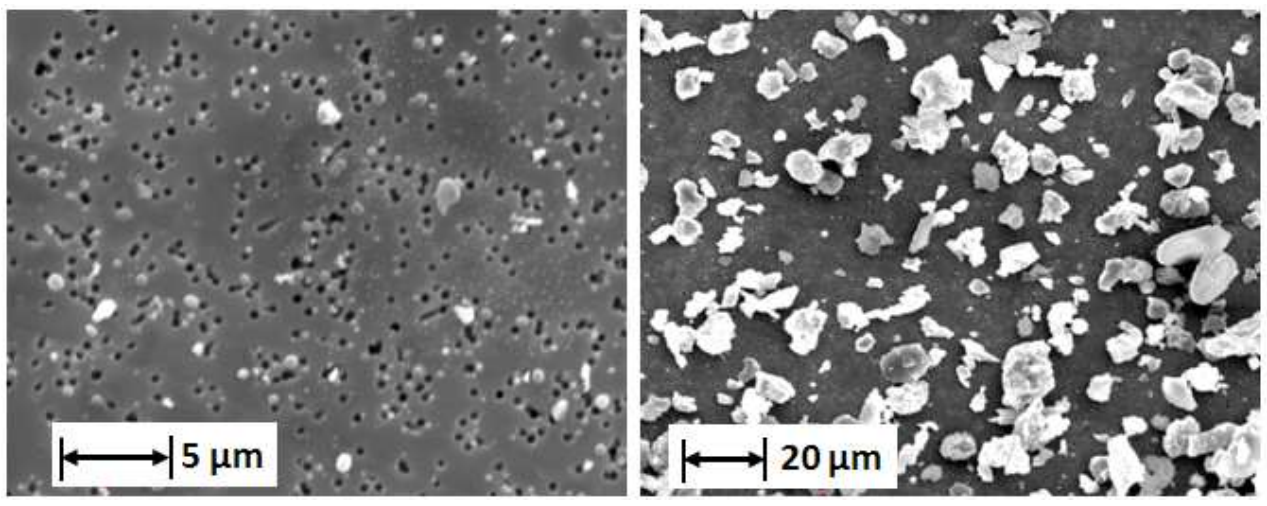

Figure 2: Scanning electron images of aerosols sampled at ground level at the Pierre Auger Observatory [10] (left: 2.5 $\mu$m filter right:10.0 $\mu$m filter).

## 2 The Raman LIDAR

In measuring $\tau(z,t)$ with elastically scattered laser light an inherent ambiguity is encountered. The measured quantity, i.e. the amount of light reaching the detector at a particular time bin (height) depends on several unknowns. These include the fraction of light transmitted to the scattering region, the fraction of light scattered in the direction of the detector by the molecular component and aerosols at the particular height, and the fraction of light transmitted back to the detector. The transmission terms can be combined if the atmosphere is assumed to be horizontally uniform, or if the receiver and laser are collocated. The molecular part of the scattering term can be determined to good accuracy from radiosonde measurements and molecular scattering theory. However the aerosol scattering term can not be modeled well because aerosol particles span a wide range of sizes and irregular shapes [10] (Fig. 2) and these properties typically vary with height.

Raman LIDARs evade this ambiguity by measuring light Raman scattered by $N_2$ molecules. The Raman scattering cross section for $N_2$ is well understood. The $N_2$ density profile can be derived from radiosonde data or through the Global Data Assimilation System (GDAS) [14] [15]. Over the past few decades, Raman LIDAR has become the standard method to measure $\tau(z,t)$.

The Raman LIDAR receiver used in these tests features a 50 cm diameter f/3 parabolic mirror pointing vertically beneath a UV transmitting window and motorized roof hatch. A liquid light guide couples the reflected light from the mirror focus to a three channel receiver (Fig. 3). Dichroic beam splitters direct this light onto 3 photomultiplier tubes (PMTs) that are located behind narrow band optical filters. These isolate the three scattered wavelengths of interest: 355 nm (Elastic scattering), 386.7 nm (Raman $N_2$ backscattering), and 407.5 nm (Raman $H_2O$ backscattering). The data acquisition system uses fast photon counting (250 MHz) modules. The LIDAR receiver and solid state UV laser were deployed 15 km south of Lamar, Colorado.

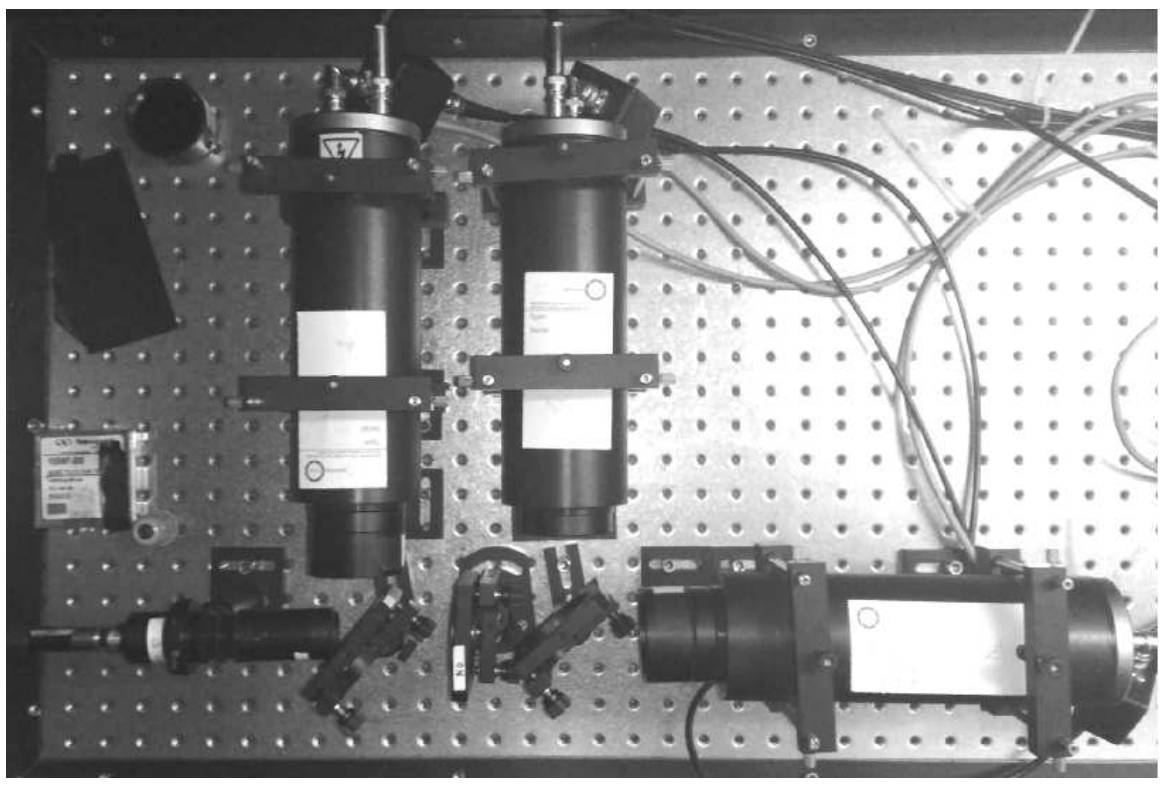

Figure 3: The three channel LIDAR receiver. Light reflected from the parabolic mirror (not shown) enters via liquid light guide seen near the lower left corner of this picture.

## 3 The AMT detector

The AMT (Fig. 4) is a modified HiRes II type telescope. The 3.5 $m^2$ mirror, camera, photomultiplier tube assemblies, and UV filter are all housed in a custom-built shelter with a roll-up door across the aperture. For these tests, the central 4 columns of $1^\circ$ pixels were instrumented. The AMT is mounted on four concrete posts and aligned so that the vertical laser track passed near the center of the field of view (FOV). The FOV at the vertical laser spans 1.54 to 10.8 km above the ground. A precipitation and ultrasonic wind sensor ensure the door was closed during rainy or windy conditions. The AMT is pointed toward the north so that direct sunlight could not damage the camera if the door is open during the day. The door and field of view can be observed remotely through a network video camera.

The PMTs were gain sorted prior to installation. Data from a temperature controlled UV LED system at the mirror center and from a vertical nitrogen laser scanned across the field of view were used to flat field and debug the camera. During routine nightly operation, the relative calibration was monitoried using the LED system.

The readout of the PMT current is performed by pulse shaping and digitization system electronics that are also implemented in the High Elevation Auger Telescope (HEAT) [11] [12] extension to the Auger Observatory. The sampling period is 50 ns. The readout is triggered externally, either by pulses from the UV LED system, or from a GPS device [13]. The laser is also triggered by the same model GPS device. The AMT GPS pulse output is delayed by 130 $\mu$s to allow for light travel time between the two instruments.

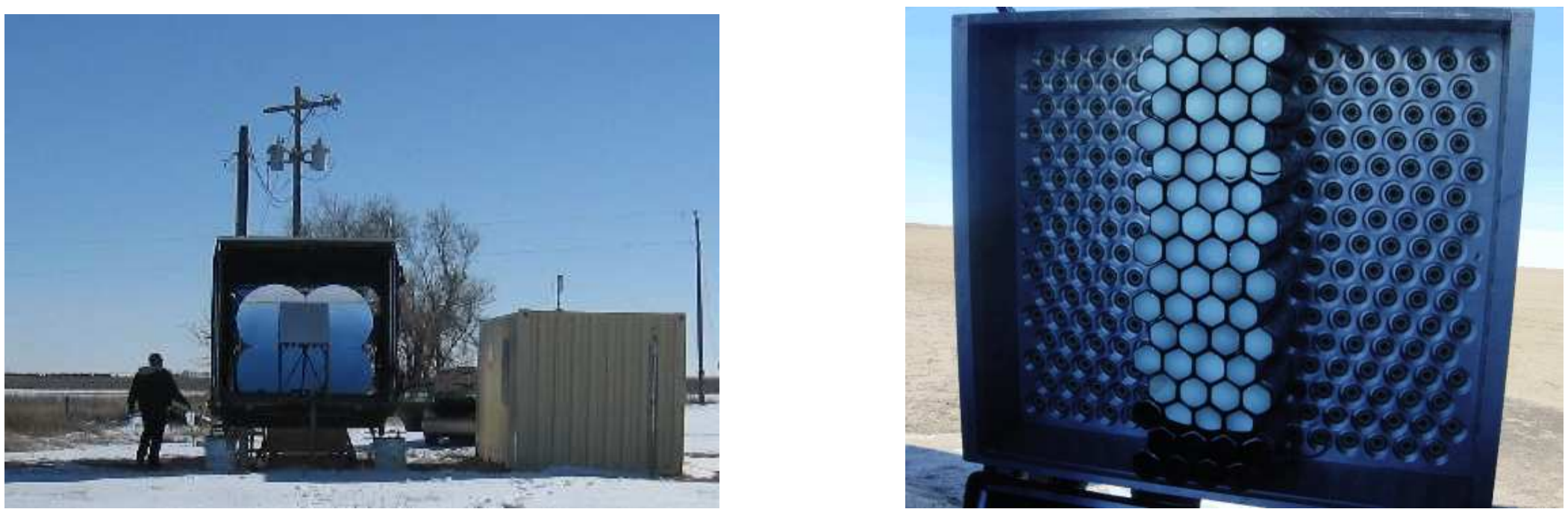

Figure 4: The remotely operated Atmospheric Monitoring Telescope (left) and its camera (right) with the central 4 columns instrumented. A rectangular UV transmitting filter (not shown) normally covers the camera surface.

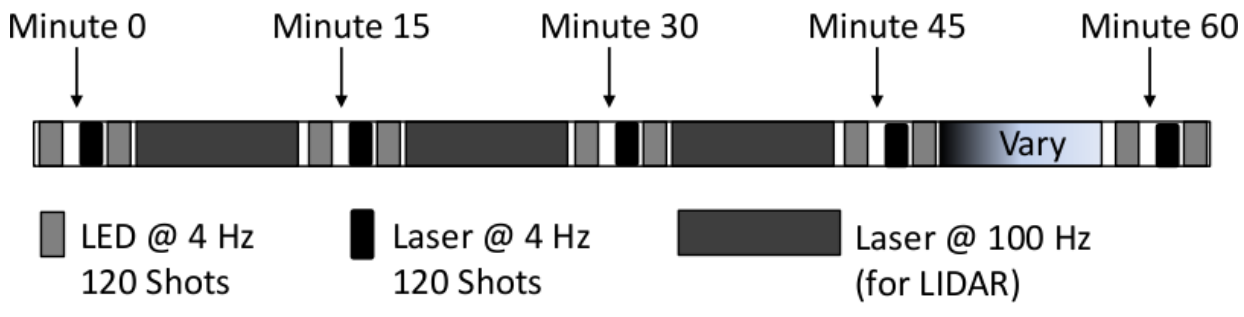

Figure 5: The hourly sequence of operations.

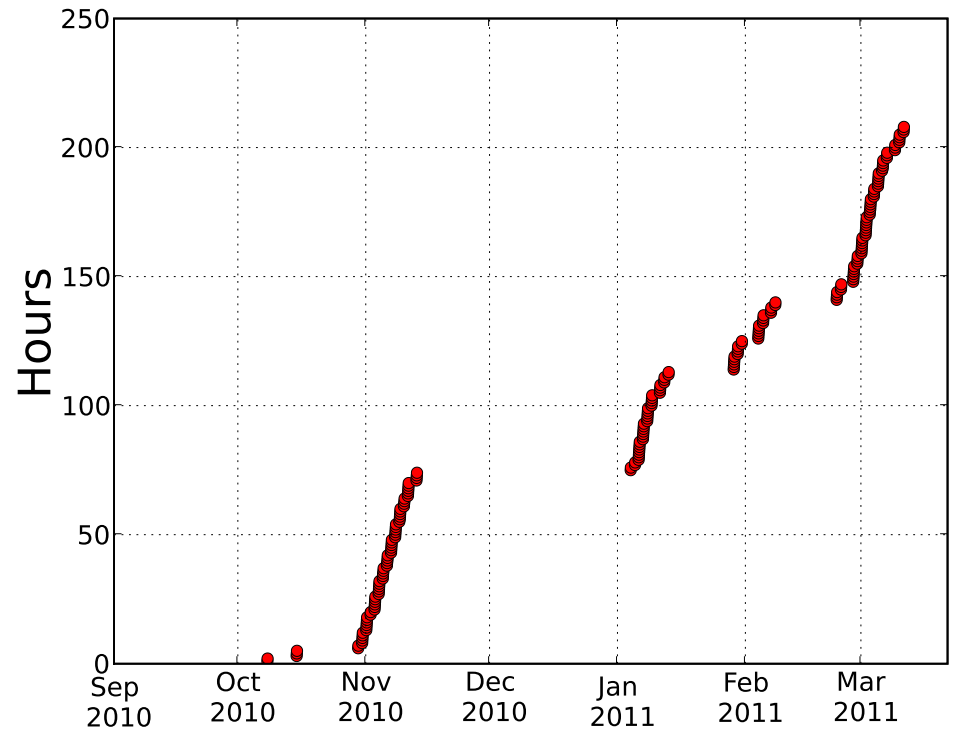

Figure 6: Accumulation of data when the AMT and the Raman LIDAR operated on the same hour

## 4 Operations and Data Analysis

The AMT, LIDAR, laser and various subsystems are all operated under computer control. Their nightly operation is sequenced by automation scripts initiated on moonless nights from the Colorado School of Mines campus. Operation and data collection are then monitored remotely by collaborators in Colorado, Germany, and Italy. The hourly sequence (Fig. 5) interleaves sets of 200 laser shots at 4 Hz for AMT measurements, sets of 120 UV LED shots for AMT relative calibration, and 12 minute sets of 100 Hz laser shots for LIDAR measurements. Between October 2010 and March 2011, more than 200 hours of data have been accumulated for which the AMT and LIDAR measured laser light during the same hour (Fig. 6).

The Raman LIDAR was benchmarked against the European LIDAR network EARLINET [7] prior to shipment

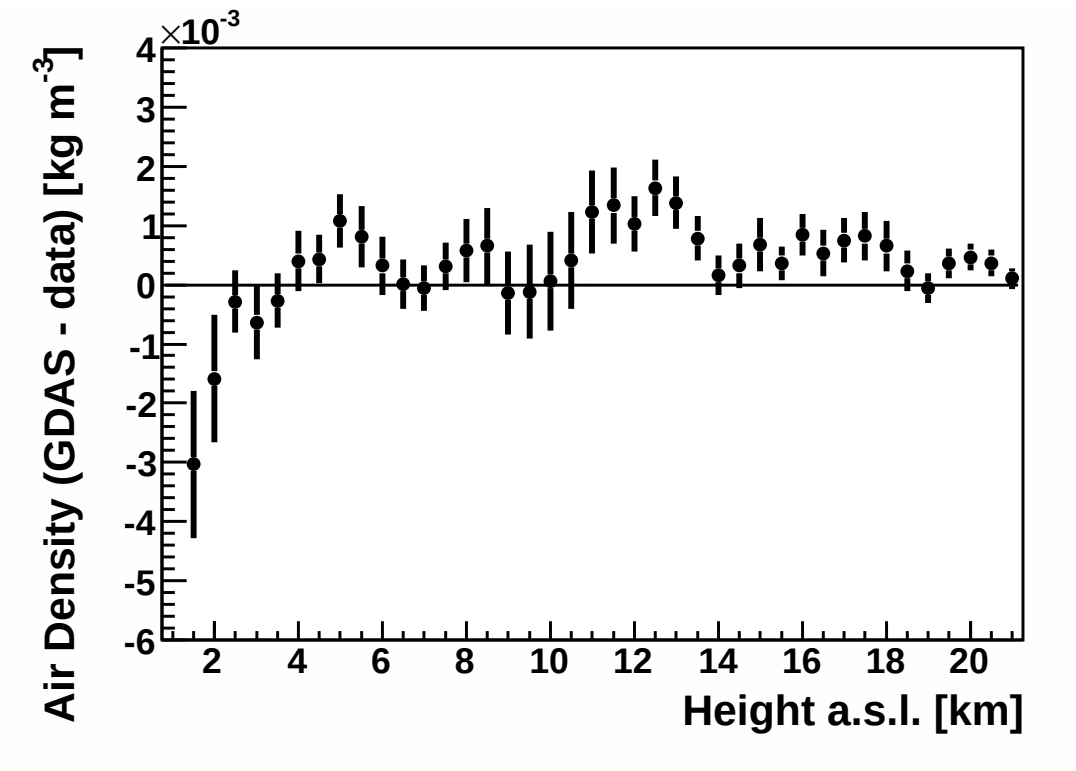

Figure 7: Average difference in atmospheric density as determined from the GDAS model and measured from 27 radiosondes launches.

from Italy. The algorithm used in this benchmark was also used to retrieve aerosol profiles in Colorado from the $N_2$ channel. The $N_2$ density was obtained from the GDAS model. The model agreed well with the radiosonde data collected at the site (Fig. 7).

The measurement of $\tau(z,t)$ from the AMT data used the data normalized retrieval algorithm adapted from the version used in Argentina to obtain $\tau(z,t)$ from FD measurements of vertical CLF laser pulses. Two reference nights were selected in the Colorado sample. The analysis included corrections for variations in the laser output and in the relative calibration of the AMT. Systematic errors of 3% were assigned to these terms and an equivalent error was assigned for the choice of reference night.

## 5 Results

A correlation is observed between the two independent measurements of aerosol optical depth (Fig. 8). Periods of obvious cloud were removed from this analysis. The smallest differences in absolute terms are observed during lower aerosol conditions, i.e. $\tau(4.5\text{ km}) < 0.05$. Horizontal nonuniformity of the aerosol distribution across the 39 km between detectors can be expected to contribute to the broad-ening of the correlation under hazier conditions. Further analysis is in progress. We note this work represents the first systematic comparison between these methods as applied to astroparticle detectors.

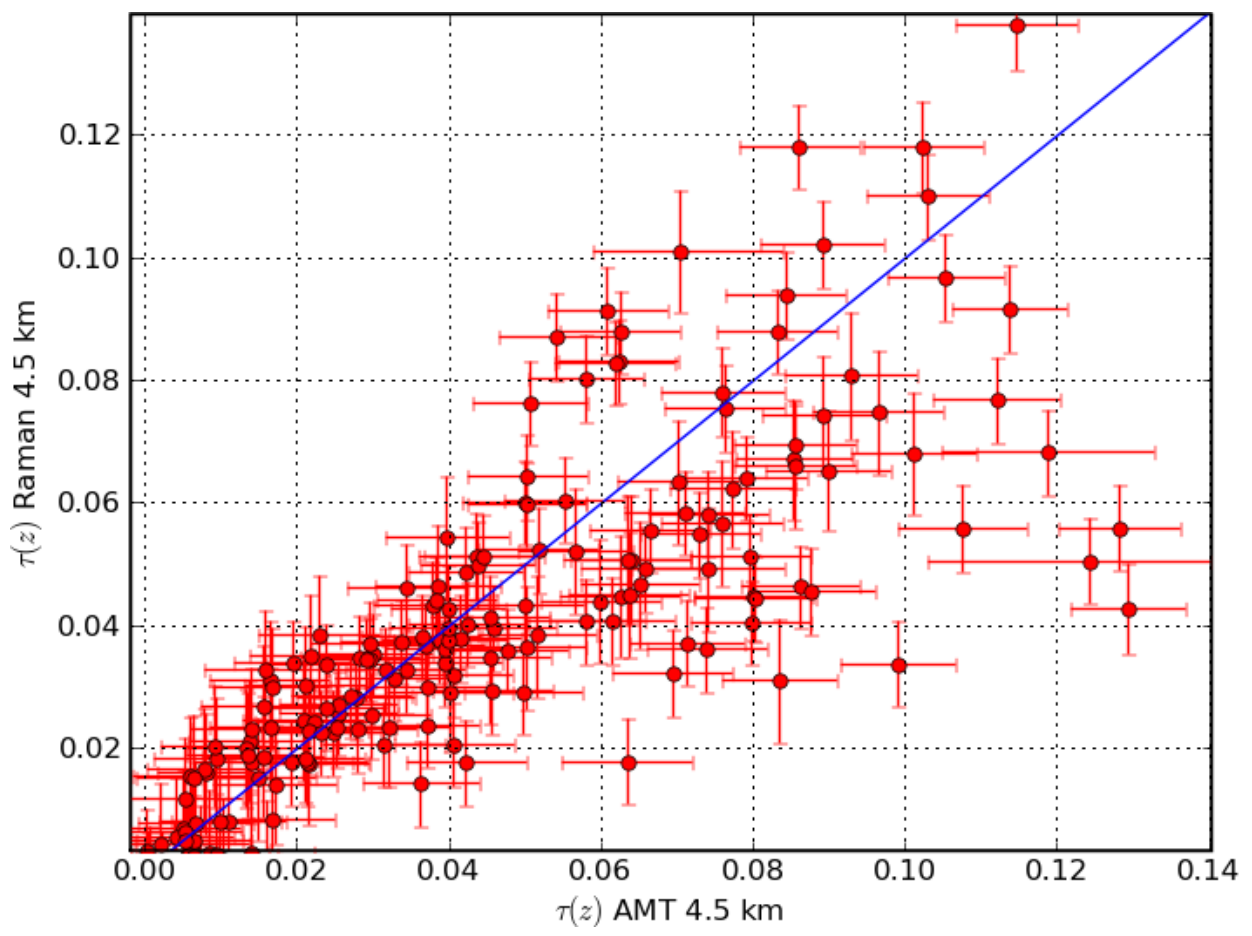

Figure 8: Comparison between the vertical aerosol optical depth at 4.5 km as measured by the AMT and the Raman LIDAR systems.

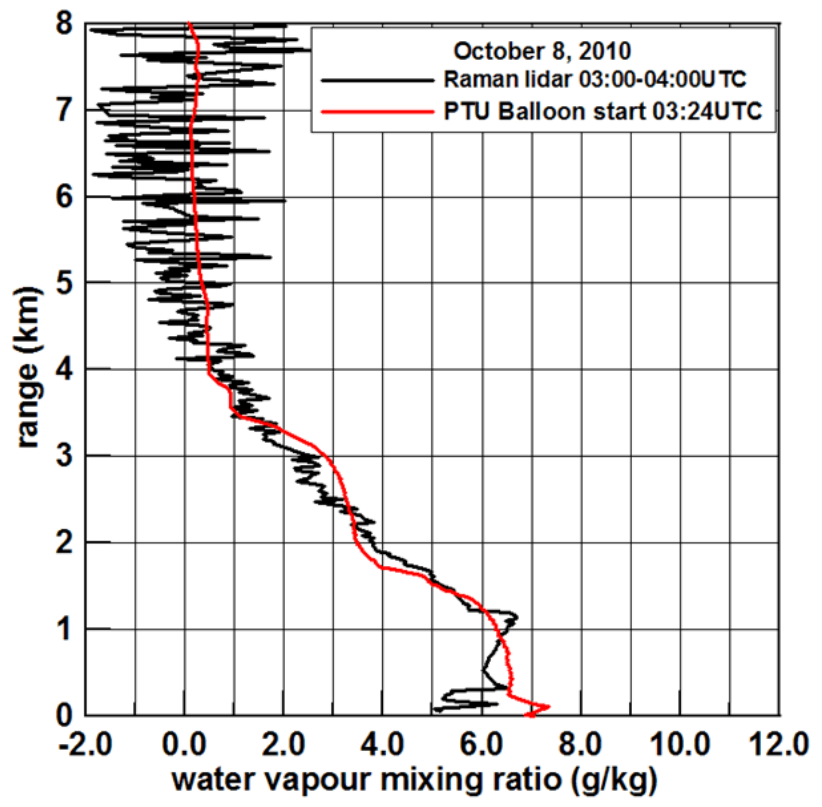

Figure 9: Water vapor profile measured by a radiosonde (smoother line) and by the LIDAR on the same evening.

## 6 Targeted applications of the CLF upgrade

Because of the relatively small size of the Raman scattering cross section, thousands of laser shots are needed to accumulate sufficient photon statistics. This has potential to interfere significantly with FD operation. However, a number of specific physics targets have been identified for which one set of Raman LIDAR measurements per night is expected to provide a valuable supplement to current methods.

1. Systematically compare the aerosol optical depth profiles measured by the Raman LIDAR and by the side-scatter method. This comparison is motivated by the elongation rate for hybrid data that suggests the particle composition may transition to heavier primaries above 10 EeV.

2. Better identify periods of extremely low aerosol concentration to reduce uncertainty in the data normalized aerosol analysis.

3. Use the super test beam to crosscheck the end-to-end photometric calibration of the FD which sets the energy scale for the observatory. The difference between the energy spectra measured by the Auger Observatory and by other experiments could be explained by a systematic difference in energy scales.

4. Precision measurement of aerosols shortly after detection of especially interesting air showers. The Raman receiver will make an independent precision measurement of the aerosol optical depth profile and water vapor profile. An example water vapor profile as measured by the LIDAR and by a radiosonde is shown in Fig. 9.

# Multiple scattering measurement with laser events

PEDRO ASSIS[1] FOR THE PIERRE AUGER COLLABORATION[2]
[1]*LIP - Laboratório de Instrumentação e Física Experimental de Partículas, Av. Elias Garcia, 14-1°, 1000-149 Lisboa, Portugal*
[2]*Observatorio Pierre Auger, Av. San Martín Norte 304, 5613 Malargüe, Argentina*
*(Full author list: http://www.auger.org/archive/authors_2011_05.html)*
*auger_spokespersons@fnal.gov*

**Abstract:** The Fluorescence Detector of the Pierre Auger Observatory performs a calorimetric measurement of the primary energy of cosmic ray showers. The level of accuracy of this technique is determined by the uncertainty in several parameters, including the fraction of the fluorescence and Cherenkov light reaching the detector after being scattered in the atmosphere through Rayleigh and Mie processes. A new method to measure this multiple scattering is presented. It relies on the analysis of the image of laser tracks observed by the fluorescence telescopes at various distances to characterize the scattering of light and its dependence on the atmospheric conditions. The laser data was systematically compared with a dedicated Geant4 simulation of the laser light propagation, allowing for any number of scatterings due to both Rayleigh and Mie processes, followed by a detailed simulation of the telescopes optics also based on Geant4.

**Keywords:** Laser; Multiple Scattering; Pierre Auger Observatory; Fluorescence detector

## 1 Introduction

The Pierre Auger Observatory [1] in Argentina has $1660$ surface detectors in a $3000\ \mathrm{km}^2$ array that is overlooked by $27$ fluorescence telescopes at four locations on its periphery. The Fluorescence Detector (FD) [2] telescopes measure the shower development in the air by observing the fluorescence light. The FD offers optical shower detection in a calorimetric way and can be calibrated with very little dependence on shower models.

The accuracy of the fluorescence technique is determined by the uncertainty in several parameters [3], among them, the fraction of shower light (both from fluorescence and Cherenkov processes) that reaches the detector after being multiply scattered in the atmosphere. The multiple scattering (MS) component, which has to be estimated and included in the reconstruction analysis to correctly derive the cosmic ray properties, depends on the atmospheric conditions, in particular on the Rayleigh and Mie scattering processes.

The atmospheric conditions at the Auger site are monitored by several devices [4]. In particular, the observatory is equipped with a set of laser systems that can shoot laser pulses into the atmosphere to be seen by the FD, allowing us to measure atmospheric conditions and monitor the performance of the telescopes. One of them is the Central Laser Facility (CLF) [5], a unit placed about $30\ \mathrm{km}$ from the FD sites emitting energy calibrated pulses of wavelength $\lambda = 355\ \mathrm{nm}$. In addition, a roving laser system, emitting vertical laser pulses at $\lambda = 337\ \mathrm{nm}$, can be positioned in front of the telescopes at distances of a few $\mathrm{km}$.

Using the large amount of laser data and profiting from the negligible width of the beam we have developed a method to extract the transverse distribution of light in the FD cameras, from which it is possible to access the MS parameters. Data is compared to a dedicated Geant4 simulation of light propagation in the atmosphere.

## 2 Extraction of the transverse light profile

The emitted laser pulses propagate upwards in the atmosphere. The direct light seen by the telescopes results from a first Rayleigh scattering, as illustrated in figure 1. The light from the first scattering can suffer additional scatterings and contribute to the signal recorded at large angles with respect to the direct light component. The method employed to extract this transverse light profile from laser data relies on the principle that identical laser events can be averaged to extract information inaccessible on an event-by-event basis.

For each acquisition time slot, the recorded image is translated into a distribution of the number of detected photons as a function of the angular distance, $\zeta$, to the direction of the direct light.

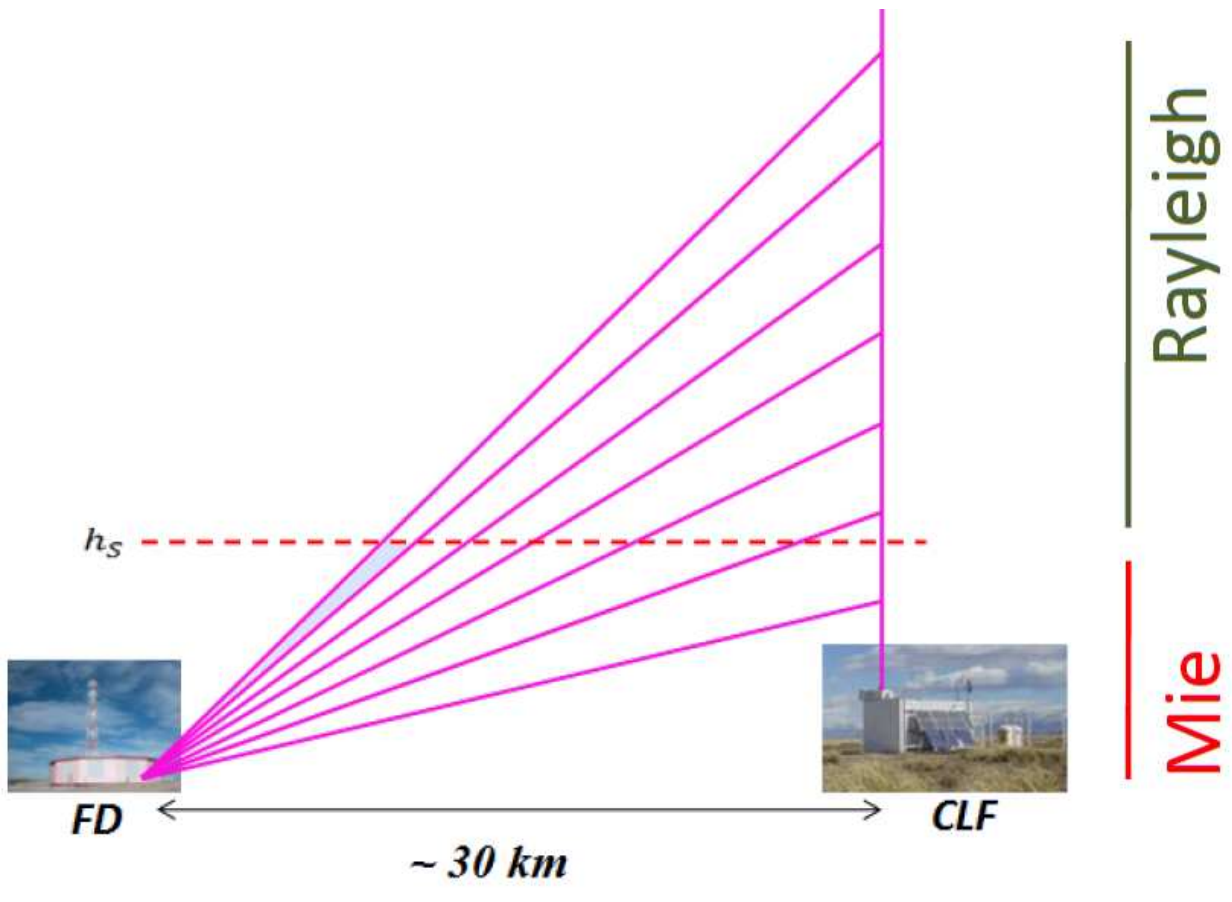

Figure 1: Schematic representation of the propagation of laser photons from the CLF to the FD site. In this model, there is an aerosol band in the area below the $h_S$ horizontal line. Rayleigh and Mie scattering dominated regions are, respectively, above and below the scale height parameter, $h_S$.

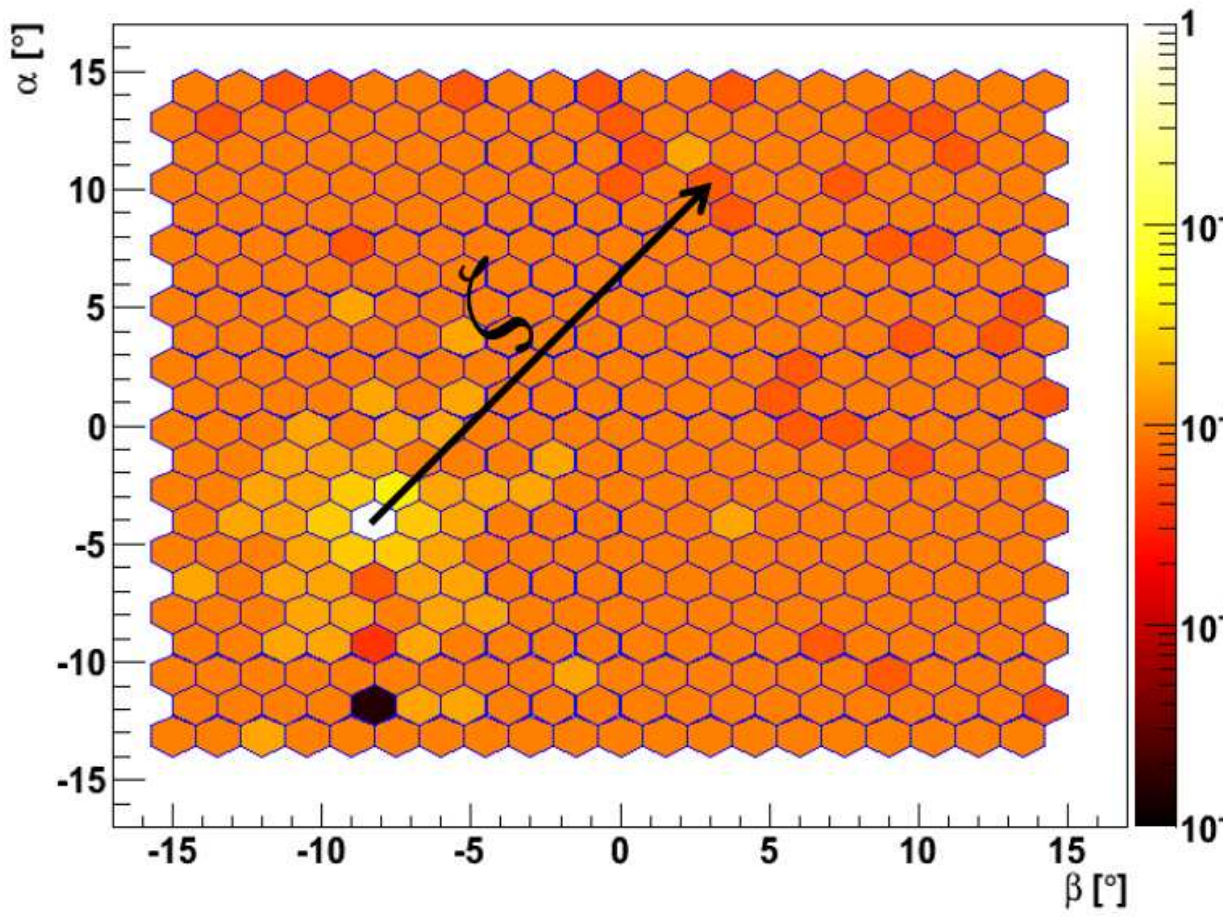

Figure 2: Camera image for a single time bin built by averaging over several CLF shots seen at Los Leones FD site with a full camera acquisition. Colors represent the accumulated charge normalized to the maximum.

In figure 2, a single time bin ($100$ ns) from the average of several CLF shots (almost one thousand) seen at Los Leones with a full camera acquisition is shown, for the case when the laser spot is at $\alpha \sim -4.9°$ and $\beta \sim -8°$, where $\alpha$ and $\beta$ are, respectively, the elevation and azimuthal angle in the camera coordinates. Here the full FD camera is represented by the $440$ hexagonal pixels and the color scale represents the number of photons detected during this time period, normalized to the maximum signal value at the spot centre. Additional structures are observed surrounding the hottest pixel: a first crown of (yellow/light-orange) pixels followed by second crown made of a bigger group of (orange) pixels, and an almost flat distribution for the rest of the camera with an intensity of $\sim 10^{-3}$ with respect to the maximum. The first crown is connected with the detector point spread function (PSF) while the farther regions are dominated by multiple scattered light. Another feature in the image are the darker pixels (corresponding to fewer detected photons) just below the spot centre. These are the first pixels crossed by the laser, for which the signal attained the largest values, and were removed from the analysis. The small signal in these pixels, some time bins after being hit by the laser direct light, is consistent with an undershoot effect: large signals affect the baseline (pulling it down), which then takes a few $\mu s$ to recover. Clouds can cause serious distortions to the light profile. Therefore, events in which clouds are identified during the reconstruction are removed from the analysis.

To build the $\zeta$ profile, $\frac{dN}{d\Omega}(\zeta)$, the number of photons in pixel $i$ and in time bin $j$, $\Delta N_{i,j}$, corrected by the pixel solid angle $\Delta\Omega_i$ is obtained as a function of $\zeta$ and averaged according to

$$\frac{dN}{d\Omega}(\zeta) = \frac{\sum_{j=1}^{N_t} \sum_{i=1}^{N_p} \frac{\Delta N_{i,j}}{\Delta\Omega_i}(\zeta)}{N_t \cdot N_p} ,$$

where $N_t$ is the number of time bins and $N_p$ is the number of acquired pixels in the camera ($440$ for a full camera acquisition). This method allows us to obtain transverse light profiles, as shown in figure 3.

The observed transverse light profile is the convolution of those of the light source (point-like at the current distances), the multiple scattering in the atmosphere and the detector PSF. The direct light convoluted with the detector PSF is expected to dominate at small $\zeta$ angles, while the multiply scattered light should dominate at large $\zeta$ (especially for a far away source like the CLF). Although some runs were performed with full camera acquisition, as in the example shown in figure 2, most available CLF data were taken with partial camera acquisition. In this case, data are taken for a small group of pixels neighbouring the pixels triggered by the laser. For vertical CLF shots the data available are contained within a band of pixels with approximately 8 degrees in $\beta$. Even in this case, the described method can be used up to values of $\zeta = 15°$.

## 3 Geant4 laser simulation

A realistic simulation capable of reproducing the features of the multiply scattered photons from production to detection was developed to support the data analysis presented in this paper. This simulation is based on Geant4[6] and performs the tracking of photons in the atmosphere. Simulation of both Rayleigh and Mie scattering processes were implemented according to [4]. Photons are individually followed through the atmosphere, allowing for any number of scatterings from both processes. The atmosphere was parametrized in layers of constant depth, 20 g cm$^{-2}$, and different atmospheric profiles can be selected. In order to simulate the camera aperture and inclination, and to improve the efficiency of the simulation, the detector was implemented as a full $2\pi$, 2 meter high cone section. The laser photons are generated at the centre of the cone section with the cone radius corresponding to the distance between the laser and the detector.

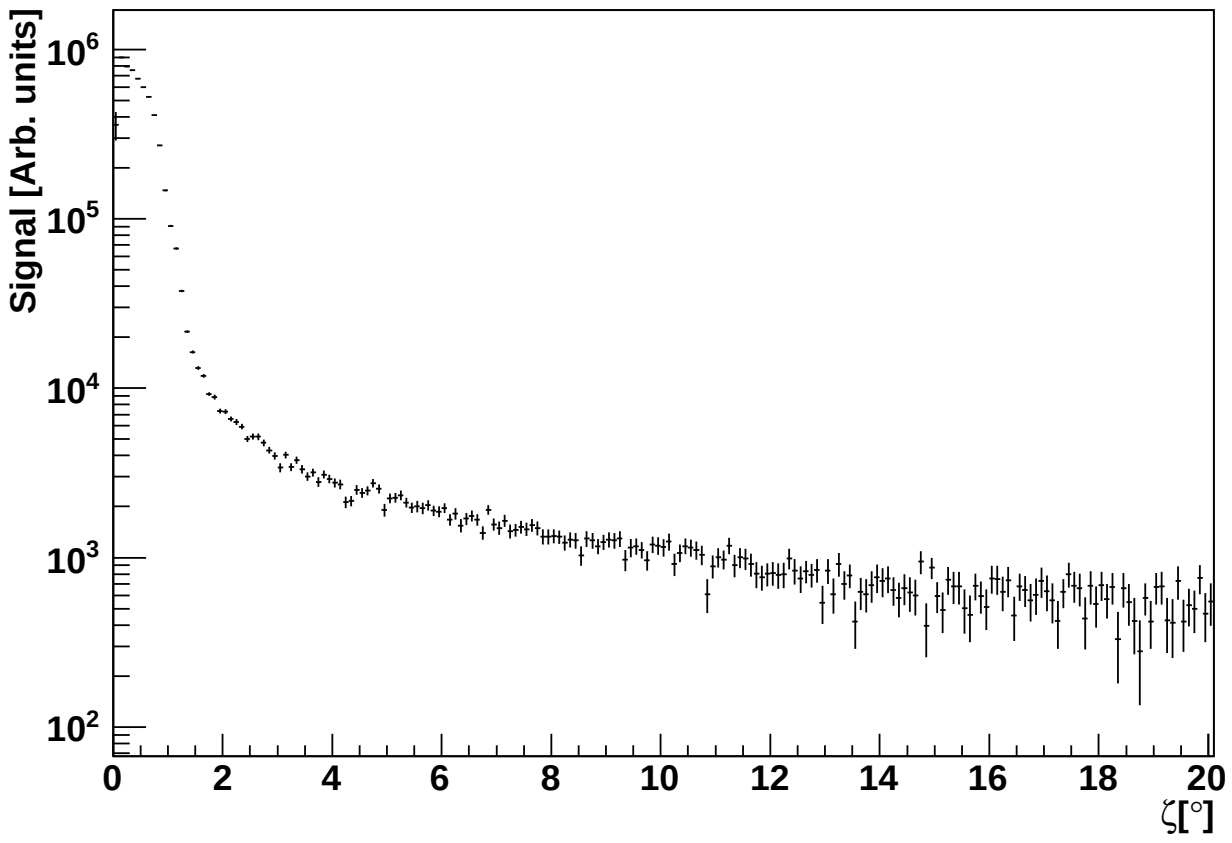

Figure 3: Transverse light profile seen at Los Lones FD site using CLF shots (full camera acquisition runs).

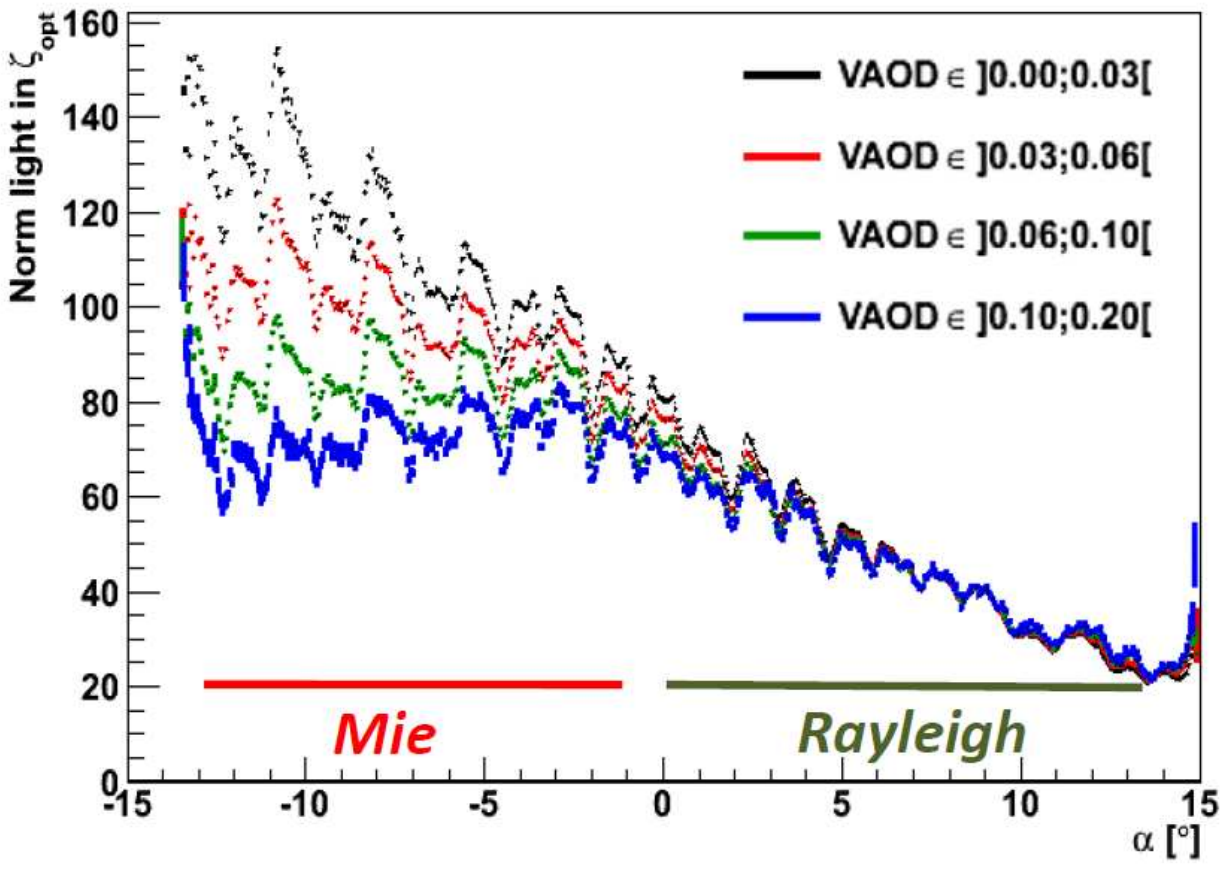

Figure 4: Average light measured at Coihueco FD site as a function of $\alpha$ for different VAOD ranges. The Mie and Rayleigh dominated regions are labeled.

The contribution to the light transverse profile from the detector was evaluated by using a full simulation of the fluorescence telescopes [2] based on Geant4. To reduce computation time maps of the optical spot at several positions on the camera were produced with this simulation. The direction of each photon at the entrance window of the telescope was smeared according to these maps.

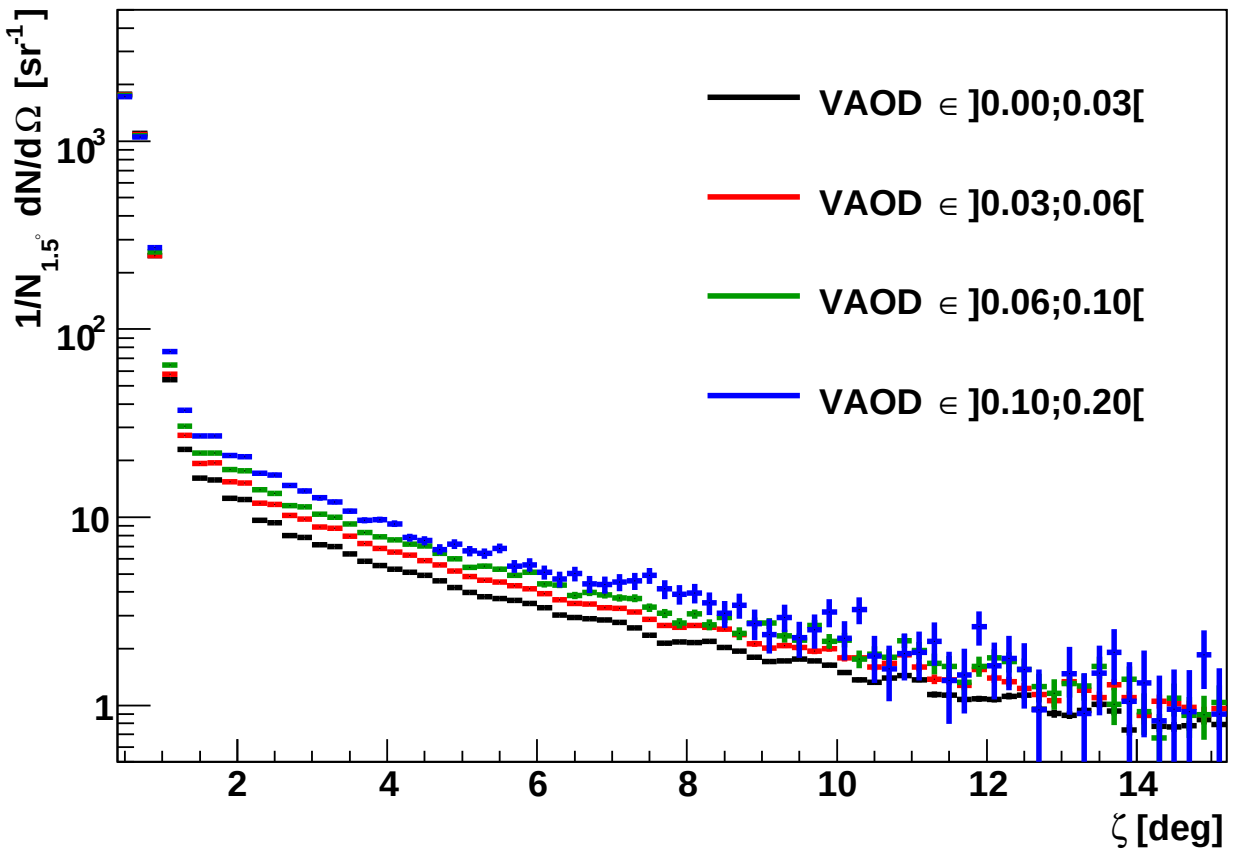

Figure 5: Average transverse light profile seen at Coihueco FD site as a function of $\zeta$ for different VAOD ranges.

## 4 Dependence on aerosol concentration

The multiple scattering processes occur in the interaction of photons with the atmosphere and therefore depend on the atmospheric conditions. In particular these processes depend on atmospheric depth but also on the quantity of aerosols. The latter is quantified in terms of the Vertical Aerosol Optical Depth profile, VAOD$(h,t)$, which is measured using CLF shots [4]. In this work the VAOD will be evaluated at a fixed reference height ($h = 3$ km, above ground level), allowing the event characterization with a single number.

To access the parameters characterizing multiple scattering processes, two distributions were considered: the total light detected by the FD as a function of $\alpha$ and the transverse light profile. The distributions were obtained with the method described in section 2, using 18 months of CLF shots recorded at the Coihueco FD site.

In figure 4 the average total light flux as a function of $\alpha$ is shown for different VAOD ranges where the expected laser attenuation dependence on the elevation angle $\alpha$ is observed. The total light flux was obtained integrating the light in $\zeta \in [0^\circ, \zeta_{\rm opt}]$, where $\zeta_{\rm opt}$ is the angle that maximizes the signal to noise ratio. The curves were normalized in the region $\alpha \in [5^\circ, 12^\circ]$ to the one with the lowest VAOD. The oscillations on the light curves are due to the FD camera non-uniformities and, since the curves result from averaging over several hundreds of thousand of laser events, the distributions show structures with high definition and small statistical errors. The normalisation and shape of the curves shows sensitivity to the aerosol content and distribution, allowing us to extract information concerning the Mie scattering process. As illustrated in figure 1, aerosols are mostly concentrated near the ground and their density can be modeled by an exponential function decreasing with a vertical scaling factor, $h_s$. Thus, photons propagating in the lower part of the atmosphere ($\lesssim h_s$) are more likely to suffer from Mie scattering than at higher altitudes. The profiles in figure 4 are in agreement with this model, where the effect of Mie scattering can be observed for low values of $\alpha$.

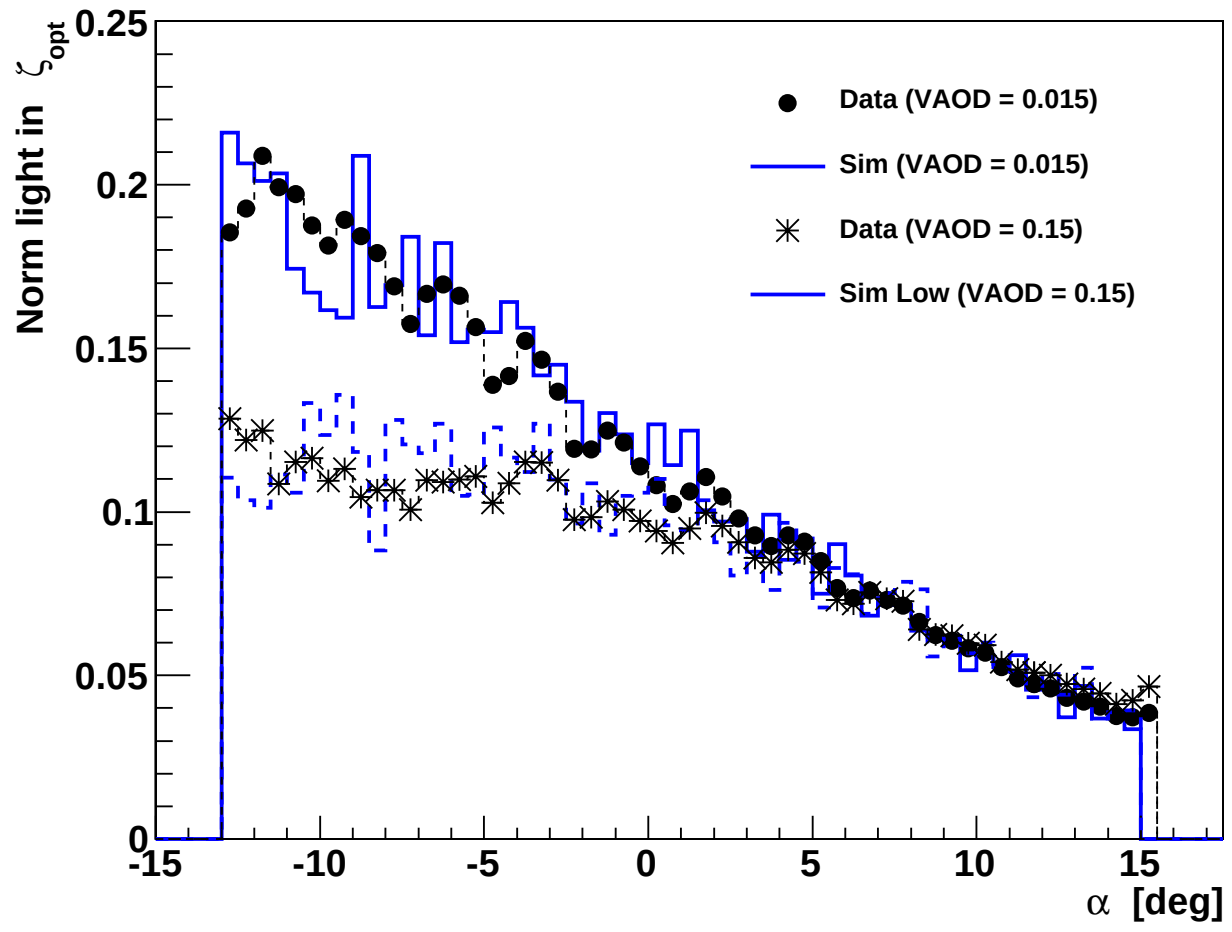

Figure 6: Average light profile seen at Coihueco FD site as a function of $\alpha$ for VAOD = $0.015$ and VAOD = $0.15$. Comparison between data and simulation.

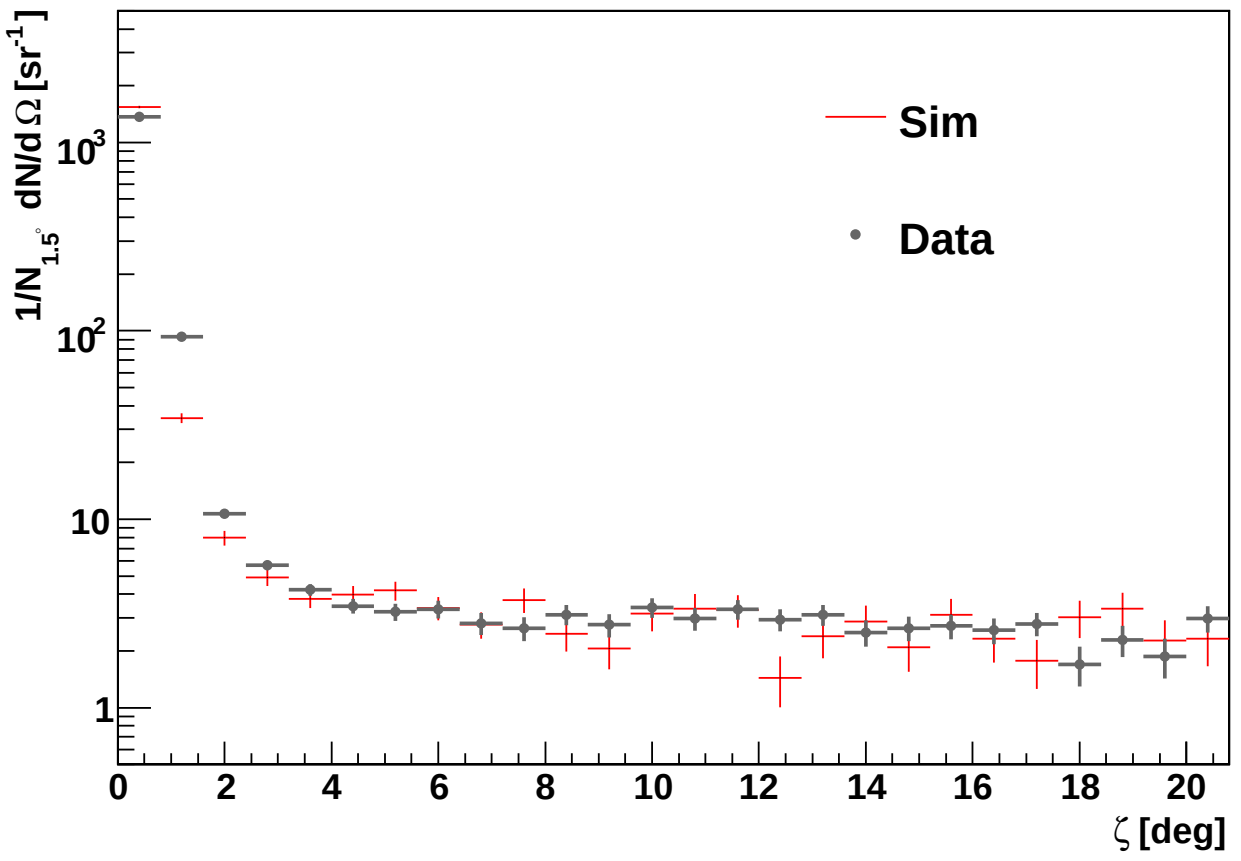

Figure 7: Average transverse light profile from CLF shots seen at FD for low VAOD. Comparison between data and simulation.

The parameters describing Mie scattering can also be constrained by the analysis of the transverse light profile. The transverse light profile distributions corresponding to different VAOD ranges are shown in figure 5. The distributions were normalized to the total signal in $\zeta < 1.5°$ ($\mathrm{N}_{1.5°}$). As described in section 2, the multiple scattering component of the signal should dominate the light transverse profile distribution for values of $\zeta$ larger than the size of the direct signal convoluted with the detector PSF ($\zeta \lesssim 1.5°$). Therefore, the dependence of the multiple scattered light component on the atmospheric conditions should be visible in the transverse profile distributions for different VAOD ranges. This is observed in figure 5, where the differences between the profiles for different VAOD ranges are clearly visible for bigger values of $\zeta$. The distributions show, as expected, that the higher the VAOD, the higher is the multiple scattering component.

The available data on the transverse light distribution from laser shots is being explored to assess the MS parameters. Both the total light profile (fig. 6) and the transverse light profile (fig. 7) show a reasonable agreement between data and simulation if the average Auger Mie parameters, which depend on the aerosol type and concentration, are used. The observed effect of higher aerosol concentrations on the light profile is well described by the simulation. For the transverse light profile, a deviation between simulation and data is observed for $\zeta < 2°$. Such effect is expected to arise from the use of a finite set of spot maps, as described in section 3.

## 5 Conclusions

A method to extract the transverse light profile using CLF laser shots was developed. The method enables the assessment of atmospheric parameters relevant for both Rayleigh and Mie scattering processes. A dedicated laser simulation based on Geant4 was developed to attain a better understanding of multiply scattered light in the atmosphere. A first comparison between data and simulation shows already a reasonable agreement. Further studies exploring the evolution of the multiple scattering component with altitude, time and distance from the FD are in progress.

# Education and Public Outreach of the Pierre Auger Observatory

GREGORY R. SNOW[1] FOR THE PIERRE AUGER COLLABORATION[2]
[1]*Department of Physics and Astronomy, University of Nebraska, Lincoln, Nebraska, USA*
[2]*Observatorio Pierre Auger, Av. San Martín Norte 304, 5613 Malargüe, Argentina*
*(Full author list: http://www.auger.org/archive/authors_2011_05.html)*
*auger_spokespersons@fnal.gov*

**Abstract:** The scale and scope of the physics studied at the Pierre Auger Observatory offer significant opportunities for original outreach work. Education, outreach and public relations of the Auger Collaboration are coordinated in a separate task whose goals are to encourage and support a wide range of education and outreach efforts that link schools and the public with the Auger scientists and the science of cosmic rays, particle physics, and associated technologies. The presentation will focus on the impact of the collaboration in Mendoza Province, Argentina. The Auger Visitor Center in Malargüe has hosted over 60,000 visitors since 2001, and a third collaboration-sponsored science fair was held on the Observatory campus in November 2010. The Rural Schools Program, which is run by Observatory staff and which brings cosmic-ray science and infrastructure improvements to remote schools, will be highlighted. Numerous online resources, video documentaries, and animations of extensive air showers have been created for wide public release. Increasingly, collaborators draw on these resources to develop Auger related displays and outreach events at their institutions and in public settings to disseminate the science and successes of the Observatory worldwide.

**Keywords:** UHECR, The Pierre Auger Observatory, education, outreach.

## 1 Introduction

Education and public outreach (EPO) have been an integral part of the Pierre Auger Observatory since its inception. The collaboration's EPO activities are organized in a separate Education and Outreach Task that was established in 1997. With the Observatory headquarters located in the remote city of Malargüe, population 23,000, early outreach activities, which included public talks, visits to schools, and courses for science teachers and students, were aimed at familiarizing the local population with the science of the Observatory and the presence of the large collaboration of international scientists in the isolated communities and countryside of Mendoza Province. As an example of the Observatory's integration into local traditions, the collaboration has participated in the annual Malargüe Day parade since 2001 with collaborators marching behind a large Auger banner (see Fig. 1). Close contact with the community fosters a sense of ownwership and being a part of our scientific mission. The Observatory's EPO efforts have been documented in previous ICRC contributions [1]. We report here highlights of recent activities.

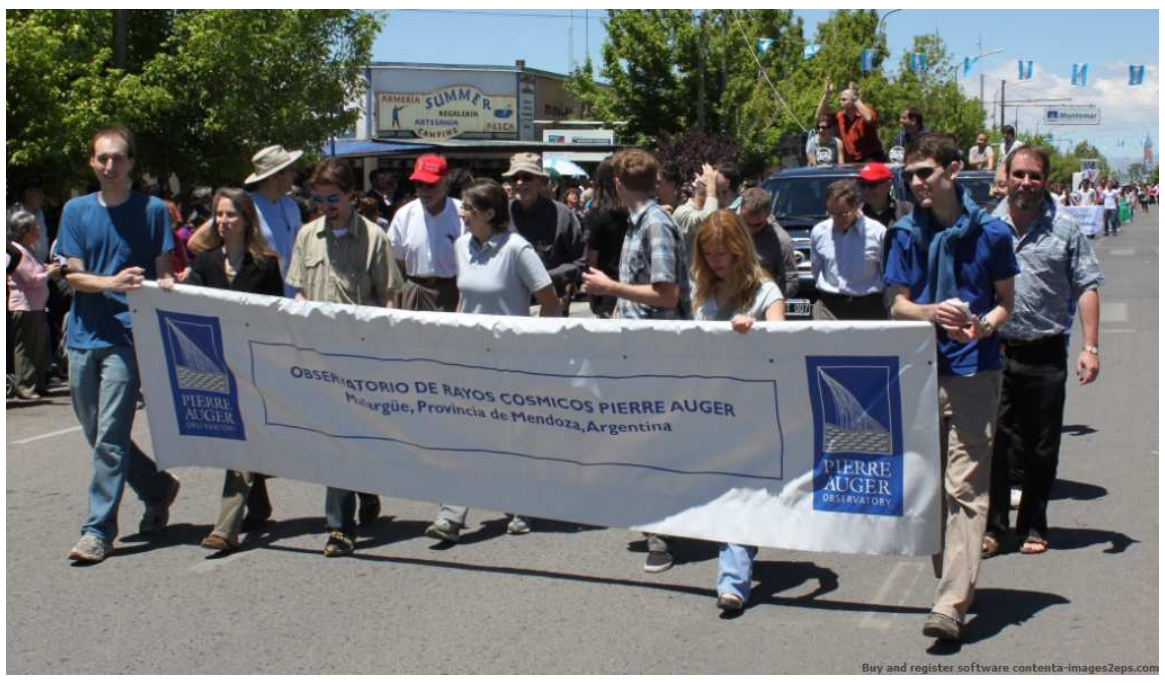

Figure 1: Auger collaborators participating in the November 2010 Malargüe Day parade.

## 2 The Auger Visitor Center in Malargüe

The Auger Visitor Center (VC), located in the central office complex and data acquisition building in Malargüe, continues to be a popular attraction. Through the end of April 2011, the VC has hosted 64,482 visitors. Fig. 2 shows the number of visitors logged per year from November 2001 through April 2011. The noticeable increase of visitors since 2008 occurred after the opening of a new, nearby planetarium [2] in August of that year. The VC is managed by a small staff led by Observatory employee Analía Cáceres; she and other collaborators share the task of giving presentations and tours to visitors and school groups. An excellent recent addition to the VC is a spark chamber, also shown in Fig. 2, produced and installed by collabo-rators from the Laboratory of Instrumentation and Experimental Particle Physics (LIP) in Portugal.

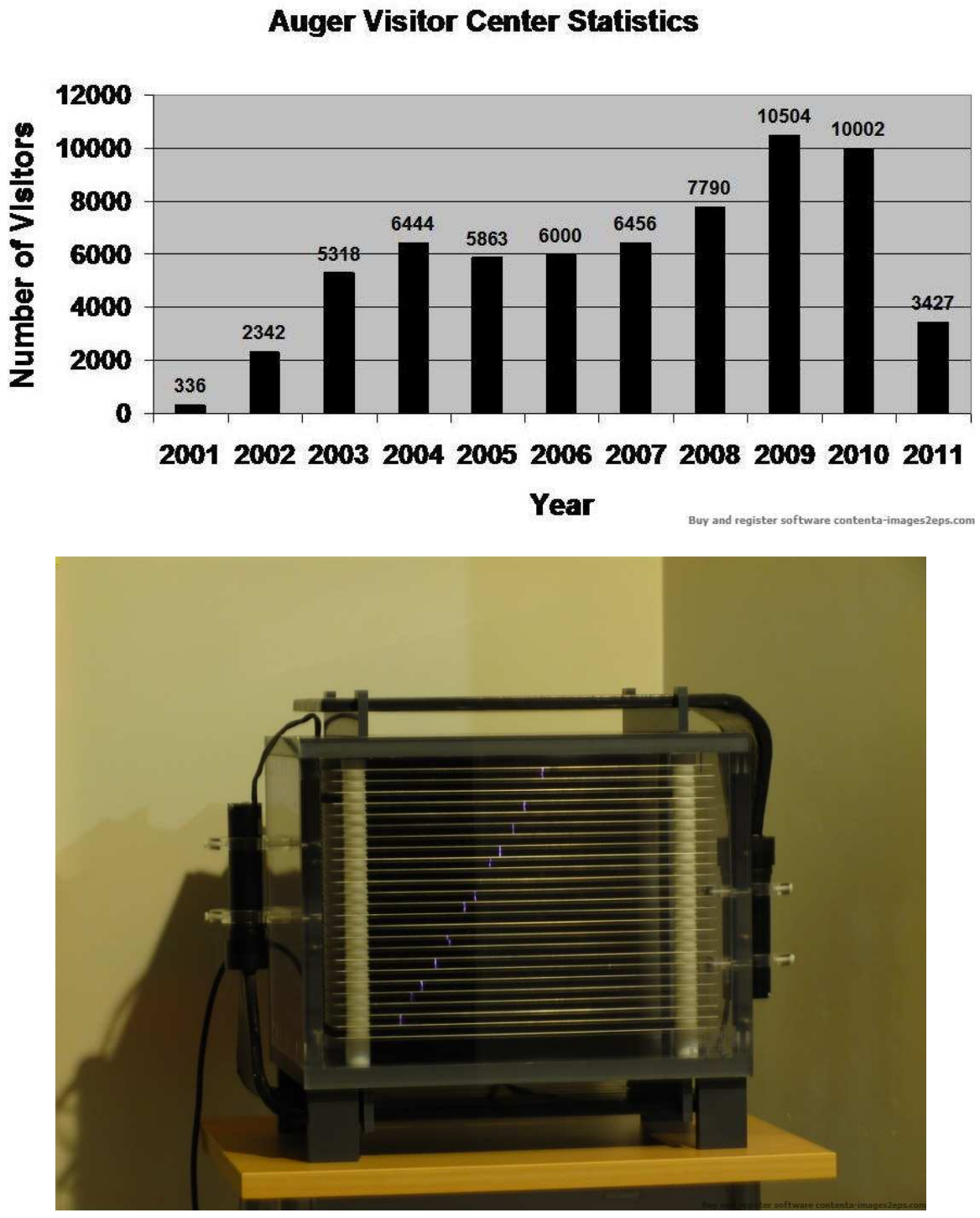

Figure 2: Top: Number of visitors logged by year at the Auger Visitor Center. Bottom: New spark chamber allows visitors to view cosmic ray muon tracks.

## 3 The Rural Schools Program and Education Fund

Many schools in the Department of Malargüe, some over 100 km away, have difficulty bringing their students to the Observatory and the Visitor Center. A dedicated team of Observatory staff members has initiated a Rural Schools Program which brings information about the Observatory and needed infrastructure improvements to remote schools. Infrastructure improvements include attention to electrical and heating systems, fences, providing school supplies, etc. The Rural Schools team consists of physicists, engineers, technicians, and office staff members who volunteer their own time. As an example, the team visited students and teachers at schools in the cities of Bardas Blancas (60 km) and Las Loicas (150 km) southwest of Malargüe on a 2-day trip in December 2010 (see Fig. 3). Analía Cáceres is shown leading a presentation and discussion with the students about cosmic ray physics and the Observatory. Other staff members are shown donating an electric heater for use in the school. The team also made personal donations which provided small Christmas gifts to the students. These efforts to reach out to the non-local community have been received with great enthusiasm. The personal contact with those involved in the Observatory has been one of the best features of the school visits. Financial support for this ongoing program is provided by contributions from collaborating institutions and generous individuals to an Education Fund managed by the Observatory staff.

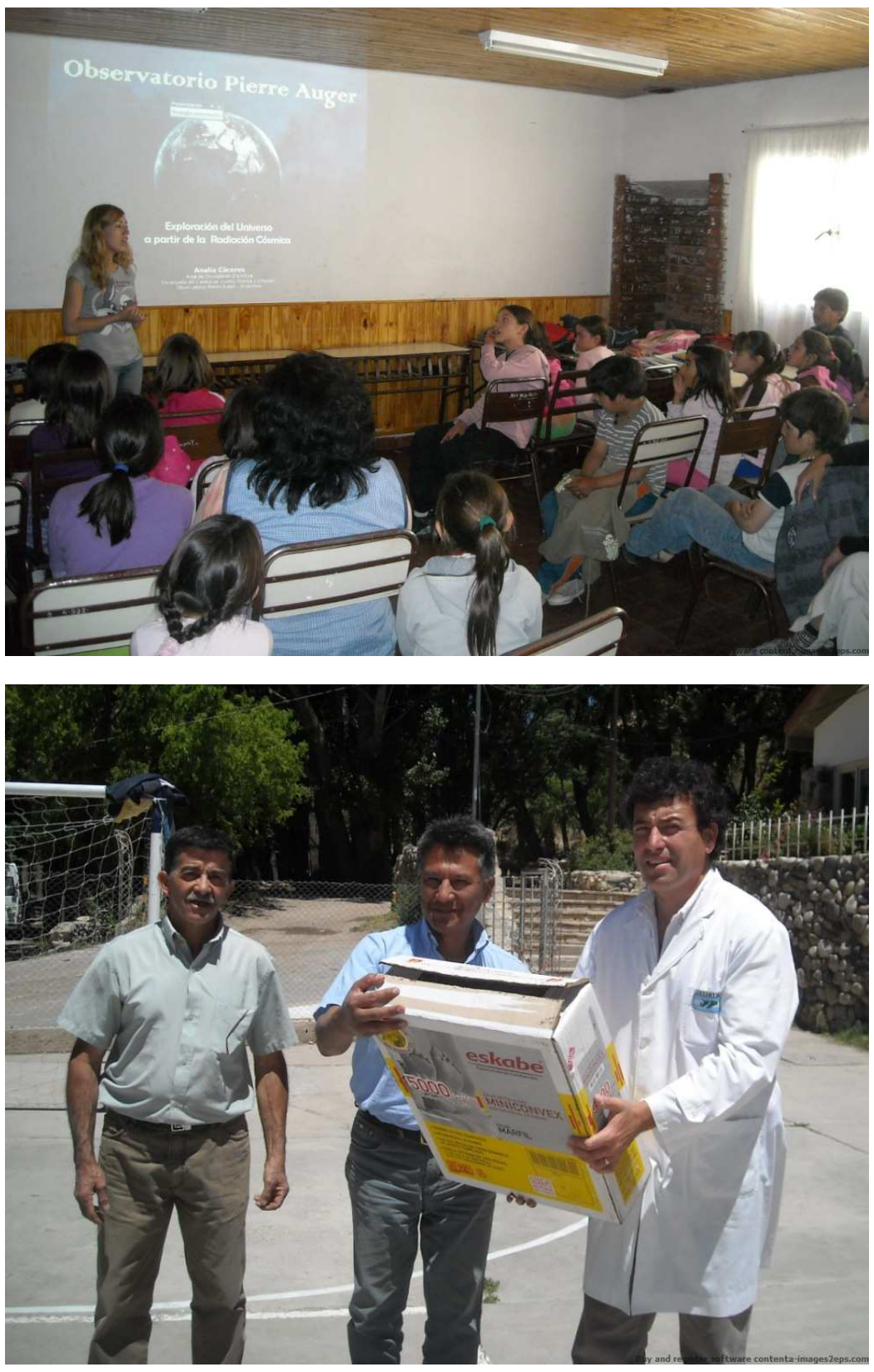

Figure 3: Top: Analía Cáceres talking to Bardas Blancas students about the Observatory, December 2010. Bottom: Observatory staff members Mario Rodríguez (left) and Gualberto Ávila (center) presenting an electric heater for use in the school.

## 4 The 2010 Auger Science Fair

Following successful Science Fairs held in November 2005 and 2007, the Collaboration sponsored a third Fair on November 19-21, 2010, that attracted the exhibition of 22 science projects in the areas of natural science and technology (see Fig. 4). This Science Fair featured the youngest group of students to date (4 first graders), as well as the team which traveled the longest distance to date to participate (from the Province of Catamarca, over 800 km north of Malargüe). The 2010 Fair spanned 3 days, which allowed participants to enjoy more activities than in previous years – a presentation in the Auger Visitor Center, a visit to Malargüe's planetarium, and an evening dinner with live music and dancing.

A team of 15 Auger collaborators judged the projects on the basis of science content, oral and visual presentation, and the written report that accompanied each project. First and second prizes were awarded in four age categories (grades 1-3, 4-7, 8-9, and 10-12), and new award categories were introduced this year: the most innovative project and a prize for the teacher whose students received the highest score. A new Challenge Cup trophy was presented to the highest-ranked team to be displayed at their school, with the expectation that a team from this school will return to the next Science Fair, foreseen to be held in November 2012. All participants received Science Fair T-shirts, and each school received a one-year subscription to *National Geographic* magazine in Spanish. The collaboration is indebted to the Observatory staff, the local organizing team of three science teachers, and the city of Malargüe for helping to make the Science Fair a success.

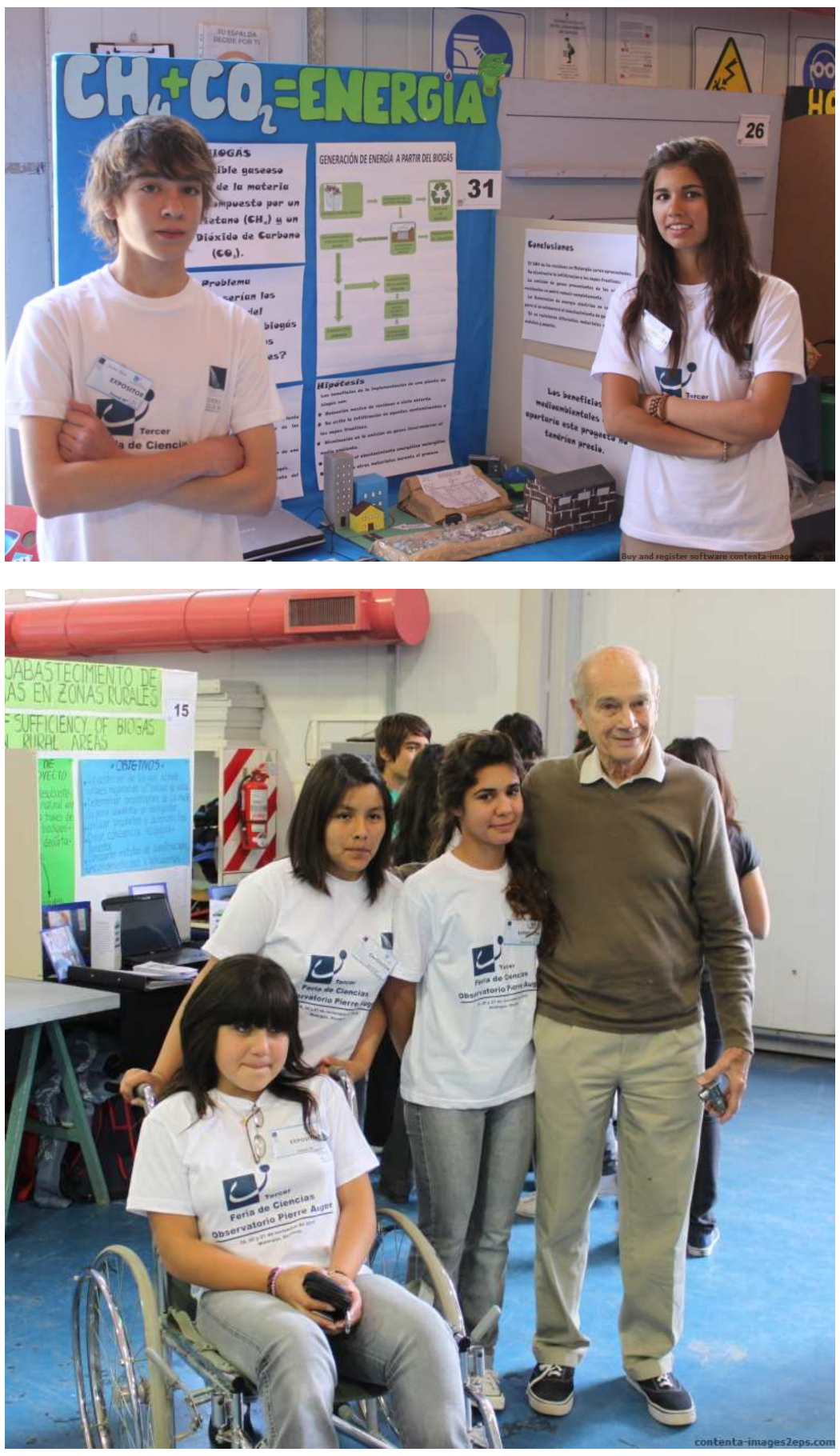

Figure 4: Participants in the 2010 Auger Science Fair, with Jim Cronin (bottom panel).

## 5 The New Auger Video Documentary

During the last year, collaborators Ingo Allekotte and Beatriz García worked with professional film producer Cristina Raschia to make a new video documentary about the Observatory entitled "Voces del Universo". The 26-minute video features interviews with collaborators from many countries speaking in their native languages, plus extensive footage of detectors and activities around the Observatory site. Several collaborating groups provided clips showing Auger activities at their home institutions. The video was produced with a choice of Spanish or English subtitles. The first public viewing of the documentary was held in Malargüe during the March 2011 collaboration meeting, and it was made available to the collaboration for dissemination in DVD format. Fig. 5 shows two still images from the video.

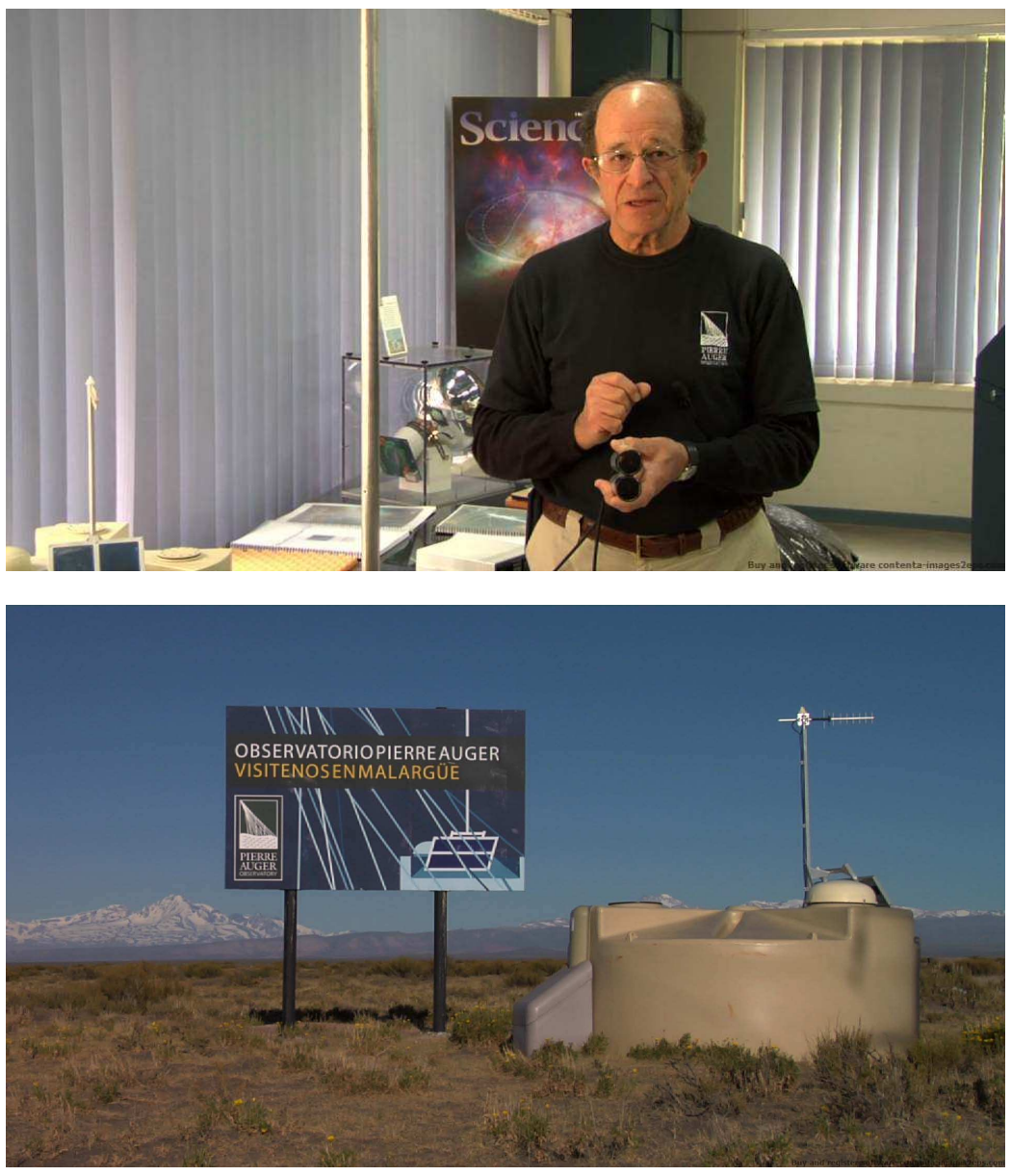

Figure 5: Sample still images from the "Voces del Universo" video. Top: Dr. Carlos Hojvat from Fermilab explaining cosmic ray particles detected with a Geiger counter. Bottom: A surface detector station and signage along Route 40 in Mendoza Province.

## 6 Other Outreach Activities

The Observatory was a co-sponsor of the 95$^{th}$ meeting of the Argentinean Physics Association held in Malargüe September 28-October 1, 2010, with several Auger collaborators serving on the Organizing Committee. Observatory staff offered tours of the Observatory campus and the fluorescence detector and HEAT installations at the Coihueco site to the 700 meeting participants.

The online release of extensive air shower data [3] for instructional purposes continues to draw attention from around the world. Between May 1, 2010, and April 30, 2011, the web site hosting the growing data set had 3920 unique visitors from 81 countries, although the bulk of the traffic is from the U.S., Argentina, and a few European countries. A number of collaborators are working on lesson plans to instruct teachers and students in the use of the data set, suggesting various plots students can make and hints on interpreting, for example, energy and arrival direction distributions.

Randy Landsberg and colleagues at the Kavli Institute for Cosmological Physics have posted high-resolution, interactive panoramas of a number of locations at the Observatory on the *Gigapan* online system [4]. Views from the Los Leones fluorescence detector building, at the central campus office building, and on the Pampa Amarilla are available.

The scholarship program which brings top Malargüe students to Michigan Technical University (MTU) has enjoyed continued success. The fifth student enrolled at MTU in the fall semester 2010. The first four students have graduated with degrees in either Mechanical Engineering or Materials Science. The first three students have embarked on engineering careers, one in the U.S. and two in Argentina.

In Madrid, Spain, the Auger Observatory and cosmic rays were featured during the Science Week held in November 2010. Visitors to the physics building at the Universidad Complutense de Madrid witnessed cosmic ray muons detected with Geiger counters and with a spark chamber like the Auger VC spark chamber shown in Fig. 2, produced by collaborators from Portugal. Visitors were filmed as a muon track passed through their hand held above the chamber. A selected still photo was copied into a take-home certificate stamped with the time and location, giving each visitor his or her personal cosmic ray to remember.

Collaborators in the Netherlands remain active in education and outreach via projects like the HiSPARC network [5], which employs high-school based scintillator detectors on school rooftops, and cosmic ray displays at open house events at NIKHEF, the National Institute for Nuclear and High Energy Physics in Amsterdam. Notably, a Dutch team that included Auger collaborators was awarded the 2008-09 Annual Academic Prize for the best translation of scientific research to the public. The generous award funded the construction of a large "Cosmic Sensation" dome (Fig. 6) [6] in a park near the Radboud University in Nijmegen. Lights and music in the dome were triggered by random cosmic ray hits detected by scintillators. Music and dance evenings held on September 30-October 2, 2010, attracted thousands of visitors and extensive press coverage.

A highlight of recent outreach activities in the Czech Republic was the "Science for the World" exhibition, September-October, 2009, which featured research activities and results from the institutes of the Academy of Sciences. Fig. 7 shows a segment of a Czech-made fluorescence detector mirror which was part of an Auger exhibit in the main building of the Academy. The exhibit attracted over 15,000 visitors and media coverage. High school students have also had the opportunity to participate in research experiences with Auger collaborators, drawing on the Auger data set released to the public, as part of the "Open Science" initiative that involves several Academy institutes and universities.

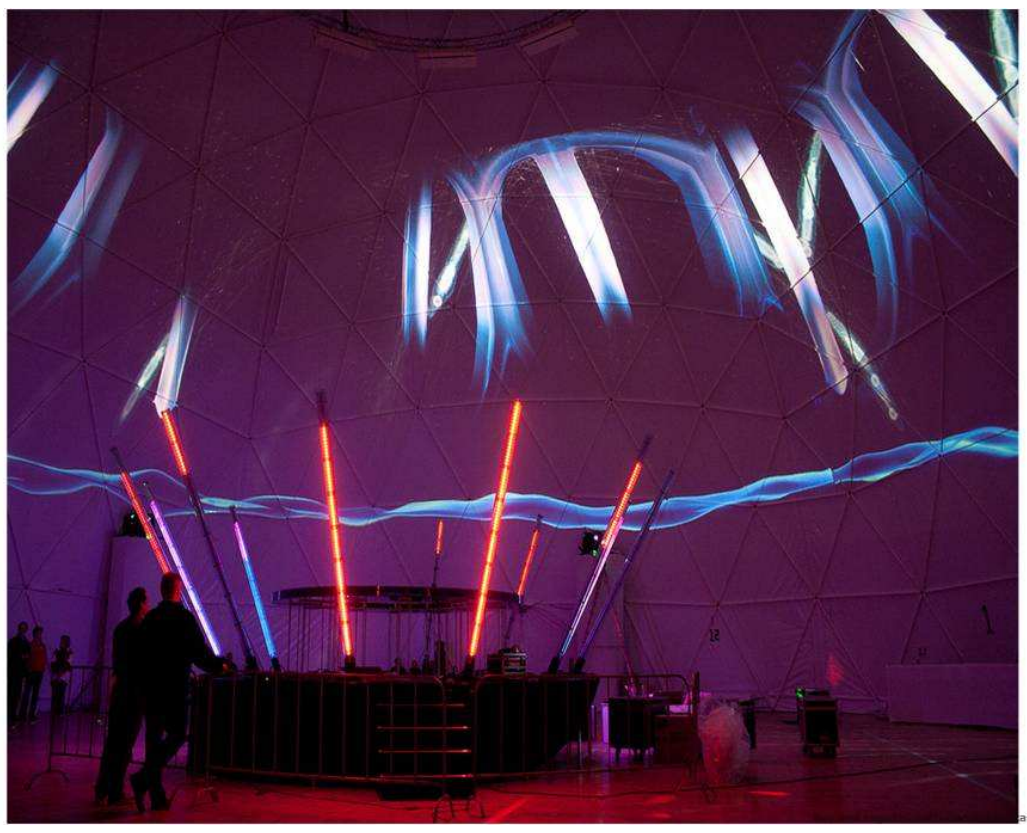

Figure 6: The Cosmic Sensation dome in Nijmegen, NL.

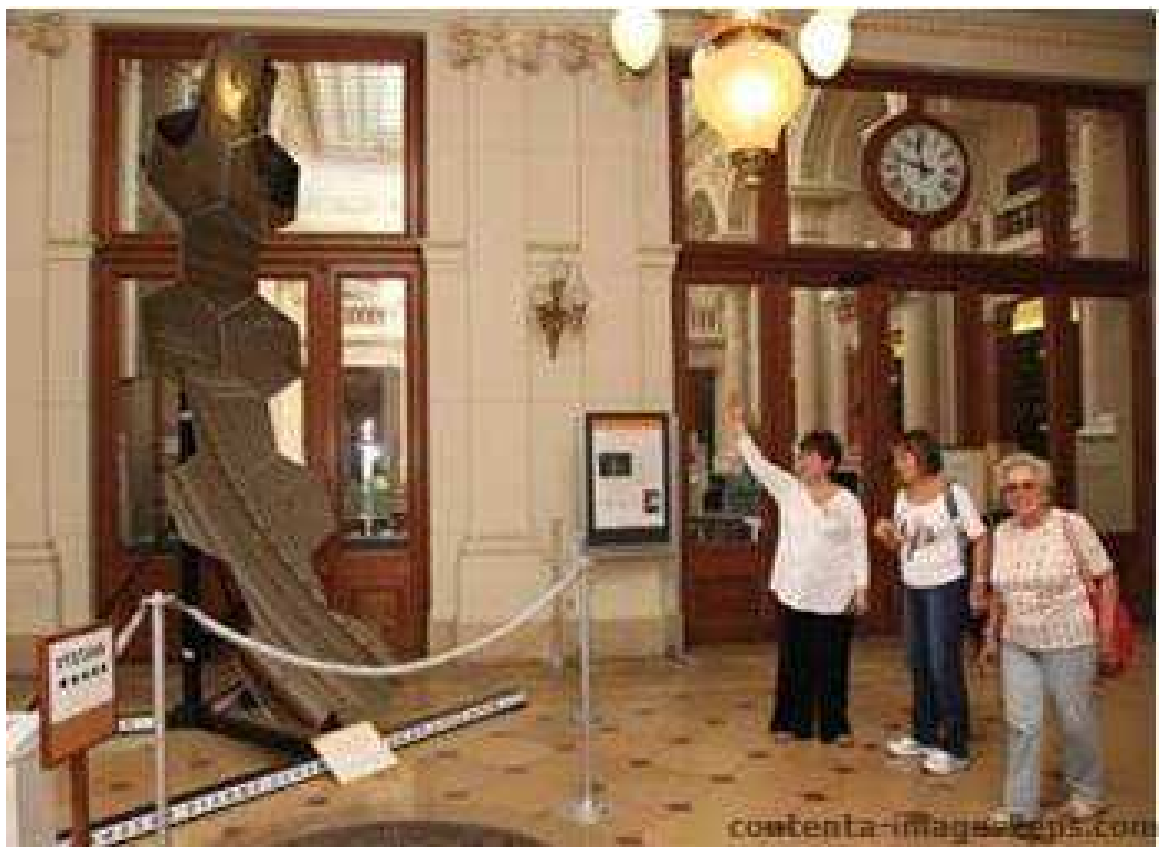

Figure 7: One-fifth size fluorescence detector mirror prototype on display at the Academy of Sciences in Prague.

# Acknowledgments

The successful installation, commissioning and operation of the Pierre Auger Observatory would not have been possible without the strong commitment and effort from the technical and administrative staff in Malargüe.

We are very grateful to the following agencies and organizations for financial support:

Comisión Nacional de Energía Atómica, Fundación Antorchas, Gobierno De La Provincia de Mendoza, Municipalidad de Malargüe, NDM Holdings and Valle Las Leñas, in gratitude for their continuing cooperation over land access, Argentina; the Australian Research Council; Conselho Nacional de Desenvolvimento Científico e Tecnológico (CNPq), Financiadora de Estudos e Projetos (FINEP), Fundação de Amparo à Pesquisa do Estado de Rio de Janeiro (FAPERJ), Fundação de Amparo à Pesquisa do Estado de São Paulo (FAPESP), Ministério de Ciência e Tecnologia (MCT), Brazil; AVCR AV0Z10100502 and AV0Z10100522, GAAV KJB100100904, MSMT-CR LA08016, LC527, 1M06002, and MSM0021620859, Czech Republic; Centre de Calcul IN2P3/CNRS, Centre National de la Recherche Scientifique (CNRS), Conseil Régional Ile-de-France, Département Physique Nucléaire et Corpusculaire (PNC-IN2P3/CNRS), Département Sciences de l'Univers (SDU-INSU/CNRS), France; Bundesministerium für Bildung und Forschung (BMBF), Deutsche Forschungsgemeinschaft (DFG), Finanzministerium Baden-Württemberg, Helmholtz-Gemeinschaft Deutscher Forschungszentren (HGF), Ministerium für Wissenschaft und Forschung, Nordrhein-Westfalen, Ministerium für Wissenschaft, Forschung und Kunst, Baden-Württemberg, Germany; Istituto Nazionale di Fisica Nucleare (INFN), Ministero dell'Istruzione, dell'Università e della Ricerca (MIUR), Italy; Consejo Nacional de Ciencia y Tecnología (CONACYT), Mexico; Ministerie van Onderwijs, Cultuur en Wetenschap, Nederlandse Organisatie voor Wetenschappelijk Onderzoek (NWO), Stichting voor Fundamenteel Onderzoek der Materie (FOM), Netherlands; Ministry of Science and Higher Education, Grant Nos. 1 P03 D 014 30, N202 090 31/0623, and PAP/218/2006, Poland; Fundação para a Ciência e a Tecnologia, Portugal; Ministry for Higher Education, Science, and Technology, Slovenian Research Agency, Slovenia; Comunidad de Madrid, Consejería de Educación de la Comunidad de Castilla La Mancha, FEDER funds, Ministerio de Ciencia e Innovación and Consolider-Ingenio 2010 (CPAN), Xunta de Galicia, Spain; Science and Technology Facilities Council, United Kingdom; Department of Energy, Contract Nos. DE-AC02-07CH11359, DE-FR02-04ER41300, National Science Foundation, Grant No. 0450696, The Grainger Foundation USA; ALFA-EC / HELEN, European Union 6th Framework Program, Grant No. MEIF-CT-2005-025057, European Union 7th Framework Program, Grant No. PIEF-GA-2008-220240, and UNESCO.